%% file: asymptotic_stability_article.tex
\author{Gideon Simpson and Michael I.~Weinstein}
\title{Asymptotic Stability of Ascending Solitary Magma Waves}
\documentclass[10pt,final]{article}
\usepackage{amssymb}
\usepackage{amsfonts}
\usepackage{amsmath}
\usepackage{graphicx}
\usepackage{cite}

\numberwithin{equation}{section}
\include{grsmacros}

\begin{document}

\maketitle
\centerline{\small{Department of Applied Physics and Applied Mathematics, Columbia University, New York, NY 10027}}
\centerline{Addresses for correspondence: {grs2103@columbia.edu}, miw2103@columbia.edu}
\begin{abstract}
Coherent structures, such as solitary waves, appear in many physical problems, including fluid mechanics, optics, quantum physics, and plasma physics.  A less studied setting is found in geophysics, where highly viscous fluids couple to evolving material parameters to model partially molten rock, magma, in the Earth's interior.  Solitary waves are also found here, but the equations lack useful mathematical structures such as an inverse scattering transform or even a variational formulation.

A common question in all of these applications is whether or not these structures are stable to perturbation.  We prove that the solitary waves in this Earth science setting are asymptotically stable and accomplish this without any pre-exisiting Lyapunov stability.  This holds true for a family of equations, extending beyond the physical parameter space.  Furthermore, this extends existing results on well-posedness to data in a neighborhood of the solitary waves.
\end{abstract}
\tableofcontents
\section{Introduction}
Coherent structures, such as solitary waves, appear in many physical problems, including fluid mechanics, optics, quantum physics, and plasma physics.  A less studied setting is found in geophysics, where highly viscous fluids couple to evolving material parameters to model partially molten rock, magma, in the Earth's interior.  Solitary waves are also found here, but the equations lack the useful structures such as an inverse scattering transform or even a variational formulation.

A important question in all of these applications is whether or not these coherent structures are stable to perturbation.  We prove that the solitary waves in this Earth science setting are asymptotically stable and accomplish this without any pre-exisiting Lyapunov stability.

\subsection{Magma--Porous Flow in a Viscously Deformable Media}
Models of magma in the Earth's interior couple Stokes flow of the viscous melt to the slow, creeping deformation of the porous rock.  These stress balance equations couple to transport equations for the volume fraction of melt, the \emph{porosity}.  Formulations may be found in \cite{McKenzie84, Scott84, Scott86, Spiegelman93a, Bercovici2001a}.  Nonlinearity appears in fluxes and through the material properties, the \emph{permeability} and \emph{viscosity} of the porous, deformable rock, which depend nonlinearly on the porosity.  Consequently, such models are known, from computations, to feature localization and generate coherent structures, see \cite{Scott84, Scott86, Barcilon86, Barcilon89, Spiegelman93a, Spiegelman93b, Aharonov95, Wiggins95, Spiegelman2001, Spiegelman03, Katz06}.  The physical assumptions and their implications will be discussed in a forthcoming review article, \cite{Simpson08}.

Under certain assumptions (small fluid fraction, absences of large-scale shear, no melting,  \emph{etc.}), such a system reduces to a single scalar equation for the porosity's evolution, \cite{Barcilon86, Barcilon89, Spiegelman93a, Spiegelman93b}.  The $d$-dimensional equation is
\begin{equation}
\label{eq:magma3d}
\dt \f + \dz\paren{\f^n}- \nabla \cdot \bracket{\f^n\nabla\paren{\f^{-m}\dt\f}}=0, \quad \bold{x}\in \R^d, t>0 
\end{equation}
with the boundary conditions that $\f(\bold{x},t) \to 1$ as $\abs{\bold{x}} \to \infty$.  $\nabla =\paren{\partial_x, \partial_y,\partial_z}$ for $d=3$ and  $\nabla =\paren{\partial_x,\partial_z}$ for $d=2$.  The nonlinearity $n$ comes from the relationship between the permeability, $K$, and the porosity of the rock, $K \propto \f^n$.  $m$  relates to the bulk viscosity, $\zeta$, to the porosity of the rock, $\zeta \propto \f^{-m}$.  In the physical regime, these exponents have values $2\leq n\leq 3$ and $0 \leq m \leq 1$, \cite{McKenzie84, Scott84, Scott86, Renner2003,Hirth95,Hirth95a,wark1998gsp,wark2003rps,zhu2003nmp}.

Equation \eqref{eq:magma3d} appears elsewhere in Earth science as a model for convective Mantle plumes.  Referred to as \emph{hot spots} at the surface, these plumes are localized regions of upwelling hot, buoyant material.  Examples include the Hawaiian Island chain and Iceland.  Modeled as the flow of a viscous fluid up a conduit embedded in a higher viscosity medium, an equation of the form \eqref{eq:magma3d} was derived in \cite{Olson86}.  There, the equation is one-dimensional ($d=1$), the exponents are $(n,m)=(2,1)$, and the depdendent variable $\phi$ is the cross-sectional area of the pipe.

\begin{figure}
\begin{center}
\includegraphics[width=3in]{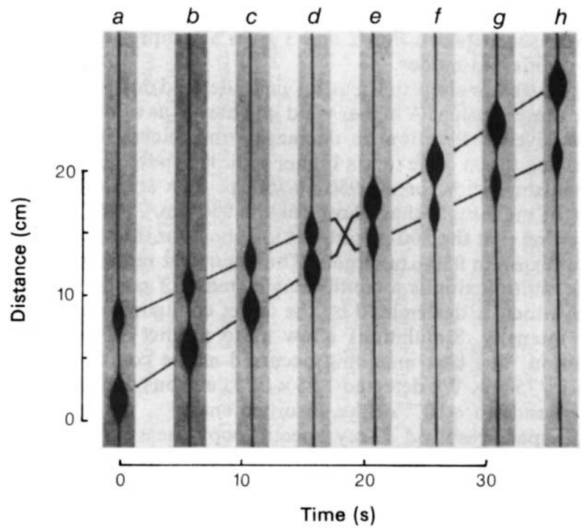}
\caption{Solitary waves colliding and propagating in an experiment with honey.  Figure 1 of   \cite{Scott86a}.}
\label{fig:conduit-flow}
\end{center}
\end{figure}

Numerical simulations of \eqref{eq:magma3d}, in one-, two-, and three-dimensions were performed in \cite{Scott84, Scott86, Barcilon86, Barcilon89, Wiggins95}, where stable, radially symmetric, \emph{solitary traveling waves} were observed.  In  \cite{Nakayama92}, it was shown that in one-dimension, solitary waves, $U_c(x-ct)$, in excess of the reference state, $\phi\equiv1$, exist if $n> 1$.  In the context of conduit flow, discussed in the preceding paragraph, analog experiments using viscous syrups appear in \cite{Olson86, Scott86a, Whitehead87}; robust solitary waves appeared as predicted, see Figure \ref{fig:conduit-flow}. 

\subsection{Stability of Solitary Waves}
We consider \eqref{eq:magma3d} in one-dimension,
\begin{equation}
\label{eq:magma}
\dt\f + \dx\paren{\f^n} - \dx\bracket{\f^n\dx\paren{\f^{-m}\dt\f}}=0, \quad \lim_{\abs{x}\to\infty} \phi(x,t) = 1
\end{equation}
where the $z$ coordinate has been relabeled $x$.  A cursory explanation for the solitary waves' stability may be found in \cite{Whitehead86}.  Under a small amplitude scaling, \eqref{eq:magma} is, to leading order, governed by the Korteweg-de~Vries (KdV) equation.  Since KdV solitons are stable, on a timescale for which KdV approximates \eqref{eq:magma} its solitary waves should also be stable.  

Based on observations of numerical experiments, we expect a slightly perturbed solitary wave to evolve into another wave with similar amplitude and phase.  It will be accompanied by some small amplitude dispersive waves and, perhaps, another solitary wave of smaller amplitude.  The leading wave will \emph{outrun} these other disturbances, cease interacting with them, and stabilize.  

\begin{figure}[t]
\begin{center}
$\begin{array}{cc}
\includegraphics[height=1.8in]{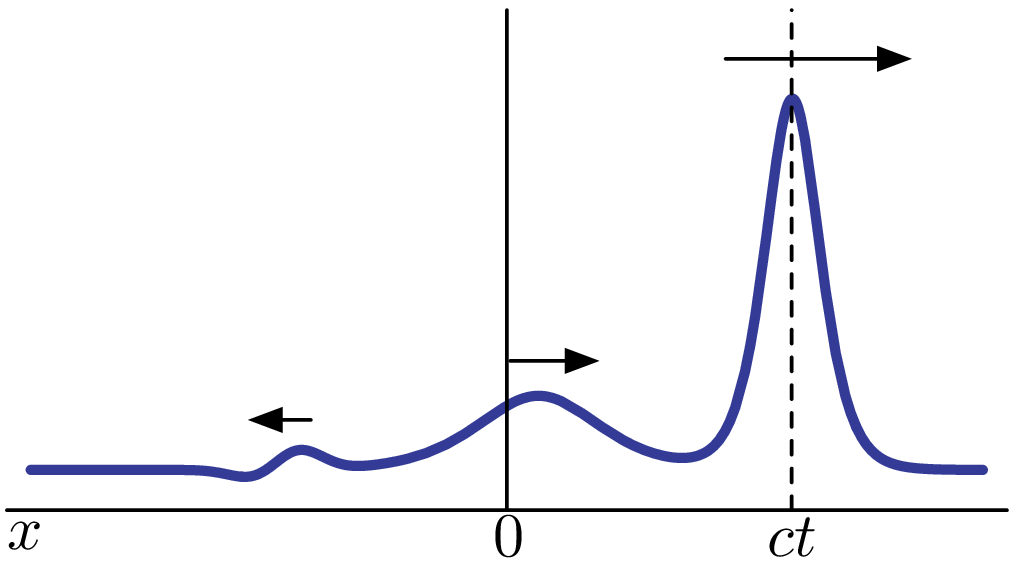}&\includegraphics[height=1.8in]{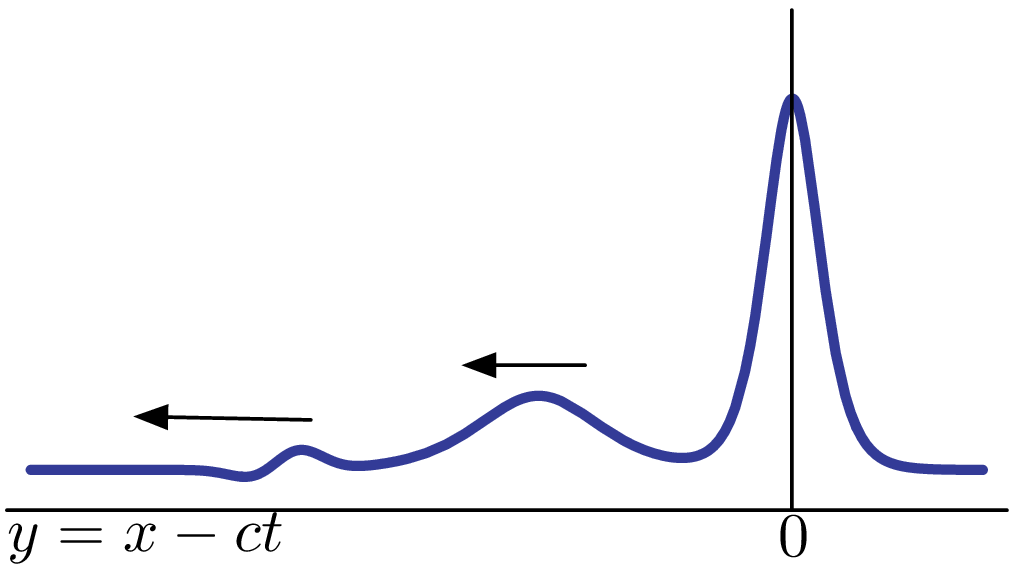}\\
\mbox{\textbf{(a)}}&\mbox{\textbf{(b)}}
\end{array}$
\caption{The largest solitary wave in \textbf{(a)} travels faster than smaller solitary waves and dispersive waves behind it.  In the frame of this wave, the rest of the solution appears to move leftward, as in \textbf{(b)}.}
\label{fig:perturbed-wave}
\end{center}
\end{figure}

Some intuition for this stability may be found in two properties.  First, taller solitary waves travel with greater speed, $c$, than smaller ones.  In the frame of the largest solitary wave, $y= x -c t$, the other waves travel leftwards.  Second, in the frame of the leading solitary wave, small perturbations of the reference state, $\phi(x,t)=1 + \psi(x-ct,t)$ and $\abs{\psi}\ll1$, evolve under the linear flow
\[
\dt\psi - c \dy \psi + n \dy \psi - \dy^2\dt \psi +c\dy^3 \psi =0
\]
The dispersion relation and group velocity are
\begin{equation*}
\omega(k) = \frac{nk}{1+k^2}-ck,\quad  \omega'(k) = n\frac{1-k^2}{(1+k^2)^2}-c
\end{equation*}
Since both phase and group velocities are negative for all $k$, small dispersive waves also travel leftwards.  These two mechanisms are diagrammed in Figure \ref{fig:perturbed-wave}, and we will exploit them to prove the main theorems.  

As the system is conservative, perturbations, such as a small solitary wave, will not vanish in a translation invariant norm.  A suitable norm will register leftward motion as decay.  We will use exponentially weighted norms, in the frame of the leading solitary wave.  These norms are defined in Section \ref{sec:notation}.

Several paths to proving stability are available.  One method is to seek constants of motion that can be combined into a metric centered at the solitary wave.  Since the metric is time independent, if the perturbation is initially small, it will remain so.  This elegant method relies on the calculus of variations and for equations such as KdV and Nonlinear Schr\"odinger (NLS) it may be used to prove \emph{orbital stability}, \cite{Benjamin72a, Bona75, Weinstein85, Weinstein86}.  A solitary wave, $U_c$, is said to be orbitally stable if for data sufficiently close to it
\[
\inf_{y\in\R} \norm{u(t)-U_c(\cdot+y)}_X <\delta\quad\mbox{for all $t$}
\]
for some $\delta>0$ and an appropriate norm $\norm{\cdot}_X$.  Typically, the norm is equivalent to $L^2(\R)$ or $H^1(\R)$.

However, in general, \eqref{eq:magma} lacks a sufficient number of conservation laws for this approach.  Indeed, in \cite{Barcilon86}, the authors searched for an additional conservation law, in hopes of proving orbital stability, \cite{Bona07}.  There are many such equations, including some of the Boussinesq systems, \cite{Pego97}, and many of the ``compacton" equations, \cite{Rosenau06a}, which also lack such structure, yet appear, in numerical experiments, to possess stable solitary waves.   Note that when $n+m=0$, \eqref{eq:magma} is Hamiltonian, and we have investigated this, proving orbital stability in \cite{Simpson07a}.  We wish to consider the general case, which includes the physically interesting cases $(n,m)=(2,1)$ and $(3,0)$.

Another approach to stability is to linearize the problem $u_t = N(u,u_x,u_{xx},\ldots)$ about a solitary wave and establish linear stability.  Then, one seeks a way to perturbatively ``boost" this to prove stability for the nonlinear flow.  This may rely on direct spectral analysis of the linearized evolution operator.  We employ this method, following \cite{Pego94, Miller96, Pego97}.  Through this, we prove that the solitary waves are \emph{asymptotically stable}, our main result.  By this we mean, that in an appropriate norm, $\norm{\cdot}_Y$,
\[
\norm{u(t)-U_c}_Y\to 0\quad\mbox{as $t\to +\infty$}
\]
for data sufficiently close to $U_c$.  

We note that another method recently appeared in \cite{Martel05}.  Without linearizing, the authors employ a virial inequality to directly prove asymptotic stability of gKdV solitary waves.

Our problem is an example of an equation for which one can prove asymptotic stability in the absence of orbital stability. Upon reflection, it is clear that the asymptotic stability of generalized KdV and BBM solitary waves could have been proven without using the orbital stability results.

\subsection{Main Results and Outline}
The main results are:
\begin{thm}
\label{thm:main}
There exists $c_\star>n$ such that for all $\co \in (n, c_\star]$, if $\f_\co(x-\theta_0)$ is a solitary wave solution of \eqref{eq:magma}, then there exist constants $K_\ast>0$, $\eps_\ast >0$ and $a>0$, such that for $\eps \leq \eps_\ast$, if  
\[
\norm{v_0}_\h{1} + \norm{e^{ax}v_0}_{H^1}\leq\eps
\]
then
\begin{description}
  \item[\textbf{(a)}] \eqref{eq:magma} has a solution with data $\f_0(x) = \f_\co(x+\theta_0)+v_0(x)$ for all time.
  \item[\textbf{(b)}] There exist $c_\infty$, $\theta_\infty$, $K_\ast$ and $\kappa>0$ such that
  \begin{align}
  \norm{\f(\cdot,t)-\f_{c_{\infty}}(\cdot-c_{\infty} t +\theta_\infty)} _\h{1} &\leq K_\ast \eps \\
  \norm{e^{ax}\bracket{\f(\cdot+ c_\infty t - \theta_\infty,t)-\f_{c_\infty}(\cdot)}}_{H^1} &\leq K_\ast \eps e^{-\kappa t}\\
  \abs{c_\infty-\co}+\abs{\theta_\infty-\theta_0}&\leq K_\ast \eps
  \end{align}
\end{description}
\end{thm}

\begin{cor}
\label{cor:mainhamiltonian}
Let $n+m=0$.  If $\partial_c \N[\f_c]>0$,  $\N$ defined in \eqref{eq:energy}, then Theorem \ref{thm:main} holds for all $c> c_\star$, except for a discrete set with no accumulation point.
\end{cor}

\begin{rem}
Theorem \ref{thm:main} and Corollary \ref{cor:mainhamiltonian} may be extended for all $n>1$ and $c>c_\star$ up to the acceptance of a time independent numerical computation.
\end{rem}

The feature of \eqref{eq:magma} that allows us to prove nonlinear stability from the linear stability is a non-negative invariant, denoted $\N[\f]$, and defined in \eqref{eq:energy}.  Taylor expanding $\N$ about a solitary wave, 
\[
\N[\f_c + v] = \N[\f_c] + \inner{\delta\N[\f_c]}{v} + \inner{\delta^2\N[\f_c]v}{v}+ O\paren{\norm{v}_\h{1}^3}
\]
The first variation does not vanish and the second variation is not a positive definite quadratic form.  However, in the frame of the solitary wave, the perturbation $v$ is migrating to $-\infty$.  Therefore
\[
\inner{\delta\N[\f_c]}{v}  \to 0 \quad\mbox{as $t\to +\infty$}
\]
The second variation may be decomposed as  $\delta^2\N[\f_c]= P+Q$, $P$ a positive quadratic form and $Q=Q(x)$ a localized function.  Then since the perturbation moves leftward
\begin{align*}
\inner{P v}{v} &\geq \kappa^2 \inner{v}{v}\\
\inner{Q v}{v} &\to 0\quad\mbox{as $t\to +\infty$}
\end{align*}
Asymptotically, 
\[
\norm{v}_\h{1} \leq K \Delta\N 
\]
giving a Lyapunov type bound on the perturbation.  

However, more is needed to formalize this into a proof, notably a sense in which the perturbation recedes from the solitary wave.  This is accomplished by analyzing the spectrum of the linearized evolution operator in a weighted space, in which the perturbation will decay.

The plan of the proof is as follows
\begin{description}
  \item[(I)]  In Section \ref{sec:prelim} we review properties of \eqref{eq:magma} and establish regularity properties of the solitary waves.
  \item[(II)] In Section \ref{sec:spectrum}, we prove that the linearized operator, $A_{a}$, has the property that there exists $\vareps>0$ such that
  \[
  \sigma\paren{A_{a}} \cap \left\{ \Re \lambda \geq -\vareps  \right\} = \{ 0 \}
  \]
 and zero is an eigenvalue of algebraic multiplicity two.
  \item[(III)] In Section \ref{sec:semigroup}, we prove  
  \[
  \norm{w(t)}_\h{1}=\norm{e^{A_{a}t} w_0}_\h{1} \leq K e^{-bt} \norm{ w_0}_\h{1}
  \]
  for appropriate $w_0$, $K$ and $b$ positive constants.
  \item[(IV)] In Section \ref{sec:estimates}, we make several estimates, including a formalization of the Lyapunov bound.  We also formulate equations for the speed and phase parameters of the solitary wave $(c(t), \theta(t))$, coupling them to the infinite dimensional system for the perturbation.
  \item[(V)] In Section \ref{sec:mainproof}, we prove  the main results, asymptotic stability and global existence of data near a solitary wave solution.
\end{description}
Some remarks are made in Section \ref{sec:discussion}, and additional details are located in the Appendices.
\subsection{Acknowledgments}
\label{sec:ack}
We thank Marc Spiegelman for his helpful comments and support, in addition to his contributions appearing in \cite{Simpson07}.  We have also benefitted from discussions with Professor J.L.~Bona and Professor P.~Rosenau.

This work was funded in part by the US National Science Foundation (NSF) Collaboration in Mathematical Geosciences (CMG), Division of Mathematical Sciences (DMS), Grant DMS--05--30853, the NSF Integrative Graduate Education and Research Traineeship (IGERT) Grant DGE--02--21041, NSF Grants DMS--04--12305 and DMS--07--07850

\subsection{Notation}
\label{sec:notation}

Generic constants will typically be denoted by the capital letters $K$, $M$, and $N$, sometimes with tildes, overlines, or primes.  Subscripts, such as $M_\gamma$, may appear to indicate that $M$ depends on $\gamma$.  We avoid using $C$ as a generic constant, as $c$ appears throughout the paper as the speed parameter, and an operator $C(\lambda)$ appears in Section \ref{sec:spectrum}.

Functions will typically live in spaces $H^k(\R)=W^{1,k}(\R)$, $k$ a non-negative integer, the spaces of square integrable functions with square integrable (weak) derivatives up to order $k$.  We will frequently omit writing $\R$.   The $L^p(\R)$ spaces will also appear, in particular $L^2$ and $L^\infty$.  While we write
$\norm{f}_\h{k}$ for the norm of a function in $H^k$, we only write $\norm{f}_{p}$
for the norm of a function in $L^p$.

We will be interested in functions in the exponentially weighted space,
\[
H^k_a = \left \{u : e^{ax} u(x) \in H^k  \right \}
\]
for $k=0,1,2,\ldots$, and $a>0$ with associated norm 
\[
\norm{u}_{H^k_a} = \norm{e^{ax}u}_{H^k}
\]
We also define the norm
\[
\norm{f}_\hinta \equiv\norm{f}_\h{1} + \norm{f}_{H^1_a}
\]
The exponential weight will \emph{always} be a positive number; we will often omit the assumption $a>0$ in statements.

Frequently, we will have an operator $T$ defined on a weighted space, $H^k_a$, but wish to make computations in the unweighted space.  To $T$ we associate $T_a = e^{ax} T e^{-ax}$, an operator on $H^k$.  For the differentiation operator, $\dx \mapsto D_a = \dx - a$.


\section{Preliminaries}
\label{sec:prelim}
\subsection{Properties of the Equation in a Weighted Space}
\label{sec:local}
Much of the analysis involves studying \eqref{eq:magma} in an exponentially weighted space.  We therefore state the following extension of the well-posedness results obtained in \cite{Simpson07} for $\h{k}$ spaces:

\begin{thm}(Local Existence in Time \& Continuous Dependence Upon Data)
\label{thm:local}

Given $0<a<1$, let $\f_0(x)$ satsify,
\begin{gather}
\norm{\f_0-1}_ \hinta \leq R< \infty\\
\label{eq:local1}
\inf_x \f_0(x) \geq \alpha_0 >0\\
\label{eq:local2}
\inf_x \f_0(x)^{m}- a^2 \f_0(x)^n \geq \beta_0> 0
\end{gather}
Then there exists $T_\local>0$ and $\f(x,t)-1 \in C^1([0,T_\local),H^1 \cap H_a^1)$, a solution of \eqref{eq:magma} with data $\f_0$, satisfying
\begin{gather}
\norm{\f(\cdot,t)-1}_ \hinta \leq  2 R\\
\inf_x\f(x,t) \geq\frac{1}{2}\alpha_0 \\
\inf_x \f(x,t)^{m}- a^2 \f(x,t)^n \geq \frac{1}{2}\beta_0
\end{gather}
for $t< T_\local$.

Moreover, there is a maximal time of existence $T_{\max}$, such that if $T_{\max}<\infty$, then
\begin{equation}
\lim_{t\to T_{\max}} \norm{\f(\cdot,t)-1}_ \hinta  + \left\|{\frac{1}{\f(\cdot,t)}}\right\|_\infty + \left\|{\frac{1}{\f(\cdot,t)^{m}-a^2\f(\cdot,t)^{n}}}\right\|_\infty = \infty
\label{eq:blowup}
\end{equation} 
\end{thm}

\begin{rem}
When $a=0$, \eqref{eq:local2} is unnecessary; this case was treated in \cite{Simpson07}.  The importance of this condition for $a>0$ will be discussed in Section \ref{sec:weightremarks}.
\end{rem}

\begin{thm}
\label{thm:continuous}
Given $0<a<1$, $\f^{(j)}-1\in C^1([0,T]; H^1 \cap H_a^1)$, $j=1,2$ be two solutions of \eqref{eq:magma} such that 
\begin{align*}
\norm{\f^{(j)}(\cdot,t)-1}_ \hinta  &\leq R< \infty\\
\inf_x \f^{(j)}(x,t) &\geq \alpha_0>0\\
\inf_x (\f(x,t)^{(j)})^{m}-a^2 (\f(x,t)^{(j)})^{n} &\geq  \beta_0>0
\end{align*}
There exists a constant $K=K(R,\alpha_0,\beta_0, a)$, such that
\[
\norm{\f^{(1)}(\cdot,t)-\f^{(2)}(\cdot,t)}_ \hinta\leq e^{Kt} \norm{\f^{(1)}_0-\f^{(2)}_0}_ \hinta\quad\text{for $t\leq T$.}
\]
\end{thm}

Additionally, \eqref{eq:magma} possesses the conservation law
\begin{equation}
\label{eq:energy}
\mathcal{N}[\f] = \begin{cases}
\int \paren{\frac{1}{2}\phi^{-2m}\phi_{x}^{2}+\f\log\paren{\f}-\f +1}dx & \mbox{if $n+m=1$,}\\
\int \paren{\frac{1}{2}\phi^{-2m}\phi_{x}^{2} + \f-1 -\log \paren{\f}}dx & \mbox{if $n+m=2$,} \\
\int \paren{\frac{1}{2}\phi^{-2m}\phi_{x}^{2}+\frac{\phi^{2-n-m}-1+(n+m-2)(\f-1)}{(n+m-1)(n+m-2)}}dx &\mbox{for all other $n$ and $m$.}
\end{cases}
\end{equation}
$\N$ is well defined for $\f$ bounded from below away from zero and $\norm{\f-1}_\h{1}<\infty$. It is also locally convex about $\phi\equiv1$. See Section 3 of \cite{Simpson07} for details.

\subsection{Solitary Waves and their Analytic Properties}
\label{sec:solwave}

Let us review the properties of the solitary waves associated with \eqref{eq:magma}.  In particular, we identify their decay and regularity properties, and introduce the KdV scaling for later use.

Substituting the traveling wave \emph{ansatz}, $\f_c(x,t) = \f_c(x- c t)$, into \eqref{eq:magma} with boundary conditions
\begin{equation}
\lim_{y\to \pm \infty} \f_c(y) = 1,\quad \lim_{y\to \pm \infty} \dy^j\f_c(y) = 0 \quad \text{for $j=1,2,\ldots$}
\end{equation}
we have, after one integration,
\begin{equation}
\label{eq:solwave1}
- c (\f_c-1) + \f_c^n - 1 + c \f_c^n \dy \paren{\f_c^{-m} \dy \f_c}=0
\end{equation}
Letting $u_c = 1 -\f_c$, $u_c$ satisfies
\begin{equation}
\label{eq:solwave2}
-c u_c + \paren{u_c+1}^n -1 + c \paren{u_c+1}^n \dy \paren{\paren{u_c+1}^{-m} \dy u_c}=0
\end{equation}

Equation \eqref{eq:solwave1} may also be integrated up to a first order equation,
\begin{equation}
\label{eq:solwave-first-order}
\frac{1}{2}\f_c^{-2m} \paren{\dx\f_c}^2 - F_1(\f_c;c)=0
\end{equation}
after applying the boundary condition $\f_c \to 1$ at $\pm \infty$.  $F_1$ depends on the particular exponents:
\begin{equation}
F_1(x; c) = \begin{cases}
\frac{x^{1-n}-1}{1-n} + \paren{1-\frac{1}{c}}\frac{x^{-n}-1}{n}-\frac{1}{c}\log\paren{x} & \text{if $m=1$,}\\
x-1 -\paren{1-\frac{1}{c}}\log\paren{x}-\frac{1}{c}\frac{x^n-1}{c} & \text{if $n+m=1$,}\\
\log\paren{x}+\paren{1-\frac{1}{c}}\paren{\frac{1}{x}-1}-\frac{1}{c}\frac{x^{n-1}-1}{n-1} & \text{if $n+m=2$,}\\
\frac{x^{2-n-m}-1}{2-n-m}-\paren{1-\frac{1}{c}}\frac{x^{1-n-m}-1}{1-n-m}-\frac{1}{c}\frac{x^{1-m}-1}{1-m} & \text{otherwise.}
\end{cases}
\end{equation}
Using \eqref{eq:solwave-first-order}, an equivalent second order, self-adjoint, equation for the solitary waves is
\begin{align}
\label{eq:solwave-a}
F_2(\f_c; c)&= - \dx^2 \f_c\\
\label{eq:solwave-b}
F_2(x;c) &= x^{m-n}\bracket{-(x-1) + c^{-1}\paren{x^n-1}-2 m x^{n-m-1}F_1(x;c)}
\end{align}

Let us introduce the KdV scaling.  Define
\begin{equation}
\label{eq:gamma}
\boxed{\gamma = \sqrt{\frac{c-n}{c}}}
\end{equation}
Applying the scalings,
\begin{equation}
\xi = \gamma\paren{x - c t},\quad u_c(y) =\frac{\gamma^2}{n-1}U(\xi(y);\gamma)
\end{equation}
\eqref{eq:solwave2} becomes
\begin{equation}
\label{eq:scaledsolwave}
-U +\frac{1}{2}U^2 + \dxi^2 U= O(\gamma^2)
\end{equation}

\begin{rem}
The parameter $\gamma$, \eqref{eq:gamma}, will be used throughout the paper.  Because it uniquely maps  $c\in \paren{n,\infty}$ onto $(0,1)$, we will use $c$ and $\gamma$ interchangeably.
\end{rem}

We summarize what is known about \eqref{eq:solwave2} and \eqref{eq:scaledsolwave} in the following two results:
\begin{thm}
\label{thm:solexist}
For any $c>n>1$, \eqref{eq:solwave2} has a unique positive, even solution $u_c$, going to zero at $\pm\infty$.  In the  KdV scaling, \eqref{eq:scaledsolwave}, $U$ is real analytic in the arguments $(\xi,\gamma)\in \R \times [0,1)$.  When $\gamma=0$
\[
U(\xi;0)=U_\star(\xi) = 3 \sech^2\paren{\frac{1}{2} \xi}
\]  

Furthermore, for $\gamma$ in any compact subset of $[0,1)$
\begin{equation*}
\dxi^j U(\xi;\gamma)e^{\pm \xi} \paren{\sign(\xi)}^j \to K_j(\gamma) \quad\mbox{as $\xi\to \pm \infty$ for $j=0,1,2$}
\end{equation*}
\end{thm}

\begin{cor}
\label{cor:soldecay}
Given a compact interval $\bracket{0,\gamma_0}\subset [0,1)$, there exists a constant $K$ such that for all $\gamma\in [0,\gamma_0]$,
\begin{align}
\abs{\dxi^j U(\xi;\gamma)}&\leq K e^{-\abs{\xi}} \quad\text{for $j=0,1,2$, $-\infty<\xi<\infty$}\\
\abs{\dy^j u_c(y)}&\leq K\frac{\gamma^{2+j}}{n-1} e^{-\gamma\abs{y}} \quad\text{for $j=0,1,2$, $-\infty<y<\infty$}
\end{align}
\end{cor}
\proof From \cite{Nakayama92}, solitary waves exist provided $c>n>1$ and $m\in \R$.  Writing the problem as a two-dimensional dynamical system, we may apply the Stable Manifold Theorem about the hyperbolic point $(0,0)$ to deduce the exponential decay, as  in Theorem 2.1 and Corollary 2.2 of \cite{Pego97}.\qed

\begin{rem} When the parameter $\gamma$ is small, $\f_c$ is in the regime of small amplitude, long waves, where KdV appears as the leading order equation in a perturbation expansion of \eqref{eq:magma}, as in \cite{Whitehead86}.
\end{rem}
 
\begin{cor}
\label{cor:solanalyticity}
For each $c>n>1$, the solitary wave solution, $\f_c-1$, lies in $H^\infty(\R)$.  Furthermore, there exists $\sigma_0>0$ such that the solitary wave $\f_c$ may be analytically continued off the real axis into the strip $\set{ z: \abs{\Im z} < \sigma_0}$.
\end{cor}
\proof This is a consequence of Corollary \ref{cor:soldecay} and Corollary 4.1.6 of \cite{Bona:1997fk}, see Appendix \ref{appendix:analyticity}.\qed

\begin{cor}
Given a solitary wave $\f_c$, $c>n$, assume $0<a<\gamma$.  Then $\f_c-1\in H^\infty_a$.
\end{cor}

\begin{cor}
\label{cor:solcontinuity}
Let $n>1$.
\begin{description}
  \item[(a)] The mapping $c\mapsto \f_c -1$ is $C^1\paren{(n,\infty); H^2}$.  In fact the mapping is analytic.
  \item[(b)] This mapping is analytic, and, for fixed x, $c\mapsto \f_c(x)$ is analytic function of $c$.
  \item[(b)] Given $a < \frac{1}{2}$, the mapping is also $C^1\paren{\paren{n/(1-4a^2),\infty}; H^2\cap H^2_a}$.
\end{description}
\end{cor}
\proof All parts are proved using the implicit function theorem, applied to the functional
\begin{equation*}
\mathcal{F}[c,u] = \dx^2 u + F_2(1+u;c)
\end{equation*}
See Appendix \ref{sec:speed-continuity} for details and \cite{Berger77} for a statement and proof of the implicit function theorem for analytic mappings.\qed
 
\subsection{Remarks and Assumptions on the Exponential Weight}
\label{sec:weightremarks}
We see in Theorem \ref{thm:local} and Corollary \ref{cor:solcontinuity} that the particular exponential weight restricts what data and which solitary waves will be permissible.  For the solitary wave result, this restriction comes from the decay rate associated with the speed; see Corollary \ref{cor:soldecay}.

In the case of the existence theorem, \eqref{eq:magma} may be written as 
\begin{equation}
\label{eq:magma2}
\begin{split}
\dt \f &= -\set{I - \dx \bracket{\f^n \dx\paren{\f^{-m} \cdot }}}^{-1}\dx\paren{\f^n}= -\f^m \set{\f^{-m} \bracket{I - \dx \paren{\f^n \dx\paren{\f^{-m} \cdot }}}^{-1}}\dx\paren{\f^n}\\
&= -\f^m H_\f^{-1}\dx(\f^n) 
\end{split}
\end{equation}
The operator $H_\f$ is
\begin{equation}
\label{eq:inv}
H_\f= \f(x)^m - \dx\paren{ \f(x)^n \dx\cdot }
\end{equation}
This is a bounded operator on $L^2 \to H^1$ provided $\f$ is continuous and bounded from below away from zero.  However, the exponential weight introduces a second constraint.  Consider solving  $H_\f u = f$, $f\in L^2_a$ for $u\in H^1_a$.  Letting $g  = e^{ax} f$ and $v = e^{ax} u$, this is equivalent to solving
\[
\bracket{\f^m -D_a\paren{\f^n D_a}} v = g, \quad v\in H^1
\]
Multiplying by $v$ and integrating by parts,
\[
\int \paren{\f^m -a^2 \f^n}v^2 + \f^n \paren{\dx v}^2 dx = \int g v dx
\]
A unique solution exists, provided $a$ and $\f$ satisfy $\inf_x \f(x)^m -a^2 \f(x)^n>0$; this is condition \eqref{eq:local2}.

We invert these restrictions; given a solitary wave of speed $c$, we will assume that $a$ is sufficiently small so that these, and other, properties hold.  There are three restrictions in what follows.  

Let 
\begin{equation}
\label{eq:abound1}
a_1 = \frac{1}{3}\gamma(c)
\end{equation}
$\f_c-1$ will then be in $H^2_{a_1}$ and $H^2_{2 a_1}$, as will all solitary waves of nearby speed.  Let
\begin{equation}
\label{eq:abound2}
a_2 = \frac{1}{2}\inf_x \f_c(x)^{(m-n)/2}
\end{equation}
Then $\f_c$ will satisfy \eqref{eq:local2}, with
\[
\inf_x \f_c(x)^m - a^2 \f_c(x)^n \geq \frac{3}{4}\inf_x \f_c(x)^m>0
\]
Hence, the solitary waves will live in a set on which the existence theorem applies.  Moreover, for all data $\f_0$ sufficiently close to $\f_c$ in the $H^1$ norm, an analogous lower bound will exist.
\begin{rem}
$\inf_x\f_c(x)^{(m-n)/2}$ is related to a physical length scale known as the \emph{compaction length}, \cite{McKenzie84}.  This length, $\delta_\comp$, is given by 
\[
\delta_\comp(x) = \sqrt{\f(x)^{n-m}}
\]
It measures the distances over which there will be geometrical rearrangement of the material, appearing macroscopically as changes in $\phi$, in response to viscous stresses.

The stipulation $a < \inf_x \f(x)^{(m-n)/2}$ may be interpreted as requiring the length scale associated with the exponential weight, $a^{-1}$, to never be smaller than this intrinsic, spatially varying, length.
\end{rem}

Finally, there is a constraints related to the \emph{essential spectrum} of a linear operator, discussed in Section \ref{sec:spectrum}.  Let
\begin{equation}
\label{eq:abound3}
a_3=\sqrt{\frac{2c}{n+2c+\sqrt{n\paren{n+8 c}}}}\sqrt{\frac{c-n}{c}}
\end{equation}
This will ensure that  for any $a\leq a_3$, the essential spectrum is located in a specific part of the complex plane.  Let
\begin{equation}
a_\star(c) =\min \left\{a_1(c), a_2(c),  \frac{1}{2} a_3(c)\right\}
\end{equation}
Then for any $a\leq a_\star(c)$, all of these properties will be satisfied for $\f$ sufficiently close in $H^1$ to $\f_c$.

 \subsection{Ansatz and Linearization}
\label{sec:ansatz}

Given a perturbed solitary wave solution, $\f$, of \eqref{eq:magma}, assume that there exists decomposition of $\f$ into a (time dependent) solitary wave of some speed $c(t)$ and phase $\theta(t)$ and a perturbation, $v$; this decomposition's existence will be proved in Section \ref{sec:decomp}. Transforming our coordinate system into the frame of this modulating solitary wave,
\begin{align}
\label{eq:coordinate-change}
y(x,t) &= x-\int_0^t c(s) ds +\theta(t)\\
\label{eq:ansatz}
\f(x,t)&=\f_{c(t)}(y(x,t)) + v(y(x,t),t)= \f_c(y,t)+v(y,t)
\end{align}

The perturbation, $v$, is governed by
\begin{equation}
\label{eq:perturbation-equation-2}
\dt v = A_c v -\thetadot \dy v -\dcdt \dc \f_c -\thetadot \dy\f_c+ \mathcal{F}_1[v; \f_c]
\end{equation}
where
\begin{align}
A_c v &= \f_c^m H_{\f_c}\inv \dy L_c v\\
L_c v&= -c \f_c^n\dy^2\paren{\f_c^{-m}v}+\bracket{c-n\f_c^{-1}+ cn(\f_c^{-1}-1)}v
\end{align}
and \eqref{eq:unweighted-nonlinear-term1} gives the defintion of $\mathcal{F}_1[v;\f_c]$, composed of terms nonlinear in $v$.  We make two remarks about \eqref{eq:perturbation-equation-2}.  First, the linear operator $A_c$, is time dependent; $c=c(t)$, and we would prefer to work with a time \emph{independent} linear operator.  Second,  the appearance of the term $\thetadot \dy v$ will prove problematic to studying the equation in $\h{1}$.  

The first problem is addressed by adding and subtracting $A_\co$, and considering the difference $A_c-A_\co$ as another perturbation of the linear flow.  To remove the $\dy v$ term, we introduce a renormalized time,
\begin{equation}
\label{eq:timechange}
\tau = \co^{-1}\bracket{\int_0^t c(s)ds -\theta(t)}
\end{equation}
The asymptotic stability proof of BBM required a similar transformation, \cite{Miller96}.  In addition, the problem will be considered in the weighted space $H^1_a$, with $w(y,t)=e^{ay} v(y,t)$.  All together, we have:
%

\begin{prop}(Perturbation Equation)
\label{prop:perturbation}

The perturbation to a solitary wave of speed $\co$ associated with the ansatz in \eqref{eq:ansatz}, $v$ and its weighted perturbation $w = e^{ay} v$, evolves according to the equations in $t$ and $\tau$-time respectively.
\begin{align}
\dt v &= A_c v -\thetadot \dy v -\dcdt \dc \f_c -\thetadot \dy\f_c+ \mathcal{F}_1[v; \f_c]\\
\dt w &= A_{c,a} w -\thetadot \paren{\dy-a} w -e^{ay}\paren{\dcdt \dc \f_c +\thetadot \dy\f_c}+ \mathcal{G}_1[w,v; \f_c]\\
\partial_\tau v &= A_\co -\frac{\co}{c-\thetadot}\paren{c \dc\f_c + \thetadot \dy \f_c}+S[\co,c,\thetadot]v + \frac{\co}{c-\thetadot}\mathcal{F}_1[v;\f_c]\\
\label{eq:weighted-perturbation-equation}
\begin{split}
\partial_\tau w &= A_{\co,a}w -\frac{\co}{c-\thetadot}e^{ay}\paren{\dcdt \dc\f_c + \thetadot \dy \f_c}+S_a[\co,c,\thetadot]w + \frac{\co}{c-\thetadot}\mathcal{G}_1[w,v;\f_c]\\
&=A_{\co,a} w+\mathcal{G}\bracket{w,v;\co,c,\thetadot}
\end{split}
\end{align}
The operator $S$ and the terms $\mathcal{F}_1$ and $\mathcal{G}_1$ are given explicitly in Appendix \ref{sec:expansions}.
\end{prop}
\proof $\mathcal{G}_1$ is obtained from $\mathcal{F}_1$ by substituting $e^{-ay} w$ for one of the $v$'s; $e^{ay}\mathcal{F}_1[v;\f_c]= \mathcal{G}_1[e^{ay} v, v;\f_c]$.  The details appear in Appendix \ref{sec:expansions}.\qed


\nb From here on, we assume $\co$ to be fixed and will suppress its appearance in the linear operators $A_{\co}$ and $A_{\co,a}$.

\section{Spectral Properties of Linearization about a Solitary Wave}
\label{sec:spectrum}

In this section we analyze the spectrum of $A$ and $A_a$.  We will use $c$ in place of $\co$ and $x$ in place of $y$ as the independent variable.
\begin{align}
\label{eq:eigen1}
AY &= \lambda Y\\
\label{eq:eigen2}
A &=\set{I - \dx \bracket{\f_c^n \dx \paren{\f_c^{-m}\cdot}}}^{-1}\dx L_c\\
\label{eq:eigen3}
L_c &= -c \f_c^n\dx^2\paren{\f_c^{-m}\cdot} + \bracket{c-n\f_c^{-1} + cn \paren{\f_c^{-1}-1}}\\
\label{eq:eigen4}
A_{a}&= e^{ax} A e^{-ax}
\end{align}
Our goal is to prove that the linearized problem is asymptotically stable; $\norm{e^{A_{a}t} w_0}_\h{1}\to 0$ as $t\to +\infty$.  We will actually show something much stronger, that this convergence to zero happens exponentially fast.  Our strategy is that of \cite{Miller96, Pego97}.  We will
\begin{enumerate}
  \item Identify the essential spectrum of $A_{a}$ by showing it to be a compact perturbation of a constant coefficient operator.
  \item Rule out point spectrum (eigenvalues of finite multiplicity) of $A_{a}$ for $\abs{\lambda}$ sufficiently large via an operator estimate.
  \item Use the \emph{Evans function}, an infinite dimensional analog of the characteristic polynomial, to rule out nonzero point spectra of $A_a$ in the set of ``small" $\lambda$, which will be compact.
  \item Show decay in time of the $C_0$-semigroup $e^{A_{a}t}$ associated with $A_{a}$.
\end{enumerate}
The spectral analysis is handled this section, and the semigroup theory in the following section.  

\begin{figure}
\begin{center}
\includegraphics[width=3in]{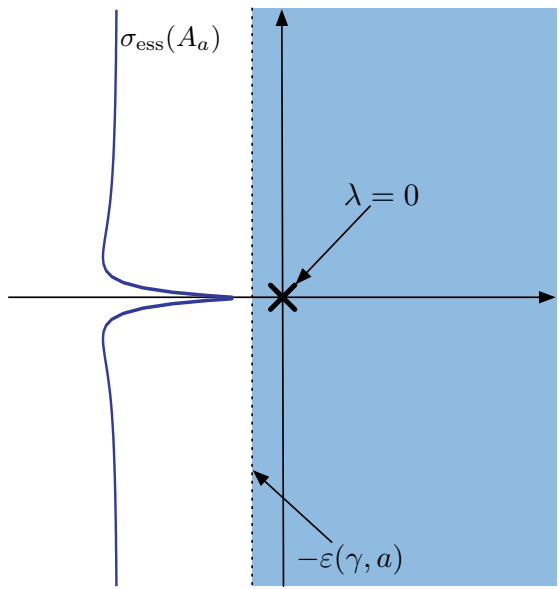}
\caption{The spectrum of the operator $A_a$.  The only eigenvalue with $\Re\lambda \geq -\varepsilon$ is $\lambda=0$.}
\label{fig:operator-spectrum}
\end{center}
\end{figure}

The principle result of this section is:
\begin{thm}(Spectrum of Linearized Operator)
\label{thm:linear}

\begin{description}
  \item[(a)] Let $a \in (0,a_\star( \gamma)]$.  The essential spectrum of $A_a$ denoted by $\sigma_{\textrm{ess}}(A_a)$ is a curve lying in the open left half-plane, with rightmost point $-\omega$, 
  \begin{equation}
  \label{eq:spectralbound}
  -\omega =\max\{\Re z | z \in \sigma_{\textrm{ess}}(A_a)\} < 0
  \end{equation}
  \item[(b)] There exists $\gamma_\star \in (0,1)$ such that for each $\gamma \in (0, \gamma_\star]$ and $a \in(0,a_\star( \gamma)]$, there exists $\varepsilon(\gamma,a)>0$ such that the only eigenvalue of $A_a$ with $\Re \lambda \geq -\varepsilon$ is $\lambda = 0$ and this is an eigenvalue of algebraic multiplicity two.
  \item[(c)] In the Hamiltonian case, $n+m=0$, part \textbf{(ii)} may extended for $\gamma\in(\gamma_\star,1)$ to all but a discrete set with no accumulation point.
\end{description}
The spectrum is pictured in Figure \ref{fig:operator-spectrum}.
\end{thm}

\subsection{Essential Specturm}
We make use of the definition of the Essential Spectrum of an operator from \cite{Schechter71,Schechter01}, that for a closed, densely defined operator $A$ on a Banach space $X$, 
\begin{equation}
\sigma_{\ess}(A) = \bigcap_{C \in \mathcal{K}(X)} \sigma(A+C)
\label{eq:definition-essential-spectrum}
\end{equation}
where $\mathcal{K}(X)$ denotes the set of compact operators on $X$.  Other definitions are possible and well known; see Chapter IX of \cite{Edmunds87} for a discussion of how these definitions relate to one another.

 $\sigma(A)\setminus \sigma_{\ess}(A)$ then consists of point spectra.  This is so because, by Theorem 7.27 of \cite{Schechter01}, if $\lambda$ is not in the essential spectrum, then $\lambda I -A$ is Fredholm with index zero.  Hence, it has closed range and a finite kernel.  Therefore, it must be an eigenvalue of finite multiplicity.


To prove Theorem \ref{thm:linear} part \textbf{(i)}, we express $A_a$ as a compact perturbation of a constant coefficient operator, $A_a^\infty$, obtained by setting $\f_c(x)$ equal to its asymptotic state, $1$.
\begin{equation}
\label{eq:weightedinfinity}
A^\infty_a= (I-D^2_a)^{-1}D_a(-cD^2_a +c-n)
\end{equation}

The difference between $A_a$ and $A_a^\infty$, given explicitly in \eqref{eq:lineardifference}, may be shown to be an $A_a^\infty$--compact operator.  Hence, $A_a$ is a compact perturbation of $A_a^\infty$ and
\[
\sigma_\ess(A_a) = \sigma_\ess(A_a^\infty)
\]


Upon examination of its Fourier symbol of $A_a^\infty$, the essential spectrum of $A_a$ is
\begin{equation}
\label{eq:ess-spectrum}
\boxed{\sigma_{ess}(A_a) = \set{ \frac{(i\ell-a)(-c(i \ell-a)^2+c-n)}{1-(i \ell-a)^2}, \quad \ell\in \mathbb{R} }}
\end{equation}
and
\[
-\omega = \max\left\{\Re z | z \in \sigma_{ess}(A_a)\right\} = -ac +\frac{a n}{1-a^2}<0
\]
and $\sigma_{ess}(A_a)$ lies in the open left half plane if $0<a<\gamma$. $\sigma_{\ess}(A_a)$ is pictured in Figure \ref{fig:operator-spectrum}.

In addition, for  $0<a\leq a_\star<a_3$, $a_3$ defined by \eqref{eq:abound3}, then the spectrum moves rightward as $a\to 0$.  This is because $a_3$ is the value for which $-\omega$ is leftmost in $\C$, maximizing the rate of decay of $e^{ay}$ as $y \to -\infty$. 

\subsection{Large Eigenvalues}
As in \cite{Miller96}, we will study the eigenvalues of $A_a$ by considering separately a large $\abs{\lambda}$ regime and a small $\abs{\lambda}$ regime. 

Rewriting the linear operator $A_a$ as
\begin{equation}
A_a  = c\f_c^m D_a \paren{\f_c^{-m}\cdot } - n \f_c^m H_{\f_c,a}^{-1}D_a \paren{\f_c^{-1}\cdot} + cm \f_c^m H_{\f_c,a}^{-1}\paren{\f_c^{-1}\dy\f_c \cdot} + cn \f_c^m H_{\f_c,a}^{-1}D_a\bracket{\paren{\f_c^{-1}-1}\cdot}
\end{equation}
we note that $\lambda$ is an $L^2$ eigenvalue of $A_a$ if and only if it is also an $L^2$ eigenvalue of $\tilde{A}_a=\f_c^{-m}A_a \f_c^m$, given by
\begin{equation}
\tilde{A}_a = c D_a-nH_{\f_c,a}^{-1}D_a \paren{\f_c^{m-1}\cdot} + cm H_{\f_c,a}^{-1}\paren{\f_c^{m-1}\dy\f_c \cdot} + cn H_{\f_c,a}^{-1}D_a\bracket{\f_c^m\paren{\f_c^{-1}-1}\cdot}
\end{equation}
We will rule out eigenvalues of $\tilde{A}_a$, thus ruling them out for $A_a$.  This is equivalent to studying $A_a$ in a space weighted by $\f_c^{-m}(x)$, a strictly positive, smooth, and bounded function.  $\tilde{A}_a$ is also a compact perturbation of $\tilde{A}_a^\infty = A_a^\infty$; they share the same essential spectrum.

Let the operator $C(\lambda)$ satisfy
\[
C(\lambda) = \paren{\lambda I-\tilde{A}_a^\infty}^{-1}\paren{\tilde{A}_a - \tilde{A}_a^\infty}
\]

\begin{prop}
\begin{description}
  \item[(a)] The operator $C(\lambda)$ is compact for $\lambda$ not in $\sigma_{ess}$.  In particular, $C(\lambda)$ is compact for all $\lambda$ with $\Re \lambda > -\omega$.
  \item[(b)] For any $\lambda \in \mathbb{C}\setminus \sigma_{\ess}(A_a)$, we have that $\lambda$ is an eigenvalue of $A_a$ if and only if $1\in \sigma(C(\lambda))$ .
  \item[(c)] Let $\lambda  \in \mathbb{C}\setminus \sigma_{\ess}(A_a)$.  A sufficient condition for $\lambda$ not to be an eigenvalue of $A_a$ is that $\norm{C(\lambda)}<1$, with norm either $\ltwo$ or $\h{1}$ depending on which space is under consideration.
\end{description}
\label{prop:compact-operator}
\end{prop}
\proof Using the equivalence of eigenvalues of $A_a$ and $\tilde{A}_a$, parts \textbf{(b)} and \textbf{(c)} will follow once \textbf{(a)} is established, see \cite{Miller96}.  The Fourier symbol of $(\lambda I-\tilde{A}_a^\infty)^{-1}$ is
\[
\frac{1-(\imath \ell -a)^2}{\lambda(1-(\imath \ell -a)^2)-(\imath \ell -a)(-c(\imath \ell - a)^2+c-n)}
\]
This operator is bounded for $\lambda$ not in the essential spectrum.  The difference, given explicitly in \eqref{eq:lsdifference},is a sum of Hilbert-Schmidt compact operators composed with bounded operators on $L^2\to L^2$, hence $C(\lambda)$ is compact on this space.

For $C(\lambda)$ to be a compact operator on $\h{1}\to \h{1}$, it will be sufficient to prove
\[
(I-\dx^2)^{1/2}C(\lambda)(I-\dx^2)^{-1/2}= (\lambda I-\tilde{A}_a^\infty)^{-1}(I-\dx^2)^{1/2}(\tilde{A}_a - \tilde{A}_a^\infty)(I-\dx^2)^{-1/2}
\]
is compact on $\ltwo\to\ltwo$.  $(\lambda I-A_a^\infty)^{-1}$ is still  bounded, and by commuting operators, it may be proven that $(I-\dx^2)^{1/2}(\tilde{A}_a - \tilde{A}_a^\infty)(I-\dx^2)^{-1/2}$ is compact.  \qed


We now rule out eigenvalues outside a rectangle that scales with $\gamma^3$.
\begin{prop}
Let $\delta \in (0,1)$ be fixed.
\begin{description}
\item[(a)] For any $c_\star>n$, there exists an $M>0$ such that for $n<c\leq c_\star$, if
\[
\Re\lambda \geq-\frac{1}{2}a c \gamma^2
\]
and either
\[
\abs{\Im\lambda}>c M \gamma^3\quad\text{or}\quad\Re\lambda >c M \gamma^3
\]
then $\norm{C(\lambda)}_{L^2\to L^2}<1-\delta$.
\item[(b)] This result also holds for $\norm{C(\lambda)}_{H^1\to H^1}$.
\end{description}
\label{prop:large-eigenvalue}
\end{prop}

\proof If $M\geq 1$, then by the assumptions on $\Re \lambda$ and $\Im \lambda$, $\lambda$ is not in $\sigma_{\ess}(A_a)$, hence we may apply Proposition \ref{prop:compact-operator} \textbf{(a)} to conclude $C(\lambda)$ is a compact operator.  If the norm of $C(\lambda)$ is less than one, part \textbf{(c)} of that proposition will imply it is not an eigenvalue; we seek an $M\geq 1$ for which $\norm{C(\lambda)}$ can be made sufficiently small. 

For all $(\gamma,a)$ in $\set{(\gamma,a)\mid \gamma \in [0,\gamma_0],\quad a\leq a_\star(\gamma)}$, there exist $K_1$ and $K_2$ such that
\[
\norm{C(\lambda)}_{L^2\to L^2}\leq \norm{\paren{\lambda I -A^\infty_a}^{-1}H_{1,a}^{-1}D_a}_{L^2\to L^2}\paren{ cK_1 \gamma^2}+ \norm{\paren{\lambda I -A^\infty_a}^{-1}H_{1,a}^{-1}}_{L^2\to L^2} \paren{cK_2 \gamma^3}
\]
This comes from expanding the difference $\tilde A_a -\tilde A_a^\infty$ and commutating operators, see \eqref{eq:lsdifference}.  $\paren{\lambda I -A^\infty_a}^{-1}H_{1,a}^{-1}D_a$ and $\paren{\lambda I -A^\infty_a}^{-1}H_{1,a}^{-1}$ are constant coefficient operators and we will treat them in Fourier space.

Thus, if we can prove that for $\lambda$ satisfying the the hypotheses
\begin{align*}
\sup_{\ell\in \R} \left|\frac{\paren{\imath \ell -a} cK_1\gamma^2}{\lambda\bracket{1-\paren{\imath \ell-a}^2}-\paren{\imath \ell -a} \bracket{-c \paren{\imath \ell -a}^2 +c \gamma^2}}\right| < \frac{1-\delta}{2}\\
\sup_{\ell\in \R} \left|\frac{ c K_2\gamma^3}{\lambda \bracket{1-\paren{\imath \ell-a}^2}-\paren{\imath \ell -a} \bracket{-c \paren{\imath \ell -a}^2 +c \gamma^2}}\right| < \frac{1-\delta}{2}
\end{align*}
we will be done.

Introducing the scalings $\lambda = c\Lambda \gamma^3$, $\ell = \gamma\xi$, and $a = \gamma \vartheta$, this is equivalent to identifying $M\geq 1$ such that when $\Re\Lambda \geq -\frac{1}{2}\vartheta$ and either $\Re\Lambda >M $ or $\abs{\Im \Lambda}>M$ then both
\begin{align}
\label{eq:lsbound1}
\sup_{\xi\in \R} \left|\frac{\paren{\imath \xi -\vartheta} K_1}{\Lambda\bracket{1-\gamma^2\paren{\imath \xi -\vartheta}^2}-\paren{\imath \xi -\vartheta} \bracket{- \paren{\imath \xi -\vartheta}^2 +1}}\right| < \frac{1-\delta}{2}\\
\label{eq:lsbound2}
\sup_{\xi\in \R} \left|\frac{  K_2}{\Lambda \bracket{1-\gamma^2\paren{\imath \xi -\vartheta}^2}-\paren{\imath \xi -\vartheta} \bracket{- \paren{\imath \xi -\vartheta}^2 +1}}\right| < \frac{1-\delta}{2}
\end{align}
are satisfied.

Squaring both sides, \eqref{eq:lsbound1} and \eqref{eq:lsbound2} may be rewritten as two polynomial inequalities, $P_1(\Im \Lambda, \Re\Lambda,\xi)>0$ and $P_2(\Im \Lambda, \Re\Lambda,\xi)>0$, respectively.  We will show that for appropriately chosen $\Lambda$, the inequalities hold for all $\xi$.  $P_1$ and $P_2$ are treated similarly.  We study $P_2$:
\begin{align*}
P_2(\Im \Lambda, \Re \Lambda, \xi) &= \alpha (\Im \Lambda)^2 + \beta \Im \Lambda + \eta_1 (\Re \Lambda)^2 + \eta_2 \Re \Lambda + \eta_3\\
\alpha&= \gamma^2\xi^4 +2 \xi^2 \gamma^2(1+\gamma^2\theta^4)+ (1-\gamma^2\theta^2)^2\\
\beta&= -2\xi^5 \gamma^2 - 2\xi^3(1+\gamma^2(1+2\theta^2))-2\xi(1+(-3+\gamma^2)\theta^2 + \gamma^2\theta^4) \\
\eta_1&=\alpha\\
\eta_2&= 2\theta\xi^4 \gamma^2 + 2\theta\xi^2 (3+ \gamma^2(-1+2\theta^2)) + 2\theta(1-\theta^2)(1-\gamma^2\theta^2)\\
\eta_3&= \xi^6 +\theta^2(1-\theta^2)^2 + \xi^4(2+3\theta^2) + \xi^2(1+3\theta^4)-\beta^2\\
\beta&={2K_2/(1-\delta)}
\end{align*}
Using similar analysis as for $P_1$ in \cite{Miller96}, we first consider $P_2$ as quadratic in $\Im \Lambda$.  Examining its discriminant,
\begin{equation*}
\begin{split}
\text{discriminant}&= \gamma^4\bracket{-4 \xi^8\paren{\theta+\gamma^2\Re\Lambda}^2 + O(\xi ^6)}\\
&\quad+\gamma^2\bracket{-24\theta^2 \xi^6 + O(\xi^4)}\\
&\quad -36 \xi^4 \theta^3 + O(\xi^2)
\end{split}
\end{equation*}
If $0\geq \Re \Lambda \geq -\theta/2$, then there exists $M_0>0$ such that for all $\Im \Lambda$, $\gamma\in[0,\gamma_0]$ and $\abs{\xi}>M_0$ the discriminant is negative and $P_2>0$.  Furthermore, since the coefficient $\alpha>0$, there exists $R_1>0$ such that $P_2>0$ when $\gamma\in[0,\gamma_0]$, $\abs{\Im\Lambda}>R_1$, $0\geq \Re \Lambda \geq -\theta/2$, and $\abs{\xi}\leq M_0$

For $\gamma \in [0,\gamma_0]$ and $\theta \in (0,1)$, both $\eta_1$ and $\eta_2$ are positive; if $P_2(\Im\Lambda, \Re\Lambda, \xi)>0$ then $P_2(\Im\Lambda, \Re\Lambda +K, \xi)>0$ for any $K>0$.  Therefore, $P_2>0$ for all $\xi$ if $\gamma\in[0,\gamma_0]$, $\abs{\Im\Lambda}>R_1$ and $ \Re \Lambda \geq -\theta/2$.  Also, $P_2>0$ for $\abs{\xi}>M_0$ and all $\Im \Lambda$ if $\Re\lambda \geq -\theta/2$.

We must still treat the case of $\abs{\Im \Lambda}\leq R_1$ and $\abs{\xi} \leq M_0$ simultaneously.  Consider $P_2$ as quadratic in $\Re\Lambda$.  $\eta_1$ and $\eta_2$ are positive for all $\xi$, and $\eta_3$ is bounded from below.  Therefore there is some $R_2>0$ such that $P_2>0$ for all $\xi$ if $\gamma\in[0,\gamma_0]$, $\abs{\Im\Lambda}\leq R_1$, and $ \Re \Lambda >R_2$.  Thus, for any $\theta \in (0,1)$, $\gamma_0 \in [0,\gamma_0]$, there exist $R_1>0$ and $R_2>0$ such that $P_2>0$ for all $\xi$ if
\begin{align*}
\gamma &\in[0,\gamma_0]\\
\Re \Lambda &\geq -\theta/2\\
\abs{\Im \Lambda} \leq R_1\quad&\text{or}\quad\Re \Lambda > R_2
\end{align*}

For $C(\lambda):\h{1}\to \h{1}$, the proof is similar, with constants $\tilde K_1$ and $\tilde K_2$ in place of $K_1$ and $K_2$.
\qed


\subsection{Small Eigenvalues--The Evans Function}
\label{sec:evans}
In this section we rule out eigenvalues of $A_a$ in the set $\abs{\lambda} \leq M \gamma^3$.  This is done using the \textit{Evans function}, an analytic function that vanishes at eigenvalues of $A_a$.  The Evans function, $D=D(\lambda;\g)$, is constructed for the eigenvalue problem using particular solutions of an associated dynamical system
\begin{align}
\label{eq:odesystem}
\dot{\bold{y}} &= B(x,\lambda,\g) \bold{y}\\
\label{eq:decayatinfty}
\bold{y} &= O(e^{\mu_1 x}) \quad\text{as $x\to +\infty$}
\end{align}
and the adjoint system,
\begin{align}
\label{eq:odesystem-adjoint}
\dot{\bold{z}} &= -\bold{z}B(x,\lambda,\g)\\
\label{eq:decayatinfty-adjoint}
\bold{z} &= O(e^{\mu_1 x}) \quad \text{as $\quad x\to -\infty$}
\end{align}
$\mu_1$ will be the eigenvalue of smallest real part of the $B^\infty$, the limit as $x\to \pm \infty$ of $B$.  When certain conditions, described in Theorem \ref{thm:evans}, are met, the Evans function exists and may be explicitly defined as
\begin{equation}
D(\lambda;\g) = \bold{z}(x;\lambda,\g)\cdot \bold{y}(x;\lambda,\g)
\end{equation}

The idea is to measure the angle between the subspace of solutions decaying at $+\infty$ with the subspace decaying at $-\infty$; hence the appearance of the dot product.  The Evans has an equivalent formualation in terms of the determinant of the fundamental solution of \eqref{eq:odesystem}.  For a more complete discussion of the Evans function, see \cite{Pego92}.

\begin{thm}
\label{thm:evans}\cite{Pego92, Pego97}

Let $\Omega$ be a simply connected subset of $\C^2$.  Suppose that the system \eqref{eq:odesystem} satisfies the following hypotheses:
\begin{description}
\item[(i)] $B: \R \times \Omega \to \C^{n\times n}$ is continuous in $(x,\lambda,\gamma)$ and analytic in $(\lambda,\gamma)$ for fixed $x$.
\item[(ii)] $B^\infty(\lambda,\gamma)=\lim_{x\to \pm \infty} B(x,\lambda,\gamma)$ exists for all $(\lambda,\gamma)\in \Omega$.  The limit is attained uniformly on compact subsets of $\Omega$.
\item[(iii)] The integral
\[
\int_{-\infty}^\infty \norm{B(x,\lambda,\gamma)-B^\infty(\lambda,\gamma)} dx
\] 
converges for all $(\lambda,\gamma)\in \Omega$ and the convergence is uniform on compact subsets of $\Omega$.
\item[(iv)] For every $(\lambda,\gamma)\in \Omega$, the matrix $B^\infty(\lambda,\gamma)$ has a unique eigenvalue of smallest real part, which is simple, denoted $\mu_1$.
\end{description}
Then $D(\lambda;\gamma)$ is well defined and analytic on $\Omega$, such that $D(\lambda;\gamma)=0$ if and only if \ref{eq:odesystem} has a solution $\bold{y}(x)$ satisfying \eqref{eq:decayatinfty} and
\begin{equation}
\bold{y}(x)=o(e^{\mu_1 x})\quad\mbox{as $x\to-\infty$}
\end{equation}
\end{thm}

\subsubsection{The KdV Evans Function}
In the case of the KdV equation, the eigenvalue problem may be scaled to
\begin{equation}
\label{eq:kdv-eigenvalue}
\dx L_{\mathrm{KdV}} Y = \dx \paren{-\dx^2Y +Y-3\sech^2\paren{{\frac{1}{2}x}} Y }=\Lambda Y
\end{equation}
Because the speed parameter has been scaled out, there is only one eigenvalue parameter, $\Lambda$.

Making the identification
\begin{equation}
\bold{y} = \threedrowvec{Y}{\dx Y}{L_{\mathrm{KdV}} Y}^T
\end{equation}
$\bold{y}$ satisfies the the dynamical system
\begin{align}
\label{eq:kdv-dynamical-system}
\dot{\bold{y}}&= B_\mathrm{KdV}(x,\Lambda) \bold{y}\\
\label{eq:kdv-B-matrix}
B_\mathrm{KdV}(x,\Lambda) &= \begin{pmatrix} 0 & 1  & 0  \\
1-  3\sech\paren{{\frac{1}{2}x}}^2 &0 & -1  \\ 
\Lambda & 0  & 0 \end{pmatrix}
\end{align}

A complete description of the associated Evans may be found in \cite{Pego94}.  We summarize:
\begin{thm}(The KdV Evans Function )

\begin{description}
  \item[(a)]The Evans function $D_{\mathrm{KdV}}(\Lambda)$ associated with \eqref{eq:kdv-eigenvalue} is given by
  \[
  D_{\mathrm{KdV}}(\Lambda) = \paren{\frac{\mu_1(\Lambda)+1}{\mu_1(\Lambda)-1}}^2
  \]
  where $\mu_1(\Lambda)$ denotes the root of $\mu^3-\mu+\Lambda=0$ of minimal real part.
  \item[(b)]The domain of $D_\mathrm{KdV}(\Lambda)$ is the slit complex plane
  \[
  \Delta_{\mathrm{KdV}} = \mathbb{C}\setminus\left(-\infty,-\sqrt{\frac{4}{27}}\right]
  \]
  \item[(c)]The essential spectrum of $A_{\mathrm{KdV}}:L_a^2\to L_a^2$ is a curve contained entirely in the domain $\{\lambda: \Re\lambda <-\eps\}$ for some $\eps>0$.  Furthermore, if $ \Delta_{\mathrm{KdV}}^+(a)$ denotes the component of $\C \setminus \sigma_{\textrm{ess}}(A_\mathrm{KdV})$ that contains the right half-plane, then $D_\mathrm{KdV}(\Lambda)$ has no zeros in $ \Delta_{\mathrm{KdV}}^+(a)$ except for a zero of multiplicity two at $\Lambda=0$.
\end{description}
\label{thm:kdv-evans}
\end{thm}

\subsubsection{The Evans Function applied}
\label{sec:evansapplied}
The eigenvalue problem $A Y = \lambda Y$,
\[
\set{I - \dx \bracket{\f_c^n \dx \paren{\f_c^{-m}\cdot}}}^{-1}\dx\set{-c \f_c^n\dx^2\paren{\f_c^{-m}\cdot} + \bracket{c-n\f_c^{-1} + cn \paren{\f_c^{-1}-1}}}Y = \lambda Y
\]
is equivalent to 
\begin{equation}
\label{eq:eigenvalue}
\boxed{\dx L_cY-\lambda\bracket{I-\dx\paren{\f_c^n\dx\paren{\f_c^{-m} \cdot}}}Y=0}
\end{equation}

Defining
\begin{equation}
\label{eq:eigenfunction-to-ODE-system}
\bold{y} = \threedrowvec{\f_c^{-m} Y}{\dx \paren{\f_c^{-m} Y}}{L_c Y + \lambda\f_c^n\dx\paren{\f_c^{-m} Y}}^T
\end{equation}
$\bold{y}$ solves the dynamical system 
\begin{align}
\dot{\bold{y}} &= B(x,\lambda,c) \bold{y}\\
\label{eq:B-matrix}
B(x,\lambda,c)&=\begin{pmatrix} 0 & 1  & 0  \\
c^{-1}\f_c^{m-n}\bracket {c-n\f_c^{-1} + cn \paren{\f_c^{-1}-1}} &\lambda /c  & -c^{-1}\f_c^{-n}  \\ 
\lambda\f_c^m & 0  & 0 \end{pmatrix}
\end{align}

The matrix $B$ may be decomposed as $B=B^\infty(\lambda,c)+R(x,\lambda,c)$.  
\begin{align}
\label{eq:B-matrix-inf}
B^\infty(\lambda,c)&=\begin{pmatrix}0 & 1  & 0  \\
\gamma^2 &\lambda /c  & -c^{-1}  \\ 
\lambda & 0  & 0 \end{pmatrix}
\\
\label{eq:remainder1}
R(x,\lambda,\g)&=\begin{pmatrix} 0 &  0 & 0  \\ 
c^{-1}\f_c^{m-n} \bracket {c-n\f_c^{-1} + cn \paren{\f_c^{-1}-1}} -\gamma^2& 0  & -c^{-1}\paren{\f_c^{-n}-1}  \\
\lambda\paren{\f_c^m-1}  &0 & 0\end{pmatrix}
\end{align}

Also note that for the corresponding adjoint eigenvalue problem
\begin{equation}
\label{eq:adj-eigenvalue}
\boxed{-L_c^\star \dx W -\lambda\bracket{I-\f_c^{-m}\dx\paren{\f_c^n\dx \cdot}}W=0}
\end{equation}
under the identifications
\begin{eqnarray}
\bold{z} = \threedrowvec{-c \dx\paren{\f_c^n \dx W}-\lambda \f_c^n \dx W}{c\f_c^n \dx W}{W}
\end{eqnarray}
$\bold{z}$ solves
\begin{equation}
\dot{\bold{z}} = -\bold{z}B(x,\lambda,\gamma)
\end{equation}

\begin{thm}(Properties of the Evans Function)
\label{thm:magma-evans}

\begin{description}
  \item[(a)]  The Evans function is defined and analytic on the set $\Omega\subset \C^2$,
   \begin{equation}
\boxed{\Omega = \left\{(\lambda,\gamma)\mid \gamma\in (0,1)\mbox{ and } \lambda \in \Omega_\gamma  \right\}}
\end{equation}
	with
	 \begin{equation}
\boxed{\Omega_\gamma = \left \{\lambda \mid \Re \lambda >-\lambda_0\right\} \setminus \left(-\lambda_0, -\lambda_\cut(\gamma)\right]}
\label{eq:evans-domain}
 \end{equation}
 where
 \begin{eqnarray}
\lambda_\cut &=&\sqrt{\frac{1}{8}}\sqrt{8 c^2 +20 cn -n^2 - 8c \sqrt{n^2 + 8cn} - n \sqrt{n^2+8c n }}\nonumber\\
&=& \frac{2}{3\sqrt{3}}n \gamma^3 + \frac{8}{9\sqrt{3}}n \gamma^5 + O(\gamma^7)
 \end{eqnarray}
 and $\lambda_0 = n \sqrt{27/16}$.
  \item[(b)]  Given $(\lambda,\gamma)\in \Omega$, and $a \leq a_\star(\gamma)$.  If $\lambda$ is to the right of $\sigma_\ess(A_a)$, then the following are equivalent
  \begin{itemize}
  \item $D(\lambda;\gamma)=0$
  \item $\lambda$ is an $L^2$ eigenvalue of $A_a$
\end{itemize}
  \item[(c)] For such zeros of $D$, the algebraic multiplicity of $\lambda$ as an eigenvalue of $A_a$ is equal to the order of $\lambda$ as a zero of $D(\lambda;\gamma)$.
  \item[(d)] $D(0;\gamma)=\partial_\lambda D(0;\gamma)=0$, hence it is an eigenvalue of algebraic multiplicity \emph{at least} two.
\end{description}
\end{thm}

\begin{rem}
\label{rem:unique-real-parts}
With this construction, for $\Re \lambda <0$, $\lambda \in \Omega_\gamma$, the characteristic polynomial, \eqref{eq:charpoly} not only has a unique root of minimal real part, all roots have distinct real part.  This is stronger than is needed.
\end{rem}

\begin{rem}
$\lambda_\cut$ is labeled as such because there is a branch cut in the Evans function there.
\end{rem}

\proof Part \textbf{(a)} requires the verification of the hypotheses of Theorem \ref{thm:evans} for this system.  Applying the properties of the $\f_c$, Corollaries \ref{cor:solanalyticity} and \ref{cor:solcontinuity}, and examination of \eqref{eq:B-matrix}, $B$ is clearly continuous in its three arguments for $\lambda\in \C$ and $c\geq n$.  In addition, for fixed $x$, it will be analytic in $(\lambda,c)$, or, equivalently, $(\lambda,\gamma)$. Thus property \textbf{(i)} holds.  

By Corollary \ref{cor:soldecay}, the limiting matrix $B^\infty$ exists and $B-B^\infty$ is in $L^1$. This will be uniform on compact subsets of $\C \times [0,1)$, establishing properties \textbf{(ii)} and \textbf{(iii)}.

Lastly, we must verify the existence of $\mu_1$, the unique eigenvalue of minimal real part.  We divide this into two parts, $\Re \lambda \geq 0$ and $\Re \lambda <0$.  The characteristic polynomial, $\mathcal{P}(\mu)$, of $B^\infty$ is 
\begin{equation}
\label{eq:charpoly}
\boxed{c\mathcal{P}(\mu)=(\lambda-c \mu)(1-\mu^2)+n\mu}
\end{equation}
Following the analysis in Section 2(c) of \cite{Pego92} for a similar polynomial in the case of gBBM, one confirms that property \textbf{(iv)} holds for $\Re \lambda \geq 0$ and all $\gamma$, hence the Evans function exists in $\left\{ \Re\lambda\geq 0\right\} \times [0,1)$.

Using the analysis in \cite{Miller96} for Theorem 2.7 of the polynomial, one can conclude the existence of some $\lambda_1>0$ and identify $\widetilde{\Omega}(\gamma)$, such that for all $\gamma\in (0,1)$, a unique root of minimal real part exists for $\lambda$ in the set
\[
\left\{\lambda:\quad \Re\lambda < -\lambda_1 \right\} \setminus (-\lambda_1,-\widetilde{\Omega}(\gamma)]
\]

Alternatively, we give a more precise analysis of the \eqref{eq:charpoly} in Appendix \ref{polynomial_proof} that yields values of $\lambda_0$ and $\widetilde{\Omega}$ given in the proposition.  This concludes the proof of part \textbf{(a)}.

To prove part \textbf{(b)}, we need a lemma regarding the location of $\sigma_\ess(A_a)$.

\begin{lem}
\label{lem:sigma-ess-location}
Let $\gamma \in (0,1)$ and $a\leq a_\star(\gamma)$.  Let $\Omega_+=\Omega_+(\gamma,a)$ denote the component of $\mathbb{C}\setminus \sigma_{\mathrm{ess}}(A_a)$ containing the origin.  
Then for $\lambda \in \Omega_+\cap \Omega_\gamma$, the roots of the characteristic polynomial satisfy the relation
\begin{equation}
\label{eq:a-root-constraint}
\Re \mu_1 < -a < \Re \mu_{j \neq 1}
\end{equation}
\end{lem}

\proof By inspection, if $\lambda \in \sigma_{\textrm{ess}}(A_a)$, there is a root $\mu_j$ of \eqref{eq:charpoly} with $\Re \mu_j=-a$.  Conversely, if there is a root with real part $-a$, then $\lambda$ is in the essential spectrum.  Hence the characteristic polynomial has a root with real part $-a$, if and only if $\lambda \in \sigma_{\textrm{ess}}(A_a)$

As noted in \cite{Pego92}, Section {2 (c)}, for large $\abs{\lambda}$, the roots of the characteristic polynomial \eqref{eq:charpoly} are
\begin{equation*}
-1 + O\paren{\abs{\lambda}^{-1}}, \quad 1 + O\paren{\abs{\lambda}^{-1}}, \quad \lambda/c + O\paren{\abs{\lambda}^{-1}}
\end{equation*}
So for large $\lambda$ in the right half plane, \eqref{eq:a-root-constraint} holds because $a<1$.  Now suppose for some $\lambda \in \Omega_+\cap \Omega_\gamma$ the inequality were false.  Because the $\Re \mu_j$ depend continuously on $\lambda$, equality would have to hold for some $\lambda$, but then it must be that $\lambda \in \sigma_\ess(A_a)$, which we have assumed is not the case.\qed

We now prove part \textbf{(b)} of Theorem \ref{thm:magma-evans}.  If $D(\lambda;\gamma)=0$, then there is a solution to the ODE, $\dot{\bold{y}} = B\bold{y}$ such that
\[
\bold{y}(x) = O(e^{\mu_1 x}) \quad\mbox{as $x\to +\infty$ and}\quad \bold{y}(x) = o(e^{\mu_1 x}) \quad \mbox{as $x\to -\infty$}
\]
Hence, Part 1(d) of Proposition 1.6 of \cite{Pego92}, for sufficiently small $\eps$,
\[
\bold{y}(x) = O(e^{\mu_\star x+\eps\abs{x}}) \quad \mbox{as $x\to -\infty$}
\]
Letting, $W(x)=e^{ax} \f_c(x)^m y_1(x)$,
\[
W(x) = O(e^{(\mu_\star +a)x + \eps \abs{x}})\quad\mbox{as $x\to-\infty$ and}\quad W(x) = O(e^{(\mu_1+a)x})\quad\mbox{as $x\to +\infty$}
\]
From \eqref{eq:a-root-constraint}, $\mu_1+a <0<\mu_\star+a$, so $W$ will decay exponentially fast at $\pm\infty$.  Hence it is an $L^2$ solution to the eigenvalue problem $A_a W = \lambda W$.  

Conversely, if we have an $L^2$ eigenfunction, then it must satisfy
\[
W(x) = O(e^{(\mu_1+a) x}) \quad\mbox{as $x\to +\infty$ and}\quad W(x) = o(e^{(\mu_1+a) x}) \quad \mbox{as $x\to -\infty$}
\]
$Y(x) = e^{-ax}W(x)$ will then satisfy \eqref{eq:eigenvalue} in a classical sense, although it may not be in $L^2$.  However it will satisfy the necessary decay estimates on $\bold{y}$, constructed from $Y$ as in \eqref{eq:eigenfunction-to-ODE-system}, such that $D(\lambda;\gamma)=0$.  This concludes the proof of part \textbf{(b)}.

The proof of part \textbf{(c)} follows that of Lemma 2.9 from \cite{Pego94}.  First, it is proved that if, for a given $\gamma$, $\lambda$ is a zero of order $k$ of $D(\lambda;\gamma)$, then $\lambda$ is an $L^2$ eigenvalue of $A_a$ of algebraic multiplicity at least $k$.  It is then shown that it cannot have algebraic multiplicity greater than $k$.  We omit repeating these details.  Part \textbf{(d)} is then a consequence of \textbf{(c)} and the calculations in Appendix \ref{sec:lambda-zero} that $D(0;\gamma) = \partial_\lambda D(0;\gamma)=0$ for all $\gamma$.
\qed

\begin{rem}  For $\Re\lambda\leq 0$ with $D(\lambda;\gamma)=0$, it is likely, but not proved, that $\lambda$ is not an $L^2$ eigenvalue of $A$.  
\end{rem}

\subsubsection{The Evans Function in the KdV Scaling}
We now introduce $D_\star(\Lambda;\gamma)$, the Evans function for \eqref{eq:eigenvalue} under the KdV scalings introduced in Section \ref{sec:solwave},
\begin{equation}
\label{eq:scalings}
\xi= \gamma x, \quad \lambda = c\Lambda \gamma^3 , \quad \f_c(x) = 1+ \frac{\gamma^2}{n-1}U(\xi(x);\gamma)
\end{equation}
The eigenvalue problem is now
\begin{equation}
\label{eq:scaled-eigenvalue}
\begin{split}
\dxi L_\gamma Y&=\dxi\bracket{-\paren{1+\frac{\gamma^2}{n-1}U}^n\dxi^2\paren{\paren{1+\frac{\gamma^2}{n-1}U}^{-m}\cdot}+ \paren{1+\frac{\gamma^2}{n-1}U}^{-1}\paren{1-U} }Y\\
&=\Lambda\bracket{I-\gamma^2\dxi\paren{\paren{1+\frac{\gamma^2}{n-1}U}^n\dxi\paren{\paren{1+\frac{\gamma^2}{n-1}U}^{-m}\cdot }}}Y
\end{split}
\end{equation}
Recall that $U = U(\xi; \gamma)$ is the solution of \eqref{eq:scaledsolwave}, and for $\gamma=0$, $U=U_\star$, the KdV soliton.

We can construct a dynamical system formulation of \eqref{eq:scaled-eigenvalue}, defining the vector $\bold{Y}$ as
\begin{equation}
\bold{Y} = \threedcolvec{\paren{1+\frac{\gamma^2}{n-1}U}^{-m} Y }{\dxi \bracket{\paren{1+\frac{\gamma^2}{n-1}U}^{-m} Y }}{L_\gamma Y + \gamma^2 \Lambda\paren{1+\frac{\gamma^2}{n-1}U}^{n} \dxi \bracket{\paren{1+\frac{\gamma^2}{n-1}U}^{-m} Y } }
\end{equation}
which satisfies
\begin{align}
\label{eq:scaled-dynamical-system}
\dot{\bold{Y}}&= B_\star(\xi,\Lambda,\gamma) \bold{Y}\\
\label{eq:B-scaled-matrix}
B_\star(\xi,\Lambda,\gamma)&=\begin{pmatrix} 0 & 1  & 0  \\
\paren{1+\frac{\gamma^2}{n-1}U(\xi;\gamma)}^{m-n-1}\paren{1-U(\xi;\gamma)} &\gamma^2\Lambda  & -\paren{1+\frac{\gamma^2}{n-1}U(\xi;\gamma)}^{-n}  \\ 
\Lambda\paren{1+\frac{\gamma^2}{n-1}U(\xi;\gamma)}^{m} & 0  & 0\end{pmatrix}
\end{align}
As $\xi \to \infty$, the matrix is
\begin{equation}
B_\star^\infty(\Lambda,\gamma)= \begin{pmatrix} 0 & 1  & 0  \\
1 &\gamma^2\Lambda & -1  \\ 
\Lambda & 0  & 0 \end{pmatrix}
\end{equation}
which has the characteristic polynomial
\begin{equation}
\label{eq:scaled-charpoly}
P_\star(\nu;\Lambda,\gamma)=\nu^3 - \gamma^2 \Lambda \nu^2 - \nu + \Lambda
\end{equation}

A few remarks about the scaled problem.  The assumptions stated in Theorem \ref{thm:evans} remain the same, except now the matrix under inspection is $B_\star$, with eigenvalue parameters $(\Lambda,\gamma)$.  $\bold{Y}$ and $\bold{y}$ are related:
\begin{equation*}
\bold{y}(x)=\threedcolvec{Y_1(\xi(x))}{\gamma Y_2(\xi(x))}{ c\gamma^2 Y_3(\xi(x))}
\end{equation*}
At $\gamma=0$, \eqref{eq:scaled-eigenvalue} is
\[
\dxi L_0 Y= \dxi\bracket{-\dxi^2Y + (1-U_\star(\xi;0)Y} = \Lambda Y 
\]
the KdV eigenvalue problem, \eqref{eq:kdv-eigenvalue} and 
\[
B_\star(\xi,\Lambda,0)=\begin{pmatrix}0 & 1  & 0  \\
1-U_\star(\xi) &0 & -1  \\ 
\Lambda & 0  & 0 \end{pmatrix}
\]
is the matrix for the KdV dynamical system.

\begin{prop}(Scaled Evans Function)
\label{prop:scaled-evans}

\begin{description}
\item[(a)] $D_\star(\Lambda;\gamma)$ is defined and analytic on the set $\Delta \subset \C^2$,
\begin{equation}
\Delta = \left\{(\Lambda,\gamma)\mid \gamma\in[0,1)\mbox{ and } \lambda \in \Delta_\gamma \right\}
\end{equation}
where, for $\gamma>0$,
\begin{equation*}
\begin{split}
\Delta_\gamma &= \left\{\Lambda \mid \Re \Lambda > -\frac{\lambda_0}{c\gamma^3}\right\} \setminus \left( -\frac{\lambda_0}{c\gamma^3}, -\frac{\lambda_\cut(\gamma)}{c\gamma^3}\right]\\
&= \left\{\Lambda \mid \Re \Lambda > -\gamma^{-3}\sqrt{\frac{27}{16}}+O(\gamma^{-1})\right\} \setminus \left(  -\gamma^{-3}\sqrt{\frac{27}{16}}+O(\gamma^{-1}), -\sqrt{\frac{4}{27}} +O(\gamma^2)\right]
\end{split}
\end{equation*}
$\lambda_0$ and $\widetilde{\Omega}$ as defined by Theorem \ref{thm:magma-evans}.  When $\gamma=0$,
\[
\Delta_0 = \Delta_\kdv = \C \setminus \left(-\infty, -\sqrt{4/27}\right]
\]
\item[(b)] For fixed $\gamma \in (0,1)$ and $\Lambda \in \Delta_\gamma$,
\[
D_\star(\Lambda;\gamma) =D (c\Lambda\gamma^3;\gamma)
\]
\item[(c)]  For $\Lambda \in \Delta_0$,
\[
D_\star(\Lambda;0) = D_\kdv(\Lambda)
\]
\end{description}
\end{prop}

\proof The proof of these statements follows that of Proposition 2.8 in \cite{Miller96} and Theorems 4.9-4.11of \cite{Pego97}.

\noindent\underline{Proof of Part \textbf{(a)}:} For $\gamma>0$, as in Theorem \ref{thm:magma-evans}, we must identify a set in $\C^2$ in which the hypotheses of Theorem \ref{thm:evans} are valid.   Parts \textbf{(i-iii)} are obvious as the solitary wave $U(\xi;\gamma)$ decays exponentially in $\xi$ and, for fixed $\xi$, will be analytic in $\gamma$. We are left to verify part \textbf{(iv)}.  The characteristic polynomial of $B_\star^\infty$ is \eqref{eq:scaled-charpoly}.  As noted in \cite{Miller96}, the roots of $P_\star$ are realted to those of $P$, \eqref{eq:charpoly}, by
\[
\mu(\lambda,\gamma)= \gamma \nu(\Lambda,\gamma)
\]
So $P_\star$ will have a unique root of minimal real part for a given $\Lambda$ and $\gamma$ when $P$ has such a unique root for $\lambda=c \Lambda \gamma^3$.  Therefore, for $\gamma\in (0,1)$, if $\Lambda$ is in the set
\[
\Delta_\gamma = \frac{1}{c\gamma^3}\Omega_\gamma
\]
\textbf{(iv)} will be satisfied.  If $\gamma=0$, \eqref{eq:scaled-dynamical-system}, \eqref{eq:B-scaled-matrix} coincides with the KdV system, for which $\Delta_0 = \mathbb{C}\setminus (-\infty, -\sqrt{4/27}]$.  Clearly, as $\gamma\to 0$, $\Delta_\gamma$ limits to $\Delta_0$.

\noindent\underline{Proof of Part \textbf{(b)}:} From part \textbf{(a)}, $\lambda=c\Lambda\gamma^3 \in \Omega_\gamma$ and, by construction,
\begin{align*}
y_1(x,\lambda,\gamma) &\sim e^{\mu_1 x}\quad\text{as $x\to+\infty$}\\
Y_1(\xi,\Lambda,\gamma) &\sim  e^{\nu_1 \xi}\quad\text{as $\xi\to+\infty$}
\end{align*}
At $\xi = \gamma x $, $\mu_1 = \gamma \nu_1$, $\lambda = c\Lambda \gamma^3$, $y_1(x,\lambda,\gamma)=Y_1(\xi,\Lambda,\gamma)$.  Using the \textbf{\emph{Transmission Coefficient}} interpretation of the Evans function, we then have
\begin{align*}
\bold{y}(x,\lambda,\gamma) &\sim D(\lambda;\gamma) e^{\mu_1 x}\threedrowvec{1}{\mu_1}{\lambda/\mu_1}^T\quad\text{as $x\to-\infty$}\\
\bold{Y}(\xi,\Lambda,\gamma)&\sim D_\star(\Lambda;\gamma) e^{\nu_1 \xi} \threedrowvec{1}{\nu_1}{\Lambda/\nu_1}^T\quad\text{as $\xi\to -\infty$}
\end{align*}
implying
\[
D(c\Lambda\gamma^3;\gamma) = D_\star(\Lambda;\gamma)
\]

\noindent\underline{Proof of Part \textbf{(c)}:}  Trivially, when $\gamma =0$, this is the KdV Evans function problem exactly.
\qed

\begin{prop} 
\label{prop:uniform-evans-convergence}

Let  $\vareps_2\in\paren{0, \sqrt{4/27}}$ and $M>0$.  Set
\[
\mathcal{O} =\left\{\Re \Lambda \ge -\vareps_2,\quad \abs{\Lambda}<M \right\}
\]
Then for all $\gamma$ sufficiently small, $\mathcal{O}\subset \Delta_\gamma$ and
\[
\lim_{\gamma\to 0} D_\star(\Lambda;\gamma) =D_\star(\Lambda;0) = D_{\mathrm{KdV}}(\Lambda)
\]
with uniform convergence for $\Lambda\in \mathcal{O}$.
\end{prop}
\proof Clearly, for $\gamma$ sufficiently close to zero, $\mathcal{O} \subset \Delta_\gamma$.  Since $\mathcal{O} $ is compact and $D_\star$ is analytic in both arguments, 
\[
\sup_{\Lambda\in \mathcal{O} } \abs{D_\star(\Lambda;\gamma)-D_\star(\Lambda;0)}
\]
may be made arbitrary small by taking $\gamma$ sufficiently close to zero.\qed

\begin{thm}
\label{thm:evans-zeros}
There exists $\gamma_\star \in (0,1)$ such that for all $\gamma \in (0,\gamma_\star]$, $a\leq a_\star(\gamma)$, there exists $\varepsilon=\varepsilon(\gamma,a)>0$, such that:
\begin{itemize}
\item The only eigenvalue of $A_a$ with $\Re \lambda \geq -\varepsilon$ is $\lambda =0$, with algebraic multiplicity two.
\item The only zero of $D(\lambda;\gamma)$ with $\Re \lambda \geq -\varepsilon$ is $\lambda =0$, a root of order two.
 \end{itemize}
\end{thm}

\proof Following the Proof of Theorem 2.1 in \cite{Miller96} we break a half-plane into two parts, a bounded set about the origin and its unbounded complement.  The operator estimates from Proposition \ref{prop:large-eigenvalue} will rule out eigenvalues in the unbounded part, and the Evans function in the KdV scaling will control eigenvalues in the bounded part.

Applying Proposition \ref{prop:large-eigenvalue} \textbf{(a)} with $\vareps_1 = 1/4$, there exists $M>0$, such that for any $c\in (n,4n]$, $a\leq a_\star(\gamma)$, and $\lambda \in \Omega_{\mathrm{U}}$
\begin{equation*}
\Omega_{\mathrm{U}} = \left\{\Re \lambda\ge  - \frac{1}{2} a c\gamma^2,\quad \abs{\lambda} > c M\gamma^3  \right\}
\end{equation*}
$\norm{C(\lambda)}_{L^2}< 1$.  By Proposition \ref{prop:compact-operator} \textbf{(c)}, such $\lambda$ cannot be $L^2$ eigenvalues of $A_a$.  Additionally, by Theorem \ref{thm:magma-evans} part \textbf{(b)}, $D(\lambda;\gamma)\neq 0$ on this set.

Let  $\vareps_2 =\frac{1}{4}$ and set
\[
\mathcal{O}  = \left\{ \Lambda : \Re \Lambda \ge -\vareps_2,\quad \abs{\Lambda} \le M\right\}
\]
Define
\[
m = \min\{ \abs{D_\kdv(\Lambda)}\min \Lambda \in \partial\mathcal{O}\}
\]
By Proposition \ref{prop:uniform-evans-convergence}, there exists a $\gamma_\star\leq 1/2$ such that for all $\gamma\in\bracket{0,\gamma_\star}$,
\[
 \abs{D_\star(\Lambda;\gamma)-D_\star(\Lambda;0)}=  \abs{D_\star(\Lambda;\gamma)-D_\mathrm{KdV}(\Lambda)}< m\quad \mbox{for all $\Lambda \in \partial \mathcal{O} $}
\] 

By Theorem \ref{thm:kdv-evans}, the only root of $D_{\mathrm{KdV}}$ in $\mathcal{O} $ is at $\Lambda=0$, with multiplicity two.  Applying Rouch\'e's Theorem (see \cite{Knopp96}  for example), for all $\gamma \in \bracket{0,\gamma_\star}$, $D_\star(\cdot,\gamma)$ and $D_\mathrm{KdV}$ have the same number of roots in $\mathcal{O}$.  By Proposition \ref{prop:scaled-evans} \textbf{(c)}, if $\gamma\in (0,\gamma_\star]$ and $\Lambda \in\mathcal{O}$, $D_\star(\Lambda;\gamma)=D(c\Lambda\gamma^3;\gamma)$; therefore on the set $\Omega_{\mathrm{B}}= c \gamma^3 \mathcal{O}$, $D(\lambda;\gamma)$ also has only two zeros.  Theorem \ref{thm:magma-evans} \textbf{(d)} asserts that $\lambda=0$ is a root of multiplicity two.  

If we now set
\begin{equation}
\vareps(\gamma,a) = \min\left\{\frac{1}{2} a c\gamma^2,\quad \frac{1}{4} c \gamma^3, \frac{1}{2}\omega \right\}
\label{eq:eps-bound}
\end{equation}
then by Theorem \ref{thm:magma-evans} \textbf{(c)}, $\lambda=0$ is the only eigenvalue of $A_a$ with $\Re \lambda \geq -\vareps$.\qed

\begin{cor}
\label{cor:functional-monotonicity}
For $\gamma \in (0,\gamma_\star]$, $\partial_c \N[\f_c]\neq 0$.
\end{cor}
\proof This is a consequence the preceding theorem and the relation $\partial_\lambda^2 D(0;\gamma) \propto\partial_c \N[\f_c]$, see Appendix \ref{sec:lambda-zero}.
\qed

\subsection{The Evans Function outside the KdV Scaling}
\label{sec:large-amplitude}

\begin{figure}[t]
\begin{center}
$\begin{array}{cc}
\includegraphics[width=3.25in]{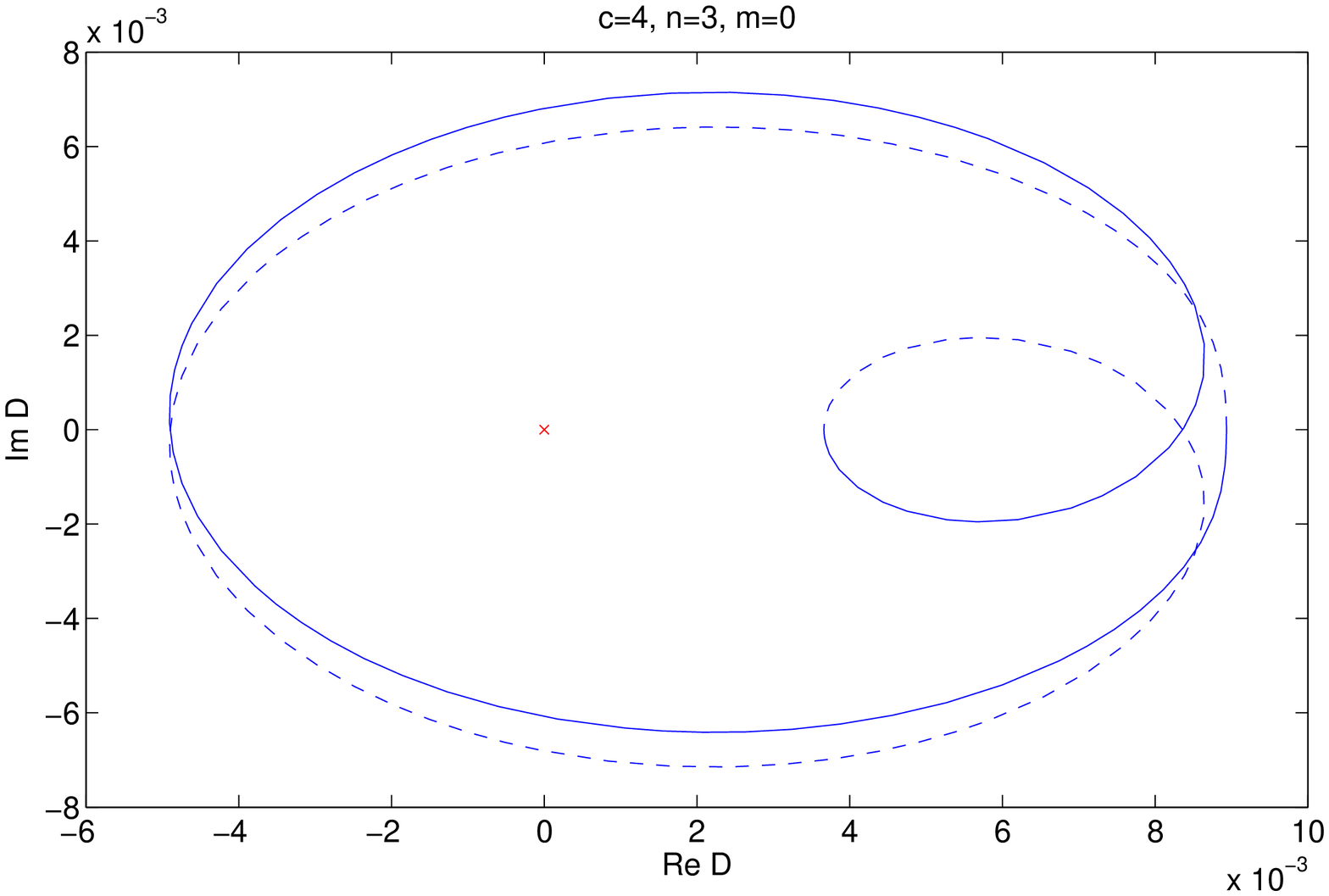}&\includegraphics[width=3.25in]{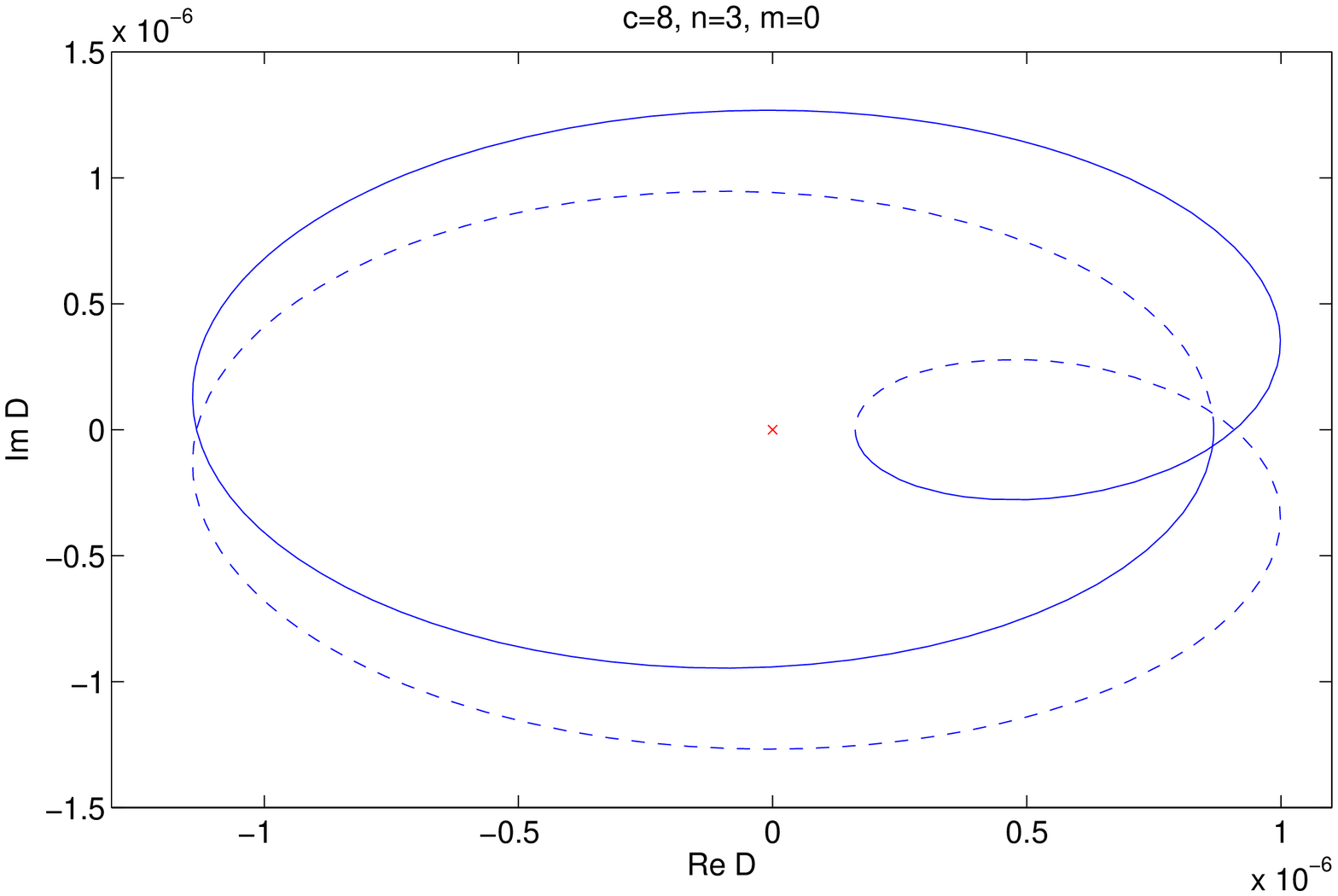}\\
\mbox{\textbf{(a)}}&\mbox{\textbf{(b)}}
\end{array}$
\caption{In \textbf{(a)} we evaluate $D(\cdot;c=4)$ on a portion of the strip $\Re\lambda = -1/5$.  In \textbf{(b)} we evaluate $D(\cdot;c=8)$ on a portion of the strip $\Re\lambda = -1$.  Since $D(\overline{\lambda})=\overline{D(\lambda)}$ we only compute $\Im \lambda >0$, and then reflect; this is the dashed curve.  Both curves wrap around the origin, marked by x, twice.  In both cases, $n=3$ and $m=0$.}
\label{fig:winding}
\end{center}
\end{figure}

For $\gamma>\gamma_\star$, one may compute the Evans function numerically to assert than a given $\lambda$ is not an eigenvalue.  Moreover, the winding number of the image of the Evans function evaluated on the line $\Re \lambda  = -\lambda_1<0$ equals the number of zeros in the set $\Re\lambda >-\lambda_1$.  Therefore, one might evaluate the Evans function on such a line and exam the plot, as we do in Figure \ref{fig:winding}.  

These plots indicate that $\lambda=0$ is the only zero in the closed right half-plane for the values of $c$, $n$, and $m$ under consideration.  Up to the acceptance of these numerics, this extends Theorem \ref{thm:evans-zeros}.  Note that we do not evaluate out to $-\lambda_1 + \imath \infty$, but merely compute at sufficiently large values of $\lambda$ such that we are near the asymptotic value of the Evans function.  It may be proven that there exists $D_\infty(\gamma)$ such that
\[
\lim_{\abs{\lambda}\to \infty, \lambda \in \Omega} D(\lambda;\gamma) = D_\infty(\gamma)
\]
A further numerical computation will reveal that this value is nonzero.

In the Hamiltonian case, $n+m=0$, an analytical result is possible for $\gamma>\gamma_\star$.  The linearized operator, $A$, may be written as $A=J_c L_c$,
\begin{align}
J_c&= \bracket{I-\dx \paren{\f_c^n\dx\paren{\f_c^n\cdot}}}^{-1}\dx & J_c^\star &= -J_c\\
L_c&=-c\f_c^n\dx^2\paren{\f_c^n}+\paren{c-n\f_c^{-1}+cn\paren{\f_c^{-1}-1}} & L_c^\star &= L_c
\end{align}
This structure permits an extension of the Theorem \ref{thm:evans-zeros} beyond the KdV regime, given below in Theorem \ref{thm:hamiltonian-extension}.  However, this is absent for general $n$ and $m$.  
\begin{rem}
This section is the only place where the analyticity of the Evans function in the $\gamma$ argument is used.  In turn, this is the only place requiring analyticity of $\f_c$ in $c$ from Corollary \ref{cor:solcontinuity} \textbf{(b)}.  For the results in the preceding section, \emph{joint continuity} of $D(\lambda;\gamma)$ in its two arguments is sufficient.
\end{rem}

\begin{lem}
\label{lem:imaginaryeigenvalues}
In the Hamiltonian case, the following are equivalent for $\Re\lambda \geq 0$, $a\leq a_\star(\gamma)$:
\begin{itemize}
\item $\lambda$ is an $L^2$ eigenvalue of $A$.
\item $D(\lambda;\gamma)=0$.
\end{itemize}
\end{lem}

\proof For $\Re \lambda >0$, we know that $\mu_\star = \min \set{\Re\mu_2,\Re\mu_3}>0$, so an eigenfunction, having submaximal growth at $-\infty$, must decay exponentially fast.  This is very similar to the relation in the case of the weighted operator from Theorem \ref{thm:magma-evans} \textbf{(b)}.

For $\Re \lambda =0$, the proof relies on the $JL$ structure of the operator $A$.  See the proof of Theorem 3.6 in \cite{Pego92}.\qed

\begin{lem}
\label{lem:embedded}
If $\lambda$ is a nonzero purely imaginary eigenvalue, then $\partial_\lambda D(\lambda;\gamma)\neq 0$.
\end{lem}
\proof The proof is by contradiction. Let $Y+$ be the corresponding eigenfunction, $A Y^+ = \lambda Y^+$.  It may be proven that the subspace $\mathcal{Y}=\mathrm{span}\set{Y^+,\overline{Y^+}}$ satisfies 
\[
\inner{L_c u}{v} = 0 \quad\text{for all $u,v\in \mathcal{Y}$.}
\]
As $\mathcal{Y} \cap \ker (L_c) = \set{0}$, we may apply Lemma 3.3 of \cite{Pego92} to conclude $\dim \mathcal{Y}\leq 1$, a contradiction. See Lemma 3.3 from \cite{Pego94} for details.\qed

\begin{lem}(Analytically confirmed for $n=2$)
\label{lem:hamiltonianhalfplane}

In the Hamiltonian case, assuming $\partial_c \mathcal{N}[\f_c]\neq 0$ for all $c$, then for all $c$, there are no eigenvalues with $\Re\lambda >0$. 
\end{lem}

\proof By Theorem \ref{thm:evans-zeros}, the result holds for $ c \leq c_\star$.  We argue by contradiction to extend it beyond $c_\star$. Assume for some $\gamma_0>\gamma_\star$, there exists $\lambda_0$, $\Re \lambda_0>0$, such that $D(\lambda_0;\gamma_0)=0$.  If $\Im\lambda_0 \neq 0$, then $D(\overline{\lambda_0};\gamma_0)=0$ and there would be two eigenvalues in the right half-plane.  But by Theorem 3.1 of \cite{Pego92}, $A=J_c L_c$ has no more eigenvalues (counting multiplicity) with $\Re \lambda >0$ than $L_c$ does with $\Re \lambda <0$.  As is discussed in \cite{Simpson07a}, $L_c$ has exactly one negative eigenvalue; therefore, this is a contradiciton, so $\lambda_0$ is real.  

Because the number of zeros in the right half-plane is at most one, counting multiplicity, we know $\partial_\lambda D(\lambda_0;\gamma_0)\neq 0$.  Applying the implicit function theorem,  we get an analytic function, $\lambda(\gamma)$, defined in a neighborhood of $\gamma_0$, such that $\lambda(\gamma_0)=\lambda_0$ and $D(\lambda(\gamma);\gamma)=0$.  

Let $\mathcal{O}$ be  the maximal domain of analyticity of $\lambda(\gamma)$.  In a sufficiently small neighborhood of $\gamma_0$, $\Re\lambda(\gamma)>0$.  For real-valued $\gamma$ in this neighborhood, we must have, by the above argument about complex-conjugates, that $\Im  \lambda(\gamma)=0$.  Considering the power series expansion of $\lambda(\gamma)$, about $\gamma_0$, $\lambda(\gamma)$ will be real-valued for real-valued $\gamma\in \mathcal{O}$.

Let 
\[
\gamma_1 = \inf\set{\gamma\in (\gamma_\star/2,\gamma_0)\cap \mathcal{O} \mid D(\lambda(\gamma);\gamma)=0}
\]
For all $\gamma\in \bracket{\gamma_1,\gamma_0}$, we must have $\lambda(\gamma)>0$.  Suppose not.  Then, by continuity, for some $\gamma$, we must have $\lambda(\gamma)=0$.  But this would imply that $\lambda=0$ was a root of multiplicity three, contradicting the assumption on $\N$, which ensures it is a root of multiplicity two.

$\lambda(\gamma)$ may be analytically continued down till at least $\gamma_\star/2$. If not, then $\gamma_1>\gamma_\star/2$ and $\partial_\lambda D(\lambda(\gamma_1);\gamma_1)\neq 0$ since this root must be simple.  Therefore we could apply the implicit function theorem again, and extend $\lambda(\gamma)$ below $\gamma_1$, contradicting its minimality.

But then $D(\lambda(\gamma_\star);\gamma_\star)=0$, and $\lambda(\gamma_\star)>0$, contradicting Theorem \ref{thm:evans-zeros}.\qed

\begin{rem}
An analogous result may be found in Theorem  \cite{Pego92} for gKdV, gBBM, and a Boussinesq equation.  However, the argument there is very different because it may be proven that $D(\lambda) \to 1$ as $\abs{\lambda}\to \infty$.  This does not hold for the Evans function associated with \eqref{eq:magma}, due to the appearance of a nonlinearity in the dispersive term.
\end{rem}

\begin{thm}(Analytically confirmed for $n=2$)
\label{thm:hamiltonian-extension}

In the Hamiltonian case, assuming $\partial_c \mathcal{N}[\f_c]\neq 0$ for all $c$, Theorem \ref{thm:evans-zeros} may be extended to all  $\gamma \in(\gamma_\star,1)$, except for a discrete set whose only possible accumulation point is $\gamma=1$.
\end{thm}

\proof By Lemma \ref{lem:hamiltonianhalfplane}, if Theorem \ref{thm:evans} were false for some $\gamma>\gamma_\star$, it would be due to a zero appearing on the imaginary axis.  We will prove by contradiction that the set
\begin{equation}
E = \set{\gamma\in [0,1)\mid \text{there exists $\beta >0$ such that $D_\star(\imath\beta;\gamma)=0$}}
\end{equation}
has no accumulation points.  We consider only positive $\beta$'s because if $\imath \beta$ is a root then so is $-\imath\beta$.  This follows the proofs of Theorem 3.6 of \cite{Pego94} and Theorem 2.1 of \cite{Miller96}.

Assuming $E$ has a limit point, there exists a sequence, $\gamma_j\in E$, $\gamma_j \to \gamma_0\in E$ as $j\to \infty$.  Taking a subsequence if neccessary, $\gamma_j$ and $\gamma_0$ are bounded away from $\gamma=1$.  We will now rule out large eigenvalues, and then argue by contradiction to rule out small eigenvalues.

Applying Proposition \ref{prop:large-eigenvalue} to this range of $\gamma$ values, there will exist $M>0$, such that the corresponding $\beta_j> 0$ must satisfy $\beta_j\leq M$.  Taking a subsequence if necessary, $\beta_j \to \beta_0$, $\beta_0\leq M$, $D_\star(\imath\beta_j;\gamma_j)=0$, and, by continuity, $D_\star(\imath\beta_0; \gamma_0)=0$.

Note that $\beta_0 \neq 0$.  $\Lambda=0$ is always a root of order at least two. The assumption $\partial_c \mathcal{N}[\f_c] \propto \partial_\lambda^2 D(0;\gamma) = (c^2 \gamma^6)^{-1}\partial_\Lambda^2 D_\star(0;\gamma)\neq 0$ forces it to be a root of order exactly two.  But $\beta_0=0$ would imply that it was a zero of at least four, a contradiction.

Applying Lemma \ref{lem:embedded},  $\partial_\Lambda D_\star(\imath\beta_j;\gamma_j)\neq 0$ and $\partial_\Lambda D_\star(\imath\beta_0;\gamma_0)\neq 0$.  By the implicit function theorem, there is an analytic function $\Lambda_0(\gamma)$ defined in a neighborhood of $\gamma_0$, such that $\Lambda_0(\gamma_0) = \imath \beta_0$, and $D_\star(\Lambda_0(\gamma);\gamma)=0$.  By considering the power series expansion of $\Lambda_0(\gamma)$  about $\gamma_0$, we see, by taking $\gamma_j$ sufficiently close to $\gamma_0$, that $\Lambda_0(\gamma)$ is purely imaginary for real $\gamma$ in its maximal domain of analyticity, $\mathcal{O}$.

Let
\[
\gamma_1 = \inf\set{\gamma\in [0,\gamma_0)\cap \mathcal{O} \mid D_\star(\Lambda_0(\gamma);\gamma)=0}
\]
For all $\gamma\in [\gamma_1,\gamma_0]$, we must have $\Im \Lambda_0(\gamma)>0$. If not, then by continuity would exist $\gamma \in (\gamma_1,\gamma_0)$, for which $\Lambda_0(\gamma)=0$, yielding a contradiction as before.  

Suppose $\gamma_1>0$.  $\Im \Lambda_0(\gamma_1)>0$ because, if not, then by continuity there would exist $\gamma \in (\gamma_1,\gamma_0)$, for which $\Lambda_0(\gamma)=0$, leading to a contradiction again.  Therefore, we may be sure that $\partial_\Lambda D_\star(\Lambda_0(\gamma_1);\gamma_1)\neq 0$.  We may then apply the implicit function theorem, allowing us to continue $\Lambda_0$ below $\gamma_1$, another contradiction.  Therefore $\gamma_1=0$.  But then $\Im\Lambda_0(0)>0$ and $0=D_\star(\Lambda_0(0); 0)=D_\kdv(\Lambda_0(0))$, a contradiction.\qed

\begin{rem}
This result is limited by our inability to analytically evaluate the functional $\N[\f_c]$.  The authors were similarly stymied in \cite{Simpson07a}, where the orbital stability of the solitary waves relies on proving $\partial_c \mathcal{N}[\f_c] >0$.  Here, as there, one may numerically evaluate the functional and observe its monotonicity in the speed argument.  See \cite{Simpson07a} for the case $n=2$.
\end{rem}

This result, along with \eqref{eq:ess-spectrum} and Theorem \ref{thm:evans-zeros}, completes the proof of Theorem \ref{thm:linear}.

\subsection{The Generalized Kernel}
\label{sec:kernel}

\begin{prop}
\label{prop:kernel}

Let $c>n$, $a \leq a_\star(c)$ and assume $\partial_c\mathcal{N}[\f_c]\neq0$.  Define
\begin{align*}
\tilde{\xi}_1&= \dx \f_c\\
\tilde{\xi}_2&= \dc \f_c\\
\tilde{\eta}_1& = \Theta\bracket{I-\f_c^{-m}\dx\paren{\f_c^n \dx\cdot}} \int_{-\infty}^{x}\paren{L_c^\star}^{-1}\bracket{I-\f_c^{-m}\dx\paren{\f_c^n \dx\cdot}} \int_{-\infty}^{x} \frac{\dx \f_c}{\f_c^{n+m}}dx\\
\tilde{\eta}_2& = \Theta\bracket{I-\f_c^{-m}\dx\paren{\f_c^n \dx\cdot}} \int_{-\infty}^{x} \frac{\dx \f_c}{\f_c^{n+m}}dx\\
\Theta &= \paren{\partial_c\mathcal{N}[\f_c]}^{-1}
\end{align*}

For $j = 1,2$, set
\begin{equation*}
\xi_j = e^{ay} \tilde{\xi}_j\quad\text{and}\quad \eta_j = e^{-ay}\tilde{\eta}_j
\end{equation*}
Then $\{\xi_1, \xi_2\}$ and $\{\eta_1, \eta_2\}$ are biorthogonal bases for $\ker_g(A_a)$ and $\ker_g(A_a^\star)$, $\inner{\xi_i}{\eta_j}=\delta_{ij}$.  They satisfy the relations
\begin{align*}
A_a \xi_1 &= 0 &A_a^\star\eta_2& = 0\\
 A_a \xi_2 &= -\xi_1  & A_a^\star\eta_1 & = -\eta_2
\end{align*}
\end{prop}

\proof It is easy to verify $A \tilde{\xi}_1=0$, $A \tilde{\xi}_2 = -\xi_1$ and $A^\star \tilde{\eta}_2 = 0$.  
\[
\ker(L_c^\star) = \mathrm{span}\set{\f_c^{-n-m}\dx \f_c }
\] 
The kernel is orthogonal to $\tilde{\eta}_2$ because $\tilde{\eta}_2$ is an even function while the kernel is odd.  Threfore, $\tilde{\eta}_1$ is well defined and $A^\star \tilde{\eta}_1 = -\tilde{\eta}_2$.
\qed

\section{Semigroup Decay}
\label{sec:semigroup}
The following result proves the convective stability of solitary waves under the \emph{linearized} flow.  As we will do in Section \ref{sec:mainproof}, this may be employed to prove full nonlinear stability.
\begin{prop}
\label{prop:semigroup}

Assume $\gamma \in (0,1)$, $a\leq a_\star(\gamma)$, and there exists $\vareps>0$ such that $\lambda = 0$ is only eigenvalue of $A_a$ with $\Re \lambda \geq -\vareps$.  Then the initial-value problem
\begin{align*}
w_t &= A_a w\\
w(0) &= w_0 \in H^1 \cap \kerg(A_a^\star)^\bot
\end{align*}
has a unique solution $w(t)=e^{A_a t} w_0 \in C_0([0,\infty);H^1)$ with
\begin{equation}
\label{eq:semigroup1}
\norm{w(t)}_\h{1} \leq C e^{-b t} \norm{w_0}_\h{1}
\end{equation}
for some $C>0$ and $b>0$.
\end{prop}

\begin{rem}
There exists a $b_{\max}>0$ such that \eqref{eq:semigroup1} will hold for all $b \in (0, b_{\max})$.  In particular, for $\gamma \in (0, \gamma_\star]$, $b_{\max} \geq \vareps$, $\vareps$ and $\gamma_\star$ as defined in Theorem \ref{thm:evans-zeros}.
\end{rem}

\proof This is based on a result due to Pr\"uss, \cite{Pruss84}:
\begin{thm}
\label{thm:pruss}

Let $B$ be the infinitesimal generator of a $C_0$ semigroup on a Hilbert space $Z$.  Let $b>0$.  If there exists $M>0$ such that 
\[
\norm{(\lambda I - B)^{-1}}_{Z\to Z} \leq M \quad \text{for all $\Re \lambda > -b$}
\]
then $\norm{e^{Bt}}_{Z\to Z} \leq e^{-b t}$.
\end{thm}

Following the approach in \cite{Miller96} for Proposition 3.1, we first show that $A_a$ is the infinitesimal generator of a $C_0$-semigroup on $H^1$. Examining the Fourier symbol of $A_a^\infty$, $A_a^\infty$ is such a semigroup.  As equation \eqref{eq:lineardifference} shows, $A_a - A_a^\infty$ is a bounded operator, so we may apply Theorem 3.1.1 of \cite{Pazy83}, that bounded perturbations of infinitesimal generators are also infinitesimal generators.

Consider the Hilbert space 
\[
Z = \h{1} \cap \kerg(A_a^\star)^\bot
\]
equipped with the $\h{1}$ norm and the operator
\[
B = A_a|_{Z}
\]
the restriction of $A_a$ to $Z$.  $B$ inherits from $A_a$ that it is the infinitesimal generator of a $C_0$ semigroup on $Z$.  Also note that $\sigma(B) = \sigma(A_a) \setminus \{0\}$ by Theorem III-6.17 of \cite{Kato:1995fk}. 

Recall from Theorem \ref{thm:linear}, $\sigma_\ess(A_a)$ is contained in left half-plane, and all points in $\sigma(A_a)\setminus \sigma_\ess(A_a)$ are eigenvalues of finite multiplicity.  By the assumption on the spectrum of $A_a$, the spectrum of $B$ is contained in the open left-half plane.  

We will now prove there exists a uniform bound on resolvent of $B$ for $\Re \lambda > -b $ for some $b>0$.  For $\lambda \in \rho(B)$,
\[
\paren{\lambda I -B}^{-1} = \paren{\lambda I -A_a}^{-1}|_Z
\]
so
\[
\norm{\paren{\lambda I -B}^{-1}}_{Z\to Z} \leq \norm{\paren{\lambda I -A_a}^{-1}}_{\h{1}\to \h{1}}
\]
The resolvent of $A_a$ may be written as,
\[
\paren{\lambda I -A_a}^{-1}  = \paren{I - C(\lambda)}^{-1} \paren{\lambda I-A^\infty_a}^{-1}
\]
$C(\lambda)$ defined in Proposition \ref{prop:compact-operator}.  Since $\Re\sigma_\ess(A_a) \leq -\omega<0$, we must have $b<\omega$.  Then, for $\Re \lambda \geq -b$, the Fourier symbol of $\paren{\lambda I - A_a^\infty}^{-1}$ is uniformly bounded, so
\[
\norm{\paren{\lambda I - A_a^\infty}^{-1}}_{\h{1}\to \h{1}} \leq M'\quad \text{for some $M'>0$.}
\]

By Proposition \ref{prop:large-eigenvalue}, for $\Re \lambda \geq -b \geq -\frac{1}{2}ac\gamma^2$, there exists $R'>0$ such that 
\[
\norm{\paren{I-C(\lambda)}^{-1} }_{\h{1}\to \h{1}} \leq 2\quad\text{for $\abs{\lambda}>R'$.}
\]
Therefore, 
\[
\norm{\paren{\lambda I -B}^{-1}}_{Z\to Z}\leq \norm{\paren{\lambda -A_a}^{-1}}_{\h{1}\to \h{1}} \leq 2  M'\quad \text{for $\Re\lambda \geq -b$ and $\abs{\lambda}> R'$.}
\]

For $\abs{\lambda}\leq R'$ and $\Re\lambda \geq -\vareps$, $B$ has no eigenvalues.  $B$ is a closed operator, therefore $\paren{\lambda I -B}^{-1}$ is holomorphic on this compact set (see Theorem III-6.7 of \cite{Kato:1995fk}) giving the bound
\[
\norm{(\lambda I -B)^{-1}}_{Z \to Z} \leq M''\quad\text{for some $M''>0$.}
\] 
Hence for all $\Re \lambda  \geq -\min\{b,\vareps\}$, 
\[
\norm{(\lambda -B)^{-1}}_{Z \to Z} \leq \max \{ 2 M', M''\}
\] 
and we may apply Theorem \ref{thm:pruss}.\qed

\section{Prelude to Nonlinear Stability}
\label{sec:estimates}
There are three results about our system needed before we prove Theorem \ref{thm:main}.  First, we establish criterion for when a decomposition of $\f$ into a modulating solitary wave and a perturbation is possible.  Then we derive equations for the evolution of the parameters associated with this modulating solitary wave.  Finally, we  relate the $\h{1}$-norm to the $\ha$-norm of the perturbation.

\subsection{Local Existence of Decomposition and Continuation Principles}
\label{sec:decomp}

In analyzing the weighted perturbation $w$, we wish to treat the nonlinear terms perturbatively, with the leading order behavior governed by the linear operator $A_a$.  As noted in Proposition \ref{prop:kernel}, the operator has a two-dimensional kernel.  To prevent the appearance of secular terms, the perturbation must be orthogonal to $\kerg(A_a^\star)$; this reveals how the decomposition of $\f$ into a perturbation and a modulated solitary wave, \eqref{eq:ansatz}, occurs.  This follows the strategy appearing in of \cite{Pego93} and \cite{Miller96}.

\begin{prop} 
\label{prop:decompexist}

Let $c_0>n$, $a\leq a_\star(c_0)$ and $t_1> 0$.  Given $\delta_1>0$, there exists $\delta_0>0$ such that for any $\f-1\in C^1([0,t_1];\hinta)$ satisfying
\begin{equation}
\label{eq:decomp1}
\sup_{ t \leq t_1} \norm{e^{a(\cdot + \theta_0)}(\f(\cdot,t)-\f_{c_0}(\cdot-c_0 t +\theta_0))}_\h{1}\leq \delta_0, \quad \text{for some $\theta_0\in \R$}
\end{equation}
there exists a unique mapping $t\mapsto (\thetat,\ct)$ in $ C^1([0,t_1];\R^2)$ such that,
\begin{gather}
\label{eq:decomp2}
\sup \abs{\theta(t)-\theta_0}+ \sup\abs{\dot{\theta}(t)}+ \sup\abs{c(t)-c_0}+ \sup\abs{\dot{c}(t)} \leq\delta_1,\quad { t \leq t_1}\\
\label{eq:decomp3}
\mathcal{T}_k[\f-1,\theta,c]=\inner{\f(x,t)-\f_{c(t)}(y)}{\tilde\eta_j(y)}=0\quad \text{ for $j=1,2$ and $t\in[0,t_1]$}
\end{gather}
where $y = x- \int_0^t c(s)ds +\theta(t)$.
 
The number $\delta_0$ may be chosen as a decreasing function of $t_1$.  
\end{prop}

\proof The proof is via the implicit function theorem.  In this context, the Banach spaces are $C^1([0,t_1];H^1_a)$, and $C^1([0,t_1];\R^2)$, the latter space equipped with the norm
\[
\sup_{t\leq t_1} \abs{\theta(t)}+ \sup_{t\leq t_1} \abs{\dot{\theta}(t)}+\sup_{t\leq t_1} \abs{c(t)}+ \sup_{t\leq t_1} \abs{\dot{c}(t)}
\]
The functional is
\[
\mathcal{T}=(\mathcal{T}_1, \mathcal{T}_2)^T :C^1([0,t_1];H^1_a) \times C^1([0,t_1];\R^2) \to C^1([0,t_1];\R^2)
\]
and it is $C^1$ in its arguments, permitting the use of the implicit function theorem.  

Setting
\[
\bold U_0 = \threedrowvec{\f_\co(x - \co t)-1}{0}{\co}
\]
we see ${\mathcal{T}}[\bold U_0]=0$.  The Fr\'echet derivative at $\bold{U}_0$ is
\begin{equation}
D\mathcal{T}[\bold U_0] \threedrowvec{\delta \f}{\delta \theta}{ \delta c} = \twodcolvec{\inner{e^{a(x- \co t)} \delta\f(x,t)}{\eta_1(x-\co t)} - \int_0^1 \delta c(s) ds  + \delta \theta(t)}{\inner{e^{a(x- \co t)} \delta\f(x,t)}{\eta_2(x-\co t )}+ \delta \ct}
\end{equation}
The derivative acting on the $(\theta,c)$ is $\bigl( \begin{smallmatrix} I&-B\\0&I\end{smallmatrix} \bigr)$, $(Bf)(t) = \int_0^t f(s)ds$.  This block operator has a bounded inverse on $C^1([0,t_1];\R^2) \to C^1([0,t_1];\R^2)$.  By the implicit function theorem, there exists $\delta_0>0$, such that if 
\begin{equation*}
\sup_{t\leq t_1} \norm{\f(\cdot-\theta_0,t) -  \f_\co(\cdot - \co t) }_\ha < \delta_0\quad\text{for some $\theta_0\in \R$}
\end{equation*}
then there exists a mapping in $C^1([0,t_1];\R^2)$, $t\mapsto (\tilde\theta(t),\ct)$, satisfying
\begin{gather*}
\sup \abs{\tilde\theta(t)} + \sup\abs{\dot{\tilde\theta}(t)}+\sup \abs{\ct - \co}+\sup\abs{\dot c(t)} < \delta_1\quad\text{ for $t\leq t_1$}\\
\quad\mathcal{T}[\f(\cdot-\theta_0,t))-1,\tilde\theta(t),c(t)]=0\quad\text{for $t\leq t_1$}
\end{gather*}
Defining $\theta(t) = \tilde\theta(t) + \theta_0$, and applying the change of variables $x\mapsto \tilde x + \theta_0$, we have the form given in \eqref{eq:decomp1}, \eqref{eq:decomp2} and \eqref{eq:decomp3}.

Finally, because our norms are taken as the supremum over $t \leq t_1$, if $\delta_0$ works for $t_1$, then it will also work for any $t_2\leq t_1$. This allows $\delta_0$ to be constructed as a decreasing function of $t_1$.\qed

\begin{prop}
\label{prop:decompcont}

Let $\co > n$, $a \leq a_\star(\co)$ and assume $\f-1 \in C^1([0,T_{\max}];\hinta)$, $T_{\max}>0$ solves \eqref{eq:magma}.

Given $\delta_1>0$, there exist $\delta_0>0$ and $\delta'_0>0$, such that for any $t_0\in [0,T_{\max})$, if the decomposition of $\f$, $v(y,t) = \f(x,t)- \f_{c(t)}(y)$, $y= x- \int_0^t c(s)ds + \theta(t)$, and $(\theta,c)\in C^1([0,t_0]; \R^2)$, satisfies
\begin{gather}
\label{eq:decomp4}
\sup_{t\leq t_0}\norm{v(t)}_\ha  \leq\delta_0/3\\
\sup_{t\leq t_0}\abs{\ct-\co} \leq \delta'_0\\
\mathcal{T}[\f-1,\theta,c](t)=0\quad\text{for $t\in[0,t_0]$}
\end{gather}
then there is a unique extension of $(\theta,c)$ in $C^1([0,t_0+t_\star];\R^2)$ for some $t_\star>0$ such that 
\begin{gather}
\mathcal{T}[\f-1,\theta,c](t)=0\quad\text{ for $t\leq t_0+t_\star \leq T_{\max}$}\\
\label{eq:decomp5}
\sup\abs{\theta(t)-\theta(t_0)}+\sup\abs{\dot\theta(t)}+ \sup\abs{c(t)-c(t_0)}+ \sup\abs{\dot c(t)}\leq \delta_1,\quad t\in[t_0,t_0+t_\star]
\end{gather}
\end{prop}

\proof  This follows the proof of Proposition 5.2 in \cite{Pego94}, although we are forced to modify it as we do not know \emph{a priori} that a solution exists for all time.

Given $\delta_1$, let $\delta_0>0$ be the value from Proposition \ref{prop:decompexist} with $t_1 = \frac{1}{2}(T_{\max}-t_0)$.  Set $\tilde\f(x,t) = \f(x,t+t_0)$, $\theta_1= -\int_0^{t_0}c(s)ds + \theta(t_0)$.  Let $\delta'_0$ be sufficiently small such that 
\begin{equation}
\norm{\f_{c'}-\f_{c_0}}_\hinta \leq \delta_0/3 \quad\text{for all $c'$ such that}\quad\abs{c'-c_0}\leq\delta'_0
\end{equation}

Then, since
\begin{equation*}
\begin{split}
\norm{e^{a(\cdot+\theta_1)}\paren{\tilde\f(0)-\f_{c_0}(\cdot+\theta_1)}}_\h{1}&\leq \norm{e^{a(\cdot+\theta_1)}\paren{\tilde\f(0)-\f_{c_1}(\cdot+\theta_1)}}_\h{1}\\
&\quad+  \norm{e^{a(\cdot+\theta_1)}\paren{\f_{c_1}(\cdot+\theta_1)-\f_{c_0}(\cdot+\theta_1)}}_\h{1}\\
& \quad= \norm{v(t_0)}_\ha + \norm{\f_{c_1}-\f_{c_0}}_\ha\leq \frac{2}{3}\delta_0
\end{split}
\end{equation*}
we have
\begin{equation*}
\begin{split}
\norm{e^{a(\cdot+\theta_1)}\paren{\tilde\f(t)-\f_{c_0}(\cdot-\co t+\theta_1)}}_\h{1}&\leq \norm{e^{a(\cdot+\theta_1)}\paren{\tilde\f(t)-\tilde\f(0)}}_\h{1}+ \frac{2}{3}\delta_0\\
& \quad= e^{a \theta_1}\norm{\f(t+t_0)-\f(t_0)}_\hinta + \frac{2}{3}\delta_0
\end{split}
\end{equation*}
As $\f$ is continuous in time, there exists a $t_\star \leq t_1$, such that 
\begin{equation*}
\sup_{t\leq t_\star} \norm{e^{a(\cdot+\theta_1)}\paren{\tilde\f(t)-\f_{c_0}(\cdot-\co t+\theta_1)}}_\h{1} \leq \delta_0
\end{equation*}
Therefore, $\tilde \f$ satisfies the hypotheses of Proposition  \ref{prop:decompexist}.  We have a unique $(\tilde \theta (t), \tilde c(t))$ with $(\tilde \theta (0), \tilde c(0))=(\theta(t_0), c(t_0))$.  This gives us the extension, $(\theta(t), c(t))=(\tilde \theta(t-t_0) -\theta_0 + \theta(t_0), \tilde c(t-t_0))$ for $ t \in[t_0, t_0+t_\star]$ and \eqref{eq:decomp5} will hold.\qed

\subsection{Modulation Equations}
\label{sec:modulation}

Given that the perturbation must be orthogonal to $\kerg(A_{a}^\star)$, the associated constraints give a pair of ODEs, coupled to the perturbation, giving a complete system of three equations for the three dependent variables.

Let $P$ denote the projection onto $\kerg(A_{a}^\star)$.  Assuming this space is two dimensional,  we use the biorthogonal bases given in Proposition \ref{prop:kernel} to define projection onto this space and its complement,
\begin{align}
P &= \inner{\eta_1}{\cdot}\xi_1 + \inner{\eta_2}{\cdot}\xi_2\\
Q&=I-P
\end{align}
The secular terms are excised from the perturbation equation, \eqref{eq:weighted-perturbation-equation}, by requiring:
\begin{align}
w_\tau &= A_{a} w + Q\mathcal{G}\\
\label{eq:nonsecular}
P \mathcal{G} &=0\\
Pw(\tau=0) &= 0
\end{align}
Constraint \eqref{eq:nonsecular} corresponds to
\begin{equation}
\inner{\eta_j}{\G}=0 \quad \text{for $j =1,2$}
\end{equation}
These two equations govern $c(t)$ and $\theta(t)$, completing our system.

Defining $p_1(y,t) = \dy \f_{c(t)}(y) - \dy \f_{c_0}(y)$ and $p_2(y,t) = \dc \f_{c(t)}(y) - \dc \f_{c_0}(y)$, the derived system is:
\begin{equation}
\label{eq:modulation1}
\twobytwo{ 1+ \inner{\tilde{\eta}_1}{p_1}}{\inner{\tilde{\eta}_1}{p_2}}{  \inner{\tilde{\eta}_2}{p_1}} {1+ \inner{\tilde{\eta}_2}{p_2}}\twodcolvec{\thetadot}{\dcdt}=\frac{c-\thetadot}{\co}\twodcolvec{\inner{\eta_1}{S_a[\co,c,\thetadot]w}  }{\inner{\eta_2}{S_a[\co,c,\thetadot]w}}+  \twodcolvec{\inner{\eta_1}{ \mathcal{G}_1}}{\inner{\eta_2}{ \mathcal{G}_1}}
\end{equation}
However, the right hand side still has $\thetadot$ dependence.  Observe,
\begin{equation}
\begin{split}
\frac{c-\thetadot}{\co}S_a[\co,c,\thetadot]w
&=cm \bracket{\dy \log\paren{\frac{\f_\co}{\f_c}}}w - \thetadot m \dy\log\paren{\f_\co} w+\Bigl[\frac{c}{\co}n \f_\co^m H_{\f_\co,a}\inv D_a\bracket{\paren{\f_\co\inv - \co(\f_\co\inv-1)}\cdot}\\
&\quad -n \f_c^m H_{\f_c,a}\inv D_a\bracket{\paren{\f_c\inv - c(\f_c\inv-1)}\cdot}\Bigr]w-\frac{\thetadot}{\co}\f_\co^m H_{\f_\co,a}\inv D_a\bracket{\paren{\f_\co\inv - \co(\f_\co\inv-1)}\cdot}w\\
&\quad- c m \bracket{\f_\co^m H_{\f_\co,a}\inv \bracket{\dy\paren{\log \f_\co}\cdot}- \f_c^m \calh_{\f_c,a}\inv \bracket{\dy\paren{\log \f_c}\cdot}}w\\
&\quad+ \thetadot m\f_\co^m H_{\f_\co,a}\inv \bracket{\dy\paren{\log \f_\co}\cdot}w
\end{split}
\end{equation}
Defining
\begin{align}
\widetilde{S}_a &=\widetilde{S}_a^1+ \widetilde{S}_a^2 +\widetilde{S}_a^3\\
\widetilde{S}_a^1 &= cm \paren{\f_\co^{-1}\dy\f_\co- \f_c^{-1}\dy\f_c} \\
\begin{split}
\widetilde{S}_a^2 &=n\Bigl\{\frac{c}{\co} \f_\co^m H_{\f_\co,a}\inv D_a\bracket{\paren{\f_\co\inv - \co(\f_\co\inv-1)}\cdot}\\
&\quad- \f_c^m H_{\f_c,a}\inv D_a\bracket{\paren{\f_c\inv - c(\f_c\inv-1)}\cdot}\Bigr\}
\end{split}
\\
\widetilde{S}_a^3 &=- c m \bracket{\f_\co^m H_{\f_\co,a}\inv \paren{\f_\co^{-1}\dy\f_\co\cdot}- \f_c^m H_{\f_c,a}\inv \paren{\f_c^{-1}\dy\f_c\cdot}}
\end{align}
and
\begin{equation}
\begin{split}
T_a &= - m \f_\co^{-1}\dy\f_\co -{\co^{-1}}\f_\co^m H_{\f_\co,a}\inv D_a\bracket{\paren{\f_\co\inv - \co(\f_\co\inv-1)}\cdot}\\
&\quad+m\f_\co^m H_{\f_\co,a}\inv \paren{\f_\co^{-1}\dy\f_\co \cdot}
\end{split}
\end{equation}
the right-hand side of \eqref{eq:modulation1} may be written as
\[
 \twodcolvec{\inner{\eta_1}{\widetilde{S}_a w}}{\inner{\eta_2}{\widetilde{S}_a w}} + \thetadot \twodcolvec{\inner{\eta_1}{T_a w}}{\inner{\eta_2}{T_a w}}+  \twodcolvec{\inner{\eta_1}{ \mathcal{G}_1}}{\inner{\eta_2}{ \mathcal{G}_1}}
\]
Equation\eqref{eq:modulation1} may be solved algebraically so that $\thetadot$ only appears on the left hand side,
\begin{equation}
\label{eq:modulation2}
\begin{split}
\mathcal{B}(t)\twodcolvec{\thetadot}{\dcdt} &=\twobytwo{ 1+ \inner{\tilde{\eta_1}}{p_1}-\inner{\eta_1}{T_a w}}{\inner{\tilde{\eta_1}}{p_2}}{  \inner{\tilde{\eta_2}}{p_1}-\inner{\eta_2}{T_a w}} {1+ \inner{\tilde{\eta_2}}{p_2}}\twodcolvec{\thetadot}{\dcdt}\\
&=\twodcolvec{\inner{\eta_1}{\widetilde{S}_a w}}{\inner{\eta_2}{\widetilde{S}_a w}}+ \twodcolvec{\inner{\eta_1}{ \mathcal{G}_1}}{\inner{\eta_2}{ \mathcal{G}_1}}
\end{split}
\end{equation}
$\mathcal{B}(t)=I+ O(\abs{c(t)-c_0}) +O(\norm{w}_\ltwo)$; for sufficiently small $\abs{c(t)-c_0}+\norm{w}_\ltwo$, $\mathcal{B}$ is invertible.  Thus we have equations for $\thetadot$ and $\dcdt$, closing the system for $\paren{v,c,\theta}$.

\begin{rem} In \eqref{eq:modulation2}, we see that, when $\mathcal{B}(t)$ is inverted, the right hand side of the system is continuous in $t$.  Therefore, provided $c(t)$, $\theta(t)$, and $w(t)$ are continuous in $t$, $c(t)$ and $\theta(t)$ will actually be $C^1$.
\end{rem}

\subsection{Lyapunov Bound}
\label{sec:lyapunov}

Using the functional, $\N[\f]$ defined in \eqref{eq:energy}, we have
\begin{prop} 
\label{prop:lyapunov-bound}

Let $\co>n$, $a\leq a_\star(\co)$, and let $\f(x,t)$ be a solution to \eqref{eq:magma} in $C^1([0,T]; \hinta)$ with data
\[
\f_0 = \f_{c_0}(x + \theta_0) + v_0(x),\quad\theta_0\in\R
\]
Assume the decomposition $\f(x,t)\to(v(y,t),\theta(t),c(t))$ exists for $t \leq T$ and
\begin{gather*}
\abs{c(t)-c_0}+ \norm{v(\cdot,t)}_\h{1}\leq \delta_1<1\quad\text{for $t \leq T$}\\
\partial_{c}\N[\f_c] \Big |_{c = \co} \neq 0
\end{gather*}
Then there exist constants $K$ and $K'$ such that
\begin{gather}
\label{eq:lyapunov-bound}
\begin{split}
\norm{v}_\h{1}^2\paren{1-K'\norm{v}_\h{1}}\leq K&\Bigl( \abs{\Delta \N}+ \norm{w}_\ltwo + \norm{w }_\ltwo^2 \\
&\quad + \abs{\ct-\co  } + \abs{\ct-\co  }^2 +  \abs{\ct-\co  }^3\Bigr)
\end{split}
\\
\Delta\N = \N[\f_\ct + v] -\N[\f_\co]\nonumber
\end{gather}
\end{prop}

\proof Taylor expanding $\mathcal{N}$ about a solitary wave with perturbation $z$.
\[
\mathcal{N}[\f_{c_0}+z] = \N[\f_{c_0}] + \inner{\delta\N[\f_{c_0}]}{z}+ \frac{1}{2}\inner{\delta^2\N[\f_{c_0}]z}{z}+ O(\norm{z}_{\h{1}}^3)
\]
From \cite{Simpson07}, the first and second variations are
\begin{gather}
\label{eq:var1}
\begin{split}
\inner{\delta\N[\f_{c_0}]}{z}&=\int \paren{\frac{1- \f_{c_0}^{1-n-m}}{n+m-1}+m \f_\co^{-2m-1}  \paren{\dx \f_{c_0}}^2-\dx^2\f_\co^{-2m} \f_\co   }z dx\\
&= \Theta^{-1}\inner{\tilde\eta_2}{z}
\end{split}
\\
\label{eq:var2}
\begin{split}
\inner{\delta^2\N[\f_{c_0}]z}{z}&=\int \paren{\f_\co^{-n-m}  -m(1+2m)\f_\co^{-2m-2}\paren{\dx\f_\co}^2 + 2m\f_\co^{-2m-1} \dx^2 \f_\co  }{z}^{2}dx\\
& \quad+\int \f_\co^{-2m}\paren{\dx z}^{2}dx
\end{split}
\end{gather}
$\tilde{\eta}_2$ and $\Theta$ as in Proposition \ref{prop:kernel}.


Take $z(y,t)= \f_\ct (y) - \f_\co(y) + v(y,t)= \f(x,t)- \f_\co(y)$.  Then 
\begin{equation}
\inner{\delta\N[\f_{c_0}]}{z}=\inner{\Theta^{-1}\tilde{\eta}_2 }{z(y,t)}= \Theta^{-1}\inner{\tilde{\eta}_2 }{\f_\ct (y) - \f_\co(y)}+  \Theta^{-1}\inner{\tilde{\eta}_2 }{v(y,t)}
\label{eq:first-variation-expansion}
\end{equation}
Using the continuity of $c \mapsto \f_c-1$,
\begin{equation}
\Theta^{-1} \inner{\tilde{\eta}_2 }{\f_\ct (y) - \f_\co(y)}\leq K \abs{\ct - \co}
\label{eq:first-variation-estimate-one}
\end{equation}
The term with the perturbation, $v$, may be bounded by
\begin{equation}
\Theta^{-1}\inner{\tilde{\eta}_2 }{v(y,t)} = \Theta^{-1}\inner{\eta_2 }{w(y,t)}\leq \Theta^{-1}  \norm{\eta_2}_\ltwo \norm{w}_{\ltwo}\leq K \norm{w}_\ltwo
\label{eq:first-variation-estimate-two}
\end{equation}

Now we bound the second variation.  For brevity, let
\begin{equation*}
\Phi_\co =  -m(1+2m)\f_\co^{-2m-2}\paren{\dx\f_\co}^2 + 2m\f_\co^{-2m-1} \dx^2 \f_\co 
\end{equation*}
Then
\begin{equation}
\label{eq:second-variation-estimate}
\begin{split}
\inner{\delta^2\N[\f_{c_0}]z}{z} &=\int \f_\co^{-2m} \paren{\dx z}^2 + \f_\co^{-n-m} z^2 +\int \Phi_\co {z}^{2}\\
&\geq K \norm{z}_\h{1}^2 +\inner{ \Phi_\co}{ \paren{\f_\ct -\f_\co}^2} + 2\inner{\Phi_\co} {\paren{\f_\ct -\f_\co}v}+\inner{\Phi_\co}{v^2}\\
&\geq K_1\norm{v}_\h{1}^2 -  K_2 \abs{\ct -\co}^2 - K_3\abs{\ct-\co} - K_4 \inner{\Phi_\co e^{-ay}}{w^2}
\end{split}
\end{equation}
We would like $\Phi_\co e^{-2ay} \in L^\infty$, so that the last term may be estimated by $\norm{w}_\ltwo^2$.  Since $a\leq a_\star(\gamma)< \frac{1}{2}\gamma$, we may do this.

Lastly, we have the remainder term $\mathcal{R}[\f_\co,z]$.  Because of the \emph{a priori} bound on $\norm{v}_\h{1}$, this may be estimated as
\begin{equation}
\abs{\mathcal{R}[\f_\co,z]}\leq K \norm{z}_\h{1}^3 \leq K \paren{\abs{\ct -\co}^3 + \norm{v}_\h{1}^3}
\label{eq:remainder-estimate}
\end{equation}
Combining these estimates, \eqref{eq:first-variation-estimate-one}, \eqref{eq:first-variation-estimate-two}, \eqref{eq:second-variation-estimate}, and \eqref{eq:remainder-estimate},
\begin{equation*}
\norm{v}_\h{1}^2(1-D \norm{v}_\h{1})  \leq K \paren{ \abs{\Delta \N} + \abs{\ct-\co  } + \abs{\ct-\co  }^2 +  \abs{\ct-\co  }^3+ \norm{w}_\ltwo + \norm{w }_\ltwo^2}
\end{equation*}
\qed

\section{Proof of Main Results}
\label{sec:mainproof}

Before proving Theorem \ref{thm:main}, we make an \emph{a priori} estimate.

\subsection{A Priori Estimates}
\begin{prop} 
\label{prop:apriori}

Let $\co>n$, $a\leq a_\star(\co)$, and assume there exists $\vareps>0$ such that 
\[
\sigma(A_{a})\cap \set{\Re\lambda \geq -\vareps  }= \set{0},\quad\text{$\lambda=0$ is an eigenvalue of algebraic multiplicity two.}
\]

Let $T>0$.  There exists $\delta_\star \in(0,1)$ and $K_\star\geq 1$ such that, if the $\f(x,t)-1 \in C^1\paren{[0,T]; \h{1}\cap H_a^1}$ solves \eqref{eq:magma} and satisfies, for $t \leq T$,
\begin{align}
\inf_x& \f(x,t)\geq\alpha_0>0\\
\inf_x & \f(x,t)^m-a^2 \f(x,t)^n \geq \beta_0>0
\end{align}
and furthermore:
\begin{description}
  \item[i.] The decomposition $\f(x,t)\mapsto(v(y,t),c(t),\theta(t))$ exists for $t \leq T$
  \item[ii.] For $t \leq T$
  \begin{equation}
  \label{eq:apriori1}
  \sqrt{\abs{\Delta\N}} + \sqrt{\norm{w(t)}_{\h{1}}} +\sqrt{\abs{c(t)-c_0}}+\abs{\theta(t)-\theta_0} +\left|1-\frac{\co}{c(t)-\thetadot(t)}\right|+\norm{v(t)}_\h{1}\leq \delta_\star
  \end{equation}
  \item[iii.] The data satisfies
  \begin{equation}
\label{eq:apriori2}
\sqrt{\abs{c(0)-\co}}+\abs{\theta(0)-\theta_0}+  \sqrt{\abs{\Delta\N}} + \sqrt{\norm{w(0)}_{\h{1}}} \leq\eps <\delta_ \star 
  \end{equation}
\end{description}
then for $t\in [0,T]$,
\begin{equation}
\label{eq:apriori3}
 \sqrt{e^{\kappa t}\norm{w(t)}_{\h{1}}} + \sqrt{\abs{\ct-\co}}+\abs{\theta(t)-\theta_0} +\left|1-\frac{\co}{c(t)-\thetadot(t)}\right|+\norm{v(t)}_\h{1}\leq K_\star \eps
\end{equation}
with $\kappa =\kappa(\alpha_0,\beta_0,\delta_\star)\in (0,b_{\max})$.
\end{prop}

\proof The strategy for proving this proposition is  to show that the  left-hand side of \eqref{eq:apriori3} may be estimated in terms of their data, \eqref{eq:apriori2}, using \eqref{eq:apriori1}.  This largely follows the proof in \cite{Miller96}, with a few changes.  The need to estimate $ \abs{1-\co/(c-\thetadot)}$ in terms of the data will require use of the modulation equations, \eqref{eq:modulation2}, to control $\thetadot$, and to control $\norm{w}_\h{1}$, we will need to work in $\tau$-time.  Thus we make the following estimates:

\underline{Temporal Change of Variables:}
First, let us assume that $\delta_\star \leq \frac{1}{2}$.  Then, since $\co>n>1$, $\frac{1}{2}\co \leq c(t)\leq \frac{3}{2}\co$.  Furthermore, this initial choice of $\delta_\star$ ensures
\[
\frac{1}{2}\leq\frac{\co}{c(t)-\thetadot(t)}= \frac{d\tau}{dt}\leq \frac{3}{2}
\]
so the change of variables $\tau =\tau(t)$ is well defined.

\underline{Time Derivatives of Modulation Parameters:}
Examining \eqref{eq:modulation2}, $\mathcal{B}(t)= I + O(\abs{\ct-\co})+O(\norm{w}_\ltwo)$; so there exists $\delta_\mathcal{B}>0$ such that for $\delta_\star\leq \delta_\mathcal{B}$, $\mathcal{B}(t)$ will be invertible.  Therefore
\begin{equation}
\label{eq:apriori4}
\abs{\thetadot(t)}+\abs{\dcdt(t)} \leq \abs{\mathcal{B}(t)\inv}\paren { \norm{\widetilde{S}_a w}_\ltwo+\norm{\G_1}_\ltwo}_\ltwo\leq K_1\paren{ \abs{c-\co}+\norm{v}_\h{1}}\norm{w}_\ltwo
\end{equation}
permitting the estimate
\begin{equation}
\label{eq:apriori5}
\abs{\thetadot(t)}+\abs{\dcdt(t)} \leq K_1\delta_\star^3
\end{equation}
As terms of the form $1/(c-\thetadot)$ will appear, we will assume that 
\[
\delta_\star \leq \paren{\frac{1}{2K_1}}^{1/3}=\delta_\theta
\]
so $\abs{\thetadot}\leq \frac{1}{2}$, and this quotient will be well defined and bounded.

\underline{Weighted Perturbation:}
In $\tau$-time,  $\norm{w(\tau)}_\h{1}$ is
\[
w(\tau) = e^{A_{a} \tau} w(\tau(t =0)) + \int_{\tau(0)}^\tau e^{A_{a} (\tau-s)} Q \G(s)ds
\]
By the semigroup decay estimate of Proposition \ref{prop:semigroup}, there exists $K_2>0$ and $b_{\max}>0$ such that
\[
\norm{w(\tau)}_\h{1} \leq K_2 e^{-b \tau}\norm{w(\tau(0))}_\h{1} + K_2 \int_{\tau(0)}^\tau e^{-b (\tau-s)} \norm{Q \G(s)}_\h{1} ds
\]
for any $b \in (0,b_{\max})$.  Estimating $\mathcal{G}$,
\begin{equation}
\begin{split}
\norm{\G}_\h{1} &\leq  K\bracket{{\left|\frac{\co}{c-\thetadot}\right|}\paren{\abs{\thetadot}+\abs{\dot{c}}} +\paren{\abs{c-\co} +\left|1-\frac{\co}{c-\thetadot}\right|} \norm{w}_\h{1} + \left|\frac{\co}{c-\thetadot}\right| \norm{v}_\h{1}\norm{w}_\h{1}}\\
&\leq K_3 \delta_\star \norm{w}_\h{1}
\end{split}
\end{equation}
We have made use of \eqref{eq:apriori4} to control $\thetadot$ and $\dot{c}$ in terms of $\norm{w}_\h{1}$.

Therefore,
\begin{equation}
\label{eq:apriori6}
\norm{w(\tau)}_\h{1} \leq K_2 e^{-b \tau} \norm{w(\tau(0))}_\h{1} + K_2 K_3\delta_\star \int_{\tau(0)}^\tau e^{-b(\tau-s)} \norm{w(s)}_\h{1} ds
\end{equation}
Defining $\psi(s) = e^{b s} \norm{w(s)}_\h{1}$, \eqref{eq:apriori6} becomes
\[
 \psi(\tau) \leq K_2 \norm{w(0)}_\h{1} + K_2 K_3 \delta_\star \int_{\tau(0)}^\tau \psi(s)ds
\] 
for which we may apply Gronwall's inequality to get
\begin{equation}
\norm{w(\tau)}_\h{1} \leq K_2 \norm{w(\tau(0))}_\h{1}e^{- (b-K_2 K_3\delta_\star)(\tau-\tau(0))}
\end{equation}
So for $\delta_\star$ small enough, $b-K_2K_3\delta_\star>0$ and we induce decay in the $\h{1}$ norm of $w$.  In particular, suppose that $\delta_\star \leq \delta_b=\frac{1}{2}b/(K_2 K_3)$ and let
\[
b'=b-K_2K_3\delta_\star
\]

We then return to $t$-time,
\begin{equation*}
\tau-\tau(0)=\frac{1}{\co} \int_0^t c(s)ds +\frac{1}{\co}\paren{\theta(0)-\theta(t)}\geq \frac{1}{\co}\paren{\co-\delta_\star}t -2\frac{\delta_\star}{\co}
\end{equation*}
Therefore,
\begin{equation}
\label{eq:apriori7}
\norm{w(t)}_\h{1} \leq \tilde K_2 \norm{w(t=0)}_\h{1}e^{-\kappa t}
\end{equation}
with $\kappa = \frac{1}{2}b'$.  We now have $e^{\kappa t}\norm{w}_\h{1}$ estimated in terms of the data.

\underline{Unweighted Perturbation:}
Applying this to Proposition \ref{prop:lyapunov-bound}, we have the estimate,
\begin{equation}
\begin{split}
\norm{v(t)}_\h{1}&\leq K \paren{\sqrt{\abs{\Delta \N}}+ \sqrt{\abs{\ct-\co}} + \abs{\ct-\co} + \abs{\ct-\co}^{3/2}+\sqrt{\norm{w(t)}_\ltwo} +\norm{w(t)}_\h{1}  }\\
&\leq K\paren{\sqrt{\abs{\Delta \N}}+ \sqrt{\abs{\ct-\co}}(1+\delta_\star+ \delta_\star^2)+\sqrt{\norm{w(t)}_\ltwo}(1+\delta_\star) }\\
&\leq K\paren{\sqrt{\abs{\Delta \N}}+ \sqrt{\abs{\ct-\co}}+\sqrt{\norm{w(0)}_\ltwo} }
\end{split}
\end{equation}
If we had control of $\sqrt{\abs{\ct-\co}}$, then we would also control $\norm{v(t)}_\h{1}$ in terms of the data.

\underline{Deviation in $c$ from $\co$:}
Estimating $\abs{c(t)-\co}$ using \eqref{eq:apriori4} and \eqref{eq:apriori7},
\begin{equation*}
\begin{split}
\abs{\ct-\co}&\leq \abs{c(0)-\co} + \int_0^t \abs{\dcdt(s)}ds\leq  \abs{c(0)-\co} + \int_0^t K_1\paren{ \abs{c(s)-\co}+\norm{v(s)}_\h{1}}\norm{w(s)}_\ltwo ds\\
&\leq  \abs{c(0)-\co} + K_1 \delta_\star \int_0^t \norm{w(s)}_\h{1}ds\leq  \abs{c(0)-\co} + K_1 \delta_\star \int_0^t K_2\norm{w(t0)}_\h{1}e^{-\kappa s}ds\\
&\leq \abs{c(0)-\co} + K_1 K_2\delta_\star  \norm{w(0)}_\h{1}/\kappa
\end{split}
\end{equation*}
So we now have $\abs{\ct-\co}$ in terms of data, which we rewrite as
\begin{equation}
\label{eq:apriori8}
\sqrt{\abs{\ct-\co}}\leq K_4\paren{ \sqrt{\abs{c(0)-\co}} + \sqrt{\norm{w(0}_\h{1}}}
\end{equation}
which in turn gives
\begin{equation}
\label{eq:apriori9}
\norm{v(t)}_\h{1} \leq K_5\paren{\sqrt{\abs{\Delta \N}}+ \sqrt{\abs{c(0)-\co}}+\sqrt{\norm{w(0)}_\h{1}} }
\end{equation}

\underline{Deviation in $\theta$ from $\theta_0$:}
As with the speed parameter,
\begin{equation*}
\begin{split}
\abs{\theta(t)-\theta_0}&\leq \abs{\theta(0)-\theta_0} + \int_0^t \abs{\thetadot(s)}ds\leq \abs{\theta(0)-\theta_0}  + \int_0^t K_1\paren{ \abs{c(s)-\co}+\norm{v(s)}_\h{1}}\norm{w(s)}_\ltwo ds\\
&\leq \abs{\theta(0)-\theta_0} + K_1 \delta_\star \int_0^t \norm{w(s)}_\h{1}ds\leq  \abs{c(0)-\co} + K_1 \delta_\star \int_0^t K_2\norm{w(t0)}_\h{1}e^{-\kappa s}ds\\
&\leq \abs{\theta(0)-\theta_0} + K_1 K_2\delta_\star  \norm{w(0)}_\h{1}/\kappa
\end{split}
\end{equation*}
This is rewritten as 
\begin{equation}
\label{eq:apriori11}
{\abs{\theta(t)-\theta_0}}\leq K_7\paren{{\abs{\theta(0)-\theta_0}} + \sqrt{\norm{w(0}_\h{1}}}
\end{equation}

\underline{Another Estimate on the Temporal Change of Variables:}
\[
\left|1- \frac{\co}{c-\thetadot}\right|\leq\frac{\abs{c-\co}+\abs{\thetadot}}{\abs{c-\thetadot}}\leq 2\paren{\abs{c-\co}+\abs{\thetadot}}\]
Then using \eqref{eq:apriori4} and \eqref{eq:apriori8}
\begin{equation}
\label{eq:apriori10}
\begin{split}
\left|1- \frac{\co}{c-\thetadot}\right|&\leq K \paren{ \sqrt{\abs{c(0)-\co}} + \norm{w(t)}_\h{1} }K \leq \paren{ \abs{c(0)-\co} +K_2\norm{w(0)}_\h{1} }\\
&\leq K_6\paren{\sqrt{\abs{c(0)-\co}} + \sqrt{\norm{w(0)}_\h{1}}}
\end{split}
\end{equation}

Combining \eqref{eq:apriori7}, \eqref{eq:apriori8}, \eqref{eq:apriori9}, \eqref{eq:apriori11}, and \eqref{eq:apriori10}, we have \eqref{eq:apriori3} with $\delta_\star = \min \{ \frac{1}{2}, \delta_\mathcal{A}, \delta_b,\delta_\tau, \delta_{\theta}\}$, $K_\star =\max\{\tilde K_2,K_4,K_5,K_6,K_7\}$.\qed


\subsection{Main Result}

We now prove Theorem \ref{thm:main}.  Let $\co \in (n, c_\star]$, $c_\star$ the value corresponding to $\gamma_\star$ from Theorem \ref{thm:linear} \textbf{(b)}, and $a\leq a_\star(\co)$.

Define $\mathcal{T}$ to be the set of non-negative numbers, $T$, such that, given $\co$, $a$ and $v_0 \in H^1 \cap H_a^1$:
\begin{itemize}
\item a solution exists, $\f-1 \in C([0,T), H^1 \cap H_a^1)$, satisfying
  \begin{align}
  \label{eq:alphabound}
   \inf_x&\f(x,t) \geq \frac{1}{4}=\alpha_0>0\\
   \label{eq:betabound}
   \inf_x &\paren{\f(x,t)^{m}-a^2 \f(x,t)^n}\geq \frac{1}{4} \inf_x\paren{\f_\co(x)^{m}-a^2 \f_\co(x)^n}=\beta_0>0
 \end{align}
  \item a decomposition of $\f$ into $\paren{v(y(x,t),t),\theta(t),c(t)}$ exists for $t\in[0,T)$
  \item \eqref{eq:apriori1} holds for $t\in [0,T)$.
\end{itemize}
Set $T_\ast = \sup \mathcal{T}$.  We will first show that there exists $\eps_\ast>0$, such that for $\eps\leq \eps_\ast$, if 
\[
\norm{v_0}_\h{1} + \norm{v_0}_{H^1_a} \leq \eps
\]
then $T_\ast>0$.  This will be proved using the continuous dependence upon the data. Using Proposition \ref{prop:apriori}, we will then prove $T_\ast = \infty$.

Let $\delta_\star$ and $K_\star$ be as in Proposition \ref{prop:apriori}.

The most difficult part of the proof will be ensuring the persistence of \eqref{eq:apriori1}.  Consider, at $t=0$, the left hand of that equation side may be estimated with
\begin{equation}
\label{eq:main1}
\begin{split}
\mathrm{LHS}(t=0)&\leq \abs{\Delta\N} + \norm{v_0}_\h{1} + \norm{\dy\f_\co}_\h{1}\abs{\theta_0-\theta(0)} \\
&\quad+ \sqrt{\abs{c(0)-\co}} +\abs{\theta(0)-\theta_0}+\norm{\f_\co-\f_{c(0)}}_\h{1} + \abs{1-\frac{\co}{c(0)-\thetadot(0)}}\\
&\quad+ \sqrt{e^{a\theta(0)} \norm{v_0}_\ha} + \sqrt{ \norm{\dy\f_\co}_\ha\abs{\theta_0-\theta(0)}}+ \sqrt{\norm{\f_\co-\f_{c(0)}}_\ha}
\end{split}
\end{equation}
There exists a choice of $\delta'$ and $\eps'$ such that if 
\begin{gather}
\abs{c(0)-\co} +\abs{\dot{c}(0)}+ \abs{\theta(0)-\theta_0}+ \abs{\dot{\theta}(0)} \leq \delta'\\
\norm{v_0}_\hinta\leq \eps'
\end{gather}
then the right hand side of \eqref{eq:main1} will be bounded by $\frac{1}{2}\delta_\star$.  Set $\delta_1 =\min\set{\frac{1}{4}, \delta'}$.   From Propositions \ref{prop:decompexist} and \ref{prop:decompcont}, let $\delta_0$, $\delta'_0$, be the corresponding values for $\delta_1$.

There exists $\eps_\exist\in (0,1)$ such that if $\norm{v_0}_\hinta\leq \eps_\exist$ then
\begin{align*}
\inf_x&\bracket{\f_\co(x+\theta_0)+v_0(x)}\geq 2 \alpha_0\\
\inf_x& \bracket{\paren{\f_\co(x+\theta_0)+v_0(x)} ^{m}-a^2 \paren{\f_\co(x+\theta_0)+v_0(x)} ^n}\geq2\beta_0
\end{align*}
By Theorem \ref{thm:local}, there exists $t_1>0$ and a solution in $C^1([0,t_1], H^1\cap H_a^1)$, satisfying \eqref{eq:alphabound} and \eqref{eq:betabound}.  Furthermore, we will have the \emph{a priori} $\hinta$ bound that,
\begin{equation*}
\sup_{t\leq t_1}\norm{\f(t)-1}_\hinta \leq 2 \norm{\f_\co(\cdot+\theta_0)+v_0-1}_\hinta\leq 2 (\norm{\f_\co-1}+1)
\end{equation*}

Set 
\begin{equation}
\eps_1 = \min\set{\eps_\exist, \eps', \frac{1}{2}e^{-a \theta_0}\delta_0}
\end{equation}
and let $\norm{v_0}\leq \eps_1$.  As noted, the solution exists, satisfying \eqref{eq:alphabound} and \eqref{eq:betabound}, until at least $t_1>0$.  At $t=0$,
\begin{equation*}
\norm{e^{a(\cdot+\theta_0)}\paren{\f_0-\f_\co(\cdot + \theta_0}}_\h{1}\leq e^{a\theta_0}\norm{v_0}_\h{1}\leq \frac{1}{2}\delta_0
\end{equation*}
so by the continuity of $\f$ in time, we have
\begin{equation*}
\norm{e^{a(\cdot+\theta_0)}\paren{\f(t)-\f_\co(\cdot -\co t + \theta_0}}_\h{1}\leq \delta_0\quad\text{for some $t_2\in (0,t_1)$}
\end{equation*}
Therefore the decomposition exists, with the $\delta_1$ bound on the modulation parameters, up till $t_2>0$.

Also at $t=0$, using the $\delta_1$ bound on the parameters,
\begin{equation}
 \sqrt{\abs{\Delta\N}} + \sqrt{\norm{w(0)}_{\h{1}}} +\sqrt{\abs{c(0)-c_0}} +\left|1-\frac{\co}{c(0)-\thetadot(0)}\right|+\norm{v(0)}_\h{1}\leq \frac{1}{2}\delta_\star
\end{equation}
All of these terms are continuous in time, there exists some $t_3\in(0, t_2)2$, for which this remains smaller than $\delta_\star$.  Therefore, for $\eps_\ast \leq \eps_1$, $t_3 \in\mathcal{T}$ , and $T_\ast>0$.

\underline{Continuing to Infinity:}  A few more constraints on $\eps_\ast$ are needed to continue out to $t=\infty$.  There exists $\eps_{\alpha\beta}$ such that for $\eps\leq \eps_{\alpha\beta}$, if
\begin{equation*}
\sqrt{\abs{c-\co}} + \norm{v}_\h{1} \leq \eps
\end{equation*}
then
\begin{align*}
\inf_y&\bracket{\f_c(y)+v(y)}\geq \alpha_0\\
\inf_y& \bracket{\paren{\f_c(y)+v_0(y)} ^{m}-a^2 \paren{\f_c(y)+v_0(y)} ^n}\geq\beta_0
\end{align*}

Let $\eps_2>0$ be so small that
\begin{align}
K_\star\eps_2 &= \min \set{\frac{\delta_\star}{3 },\sqrt{\delta_0/3}, \sqrt{\delta'_0}, \frac{1}{2}\eps_{\alpha\beta} }\quad\text{and set}\\
\eps_\ast &=\min\set{\eps_1,\eps_2}
\end{align}
Now, assume $\norm{v_0}_\hinta \leq\eps\leq\eps_\ast$.  As above, for this data we will have $T_\ast>0$.   Assume $T_\ast <\infty$.  For any $T<T_\ast$, on the inteval $[0,T]$, the solution exists with \eqref{eq:alphabound} and \eqref{eq:betabound}, as does the decomposition, and \eqref{eq:apriori1} holds.  

Then
\begin{eqnarray*}
\norm{\f(t)-1}_\hinta&\leq& \max\set{1, e^{a\paren{\int_0^t c(s)ds -\theta(t)}}}\paren{\norm{\f_\ct-1}_\hinta + \norm{v(t)}_\hinta}\\
&\leq& e^{a \paren{ (c_0+\delta_\star)T_\ast +\abs{\theta_0}+\delta_\star }} \paren{\sup_{\abs{c-\co}\leq \delta_\star} \norm{\f_c-1}_\hinta + \delta_\star + \delta_\star^2}< \infty
\end{eqnarray*}
and this bound is uniform in $T< T_\ast$.  By assumption, equations \eqref{eq:alphabound} and \eqref{eq:betabound} hold for $t\in [0,T]$, uniformly in $T<T_\ast$, which may written as 
\begin{equation*}
\left\|\frac{1}{\f(\cdot,t)}\right\|_\linf \leq \frac{1}{\alpha_0}<\infty\quad\text{and}\quad\left\|\frac{1}{\f(\cdot,t)^{m}-a^2\f(\cdot,t)^{n}} \right\|_\linf \leq \frac{1}{\beta_0}<\infty
\end{equation*}
Therefore, according to \eqref{eq:blowup}, $\f(x,t)$ may be extended beyond $T_\ast$ by some amount $t_2>0$.  Hence,if $T_\ast \neq \infty$  it must either be a failure for the decomposition to continue to exist or for \eqref{eq:apriori1} to hold.

Again using the Proposition \ref{prop:apriori} and our choice of $\eps_\ast$,
\[
\norm{w(\cdot,t)}_{H_a^1}\leq \delta_0/3 \quad \text{and} \quad \abs{c(t)-\co} \leq \delta'_0\quad\mbox{for all $t\leq T$, uniformly in $T<T_\ast$.}
\]
Since $\f$ exists until at least $T_\ast + t_2$, we may apply Proposition \ref{prop:decompcont} to extend the decomposition for some amount $t_\star\leq t_2$ also beyond $T_\ast$.  

By assumption,
\begin{equation*}
\sqrt{\abs{c(t)-\co}}+\norm{v(t)}_\h{1} \leq K_\star \eps_\ast\leq \frac{1}{2}\eps_{\alpha\beta}\quad \text{for $t< T_\ast$}
\end{equation*}
Again, by continuity, these remains bounded by $\eps_{\alpha\beta}$ until some time $t_3\in (0,t_\star)$ beyond $T_\ast$, so \eqref{eq:alphabound} and \eqref{eq:betabound} also persist beyond $T_\ast$.

We may now apply Proposition past $T_\ast$.  This gives
\begin{equation*}
\sqrt{\norm{w(\cdot,t)}_\h{1}} +\sqrt{\abs{c(t)-\co}} +\abs{\theta(t)-\theta_0}+ \abs{1-\frac{\co}{c(t) - \thetadot(t)}}+ \norm{v(\cdot,t)}_\h{1}\leq K_\star \eps_\ast \leq \frac{1}{3}\delta_\star
\end{equation*}
for $t\leq T< T_\ast$. As $\sqrt{\abs{\Delta\N}}$ is time invariant and smaller than $\frac{1}{2}\delta_\star$,
\begin{equation*}
\sqrt{\abs{\Delta\N}}+\sqrt{\norm{w(\cdot,t)}_\h{1}} +\sqrt{\abs{c(t)-\co}} +\abs{\theta(t)-\theta_0}+ \abs{1-\frac{\co}{c(t) - \thetadot(t)}}+ \norm{v(\cdot,t)}_\h{1}\leq K_\star \eps_\ast \leq \frac{5}{6}\delta_\star
\end{equation*}
for $t\leq T$, uniformly in $T<T_\ast$.  But all of these functions are continuous for $t\in [0, T_\ast+ t_3]$; so for some $t_4>0$, this expression remains bounded by $\delta_\star$.  This contradicts $T_\ast <\infty$.  So a solution exists for all time, satisfying \eqref{eq:alphabound}, \eqref{eq:betabound}, along with a decomposition and \eqref{eq:apriori1}.

Since we may then apply Proposition \ref{prop:apriori} for all time, we will always have \eqref{eq:apriori3}. By virtue of $K_\star\eps_\ast< \delta_\star\leq \delta_\mathcal{B}$, we will be able to invert the matrix $\mathcal{B}(t)$ for the modulation equations, \eqref{eq:modulation2}.  Therefore $\abs{\dcdt(t)}+\abs{\thetadot(t)}\leq K \eps e^{-\kappa t}$ and
\[
\lim_{t\to \infty} c(t) = c_\infty
\]
exists, and if define
\[
\lim_{t\to\infty} \paren{\theta(t) + \int_0^t \paren{c(s) - c_\infty }ds} = \theta_\infty
\]
then
\begin{equation*}
\begin{split}
\norm{\f(\cdot,t)-\f_{c_\infty}(\cdot - c_\infty t + \theta_\infty)}_\h{1} &\leq K_\star \eps + \norm{\f_{c_\infty}\paren{\cdot - c_\infty t + \theta_\infty}-\f_{c(t)}\paren{\cdot - \textstyle\int_0^t c(s)ds +\theta(t)}}_\h{1}\\
&\leq K_\star \eps +K\abs{ c(t)-c_\infty} + \norm{\dy\f_{c_\infty}}_\h{1}\abs{\theta(t) + \int_0^t \paren{c(s) - c_\infty }ds-\theta_\infty}\\
&\leq K_\ast \eps
\end{split}
\end{equation*}
Similarly 
\[
\norm{\f(\cdot +c_\infty t - \theta_\infty,t)-\f_{c_\infty}}_{H_a^1} \leq K_\ast \eps e^{-\kappa t}
\]\qed
\subsection{Remarks}
This proof is equally applicable in the Hamiltonian case, $n+m=0$, for values of $c$ not in the discrete set, $E$, of points for which $A_a$ has an imaginary eigenvalue.  

\section{Summary and Discussion}
\label{sec:discussion}

We have thus shown that in the space $H^1 \cap H_a^1$, the solitary waves are asymptotically stable.  This dovetails with an extension of global existence to data in a neighborhood of the solitary waves.  In the Hamiltonian case, we can extend it beyond $c_\star$ via analytic continuation, as was done in \cite{Miller96}, and this is analytically verified for $n=2$, with computations in \cite{Simpson07a} suggesting it is true for all $n>1$.  Furthermore, to the extent that we will accept a computation of the Evans function as proof, our result generalizes to large amplitude solitary waves with $c> c_\star$.    

To our knowledge, this is the first result for which asymptotic stability is established for a conservative PDE in the absence of a variational principle.

Open problems include a weakening of the assumption of exponential decay on the perturbation.  This might be accomplished through the use of an algebraic spatial weight, which would require the perturbation to decay algebraically rapidly.  Yet less restrictive would be to use the approach of F. Merle and his colleagues, \cite{Martel00,ElDika03,Mizumachi04,Martel05}.  However, there is a tradeoff in both of these approaches; weakening assumption on the spatial decay rate of the perturbation, weakens what can be proved about the rate at which the perturbation decays in time.

Finally we remark that the multi-dimensional case of \eqref{eq:magma3d} is wide open.  While there is no existence theory for the two- and three-dimensional problems, perhaps a similar approach, of working in a neighborhood of a solitary wave, could be applied, proving both existence and stability.

\appendix

\section{Properties of Solitary Waves}

\subsection{Analyticity}
\label{appendix:analyticity}
Here we provide a proof of Corollary \ref{cor:solanalyticity}.  Let us restate the crucial theorem,
\begin{thm}Corollary 4.1.6 of \cite{Bona:1997fk}
\label{thm:bona}
Suppose that $f$ is a solution of the convolution equation $f = K \ast G(f)$ such that $f\in L^2 \cap L^\infty$ and $\lim_{\abs{x} \to \infty} f(x) = 0$.  If the Fourier transform $\hat{K}$ of the integral kernel $K$ satisfies the decay condition $\abs{\hat{K}(\xi)}\leq A_1(1+ A_2 \abs{\xi}^m)$ for some constants $A_1, A_2 >0$ and $m\geq 1$, and $G$ is infinitely differentiable function whose domain contains the range $R(f)$ of $f$, having all its derivatives bounded on $R(f)$ and satisfying the condition $G(0)=0$, then $f, G(f) \in H^\infty$.  In addition, if $G$ is an analytic function on an open set $U$ containing $R(f)$, $G$ is continuous up to the boundary of $\partial U$ of $U$ and 
\begin{equation}
\dist \paren{\partial U, R(f)} > \norm{K}_{L^2}
\label{eq:analyticity-distance}
\end{equation}
then there exists a constant $\sigma_0>0$ such that $f$ and $G(f)$ both have analytic extensions to the strip $\left \{ z \in \mathbb{C} : \abs{\Im z}< \sigma_0\right\}$.
\end{thm}

 Let $u_c = \f_c -1$.  By Theorem \ref{thm:solexist} and Corollary \ref{cor:soldecay}, $u_c$ is positive, in $L^\infty \cap L^2$, and decays exponentially fast at $\pm \infty$.  Using equations \eqref{eq:solwave-a} and \eqref{eq:solwave-b}, the equation may be written as
\begin{equation}
\label{eq:ansolwave1}
-\gamma^2 u_c + \dx^2 u_c + \int_0^{u_c} \partial_{\tau}^3 F_2(1+\tau;c)\frac{\paren{u_c+1-\tau}^2}{2}d\tau  =0
\end{equation} 
Define
\begin{equation}
G(z) =\int_0^{z} \partial_{\tau}^3 F_2(1+\tau;c)\frac{\paren{z+1-\tau}^2}{2}d\tau
\end{equation}
Taking a Fourier transform of \eqref{eq:ansolwave1}, the equation is
\[
-\gamma^2 \widehat{u_c}(\xi) - \xi^2 \widehat{u_c}(\xi) + \widehat{G(u_c)}(\xi) = 0
\]
This becomes the nonlinear convolution equation
\begin{align}
u_c(x) &= K\ast G(u_c) (x)\\
\hat{K}(\xi) &= \frac{1}{\gamma^2+ \xi^2}
\end{align}
For purposes of satisfying \eqref{eq:analyticity-distance}, let us take
$\tilde{K}(x) =\alpha K(x)$ and $\tilde{G}(z) = \alpha^{-1} G(z)$ for $\alpha>0$, $\alpha$ to be determined.  Under this trivial scaling, $u_c = \tilde{K} \ast \tilde{G}(u_c)$.

$\widehat{\tilde{K}}$ satisfies the decay estimate for Theorem \ref{thm:bona} .  $\tilde{G}(z)$ will have a singularity at $z=-1$, but is otherwise analytic.  The range of $u_c$ is the finite segment
\[
R(u_c) = [0, u_{\max}]
\]
and $\tilde G$ is infinitely differential there, with all derivatives bounded.  $G(0)=0$.  Hence, the first part of the Theorem \ref{thm:bona} applies; $u_c$ and $\tilde{G}(u_c)$ are in $H^\infty$.

\begin{figure}
\begin{center}
\includegraphics[width=2in]{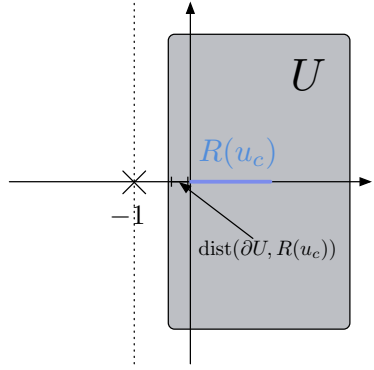}
\caption{A plot of a possible domain $U$, and the range of $u_c$, $R(u_c)$.  Note that the distance between these sets as drawn is the distance into the left hand side of the complex plane that $U$ extends, and that any such ovoid will be acceptable so long as it stays to the right of $\Re z =-1$.}
\label{fig:analyticity-fig}
\end{center}
\end{figure}

Now, consider the set $U$ in figure \ref{fig:analyticity-fig}.  In this figure, the $\dist(\partial U, R(u_c))$ is the distance $U$ stretches into the left half-plane.
\[
\norm{\tilde K}_2= \alpha \sqrt{\frac{\pi}{2}}\gamma^{3/2}
\]
Picking $\alpha$ so small that the norm is less than $1$, we can find a $U$ such that the distance between $\partial U$ and $R(u_c)$ exceeds $\norm{\tilde K}_2$, satisfying \eqref{eq:analyticity-distance}, and proving analyticity in a strip.\qed

\subsection{Continuity as a Function of Speed}
\label{sec:speed-continuity}

Consider the functional
\begin{equation}
\mathcal{F}\bracket{c,u} = \dx^2 u + F_2(1+u;c)
\end{equation}
as a mapping from $\h{2} \times \R \to L^2$.  The solitary wave $u_c =\f_c-1$ satisies $\mathcal{F}\bracket{c,u_c}=0$.  
Using this functional, we prove Corollary \ref{cor:solcontinuity} via the implicit function theorem.  Given a particular $\hat{c}>n$, set $\hat{u}=u_{\hat{c}}$.

Let $H^2_\even$ and $L^2_\even$ be the subspaces of $H^2$ and $L^2$, respectively, of only even functions.  Define the sets
\begin{align*}
M_0&= \paren{\hat{c}-\frac{\hat{c}-n}{2}, \hat{c}+\frac{\hat{c}-n}{2}}\subset \R \\
N_0&= \left \{u \in H^2_\even: \quad \norm{u-\hat{u}}_\h{2} \leq \frac{1}{2} \right\}\subset H^2_\even\\
Z&=  L^2_\even
\end{align*}
Note that set $ M_0 $ is bounded away from zero and all of the functions in $N_0$ are uniformly bounded from below by $\frac{1}{2}$.  Thefore $\mathcal{F}$ is well defined on $M_0 \times N_0$ and will be a $C^1$ mapping on this set into $Z$.

Set
\begin{align*}
T &= \frac{\delta \mathcal{F}}{\delta c}\bracket{\hat{c}, \hat{u}} = \partial_c F_2(1+\hat{u}; \hat{c})\\
S &= \frac{\delta \mathcal{F}}{\delta u}\bracket{\hat{c}, \hat{u}} = \dx^2 +\partial_u F_2(1+\hat{u}; \hat{c})
\end{align*}
$T$ and $S$ are bounded operators on $\R \to L^2_\even$ and $H^2_\even \to L^2_\even$, respectively.  

Let $f \in L^2_\even$ and  consider the problem $S u = f$.  As an elliptic problem, this has a solution provided $f \bot \ker \paren{S^\dagger}$.  Note that  $S \dx \hat{u} = 0$.  $S$ is self-adjoint, has smooth coefficients, and is in one spatial dimension this is the unique element of the kernel.  But $\dx \hat{u}$ is an odd function, hence $f$ is orthogonal to it and the equation has a solution $u$ satisfying a bound
\[
\norm{u}_{H^2} \leq K \norm{u}_{L^2}
\]
Because the coefficients in $S$ and the right-hand side, $f$, are all even functions, $\tilde{u}(x) =u(-x)$ also solves $Su = f$.  By the uniqueness of the solution, $u = \tilde u$, so $u$ is an even function.  Therefore $u\in H^2_\even$ and the map $S: H^2_\even \to L^2_\even$ is onto with bounded inverse.

The kernel of $S$, restricted to $u\in H^2_\even$ is trivial, so the implicit function theorem may be applied to conclude the existence of a function $\mathcal{G}: M_1\to H^2_\even$, $\hat{c} \in M_1 \subset M_0$, such that
\[
\mathcal{F}\bracket{c, \mathcal{G}(c)}=0
\]
for all $c\in M_1$.  The mapping $\mathcal{G}$ is $C^1$.  For any such $c$, 
\[
\dx^2 \mathcal{G}(c) + F_2(\mathcal{G}(c)+1;c)=0
\]
the solitary wave equation.  Therefore $\mathcal{G}(c)= u_c=\f_c-1$, and the mapping $c\mapsto \f_c-1$ is $C^1(\R; \h{2})$.  The analyticity of the mapping may be proven by checking the analyticity of $\mathcal{F}$ in a neighborhood of $(\hat{c},\hat{u})$.

To prove continuity in $H^2 \cap H^2_a$, the proof is similar.  Fixing $a< \frac{1}{2}$, and taking $\hat{c}\in (n/(1-4a^2),\infty)$, we let $M_a = M_0 \cap(n/(1-4a^2),\infty)$, $N_a = N_0 \cap H^2_a$, and $Z_a = Z \cap L^2_a$.  We must check that $S: H^2_\even \cap H^2_a \to L^2_\even \cap L^2_a$ is onto with bounded inverse.  This is accomplished using the previous result and studying $S_a = D_a^2 + \partial_u F_2(1+\hat{u}; \hat{c})$ on $H^2_\even \to L^2_\even$.

\section{Perturbation Expansions}
\label{sec:expansions}
Here we provide some explicit calculations, including those for Proposition \ref{prop:perturbation}.

Note the expansions
\begin{align*}
\f^n &= \f_c^n + n \f_c^{n-1} v + f_n[\f_c,v]v\\
\f^m &= \f_c^m + m \f_c^{m-1} v + f_m[\f_c,v]v\\
f_p[a,b] &= \int_0^1 \bracket{p\paren{a + \tau b}^{p-1} - pa^{p-1}} d\tau
\end{align*}
and
\begin{align}
\label{eq:operatorexpansion}
\begin{split}
H_\f\inv&= H_{\f_c}^{-1}  - H_{\f_c}^{-1} B[n\f_c^{n-1}v+ f_n[\f_c,v]v,m\f_c^{m-1} v+ f_m[\f_c,v] v]H_{\f_c}^{-1}\\
&\quad+ H_{\f_c}^{-1}B[\f^n-\f_c^n, \f^m-\f_c^m]\paren{H_{\f}^{-1}-H_{\f_c}^{-1}}
\end{split}
\\
\label{eq:variablelinearoperator}
B[a,b]u &=-\dx\paren{a \dx u} + b u
\end{align}

Recall \eqref{eq:magma}
\begin{equation*}
\dt \f=\dcdt \dc \f_c + \paren{\thetadot-c}\dy\f_c + \paren{\thetadot-c}\dy v + v_t=-\paren{\f_c+v}^n H_{\f_c+v}^{-1} \dy \paren{\f_c+v}^n
\end{equation*}
Using the above expansions and the solitary wave equation, $c\dy\f_c = \f_c^m H_{\f_c}\inv\dy \paren{\f_c^n}$, this may be expanded into:
\begin{align}
\label{eq:linear-part}
v_t &= \f_c^m H_{\f_c}\inv \dy\bracket{-c\f_c^n \dy^2\paren{\f_c^{-m} v}+ cv -n\f_c^{n-1}v - cn\f_c^{n-1}\dy\paren{\f_c^{-m}\dy\f_c}v}\\
\label{eq:modulation-part}
&\quad-\thetadot \dy v -\dcdt \dc \f_c -\thetadot \dy\f_c\\
\label{eq:nonlinear-part}
\begin{split}
&\quad-f_m[\f_c,v]v H_{\f_c}\inv \dy (\f^n) +m\f_c^{m-1} v H_{\f_c}\inv\paren{H_\f -H_{\f_c}}H_\f\inv\dy(\f^n)  \\
&\quad-m\f_c^{m-1}v H_{\f_c}\inv\dy\paren{\f^n-\f_c^n}-\f_c^m H_{\f_c}\inv \dy \paren{f_n[\f_c,v]v} \\
&\quad+\f_c^mH_{\f_c}\inv B[f_n[\f_c,v]v,f_m[\f_c,v] v]H_{\f}\inv\dy\paren{\f^n}\\
&\quad-\f_c^m  H_{\f_c}\inv B[n\f_c^{n-1}v,m\f_c^{m-1} v]H_{\f_c}\inv\paren{H_\f-H_{\f_c}}H_\f\inv\dy\paren{\f^n}\\
&\quad+\f_c^m{H_{\f_c}\inv B[n\f_c^{n-1}v,m\f_c^{m-1} v]H_{\f_c}\inv}\dy\paren{\f^n-\f_c^n}
\end{split}
\end{align}
\eqref{eq:linear-part} is a linear term. \eqref{eq:modulation-part} will decay to zero as $\theta(t)$ and $c(t)$, the modulating parameters, approach their asymptotic limits.  \eqref{eq:nonlinear-part} is purely nonlinear in $v$.

We define $\mathcal{F}_1[v;\f_c]$, the term nonlinear in $v$, as
\begin{equation}
\label{eq:unweighted-nonlinear-term1}
\begin{split}
\mathcal{F}_1[v;\f_c]&=-f_m[\f_c,v]v H_{\f_c}\inv \dy (\f^n) +m\f_c^{m-1} v H_{\f_c}\inv\paren{H_\f -H_{\f_c}}H_\f\inv\dy(\f^n)  \\
&\quad-m\f_c^{m-1}v H_{\f_c}\inv\dy\paren{\f^n-\f_c^n}-\f_c^m H_{\f_c}\inv \dy \paren{f_n[\f_c,v]v} \\
&\quad+\f_c^mH_{\f_c}\inv B[f_n[\f_c,v]v,f_m[\f_c,v] v]H_{\f}\inv\dy\paren{\f^n}\\
&\quad-\f_c^m  H_{\f_c}\inv B[n\f_c^{n-1}v,m\f_c^{m-1} v]H_{\f_c}\inv\paren{H_\f-H_{\f_c}}H_\f\inv\dy\paren{\f^n}\\
&\quad+\f_c^m{H_{\f_c}\inv B[n\f_c^{n-1}v,m\f_c^{m-1} v]H_{\f_c}\inv}\dy\paren{\f^n-\f_c^n}
\end{split}
\end{equation}
The operator $S$ is given by:
\begin{equation}
\label{eq:modulation-operator}
\begin{split}
S[\co,c,\thetadot] &=m\co \paren{ \f_\co^{-1}\dy \f_\co- \frac{c}{c-\thetadot}\f_c^{-1} \dy\f_c }\\
&\quad+ n\set{\f_\co^m H_\co\inv \dy\bracket{\paren{\f_\co\inv - \co(\f_\co\inv-1)}\cdot}-\frac{\co }{c-\thetadot} \f_c^m H_{\f_c}\inv \dy\bracket{\paren{\f_c\inv - c(\f_c\inv-1)}\cdot}}\\
&\quad-\co m\set{\f_\co^m H_{\f_\co}\inv \bracket{\f_\co^{-1}\dy \f_\co\cdot}-\frac{c}{c-\thetadot} \f_c^m H_{\f_c}\inv \bracket{\f_c^{-1} \dy\f_c\cdot}}
\end{split}
\end{equation}
Finally, the terms making up $\mathcal{G}_1$ from \eqref{eq:weighted-perturbation-equation}:
\begin{align}
\label{eq:weighted-nonlinear-term1}
\mathcal{G}_1 &= \tilde\G_1+ \tilde\G_2 + \tilde\G_3 + \tilde\G_4 + \tilde\G_5 + \tilde\G_6 + \tilde\G_7\\
\tilde\G_1&= -f_m[\f_c,v]w H_{\f_c}\inv \dy (\f^n)\\
\tilde\G_2&=m \f_c^{m-1} w H_{\f_c}\inv\paren{H_\f - H_{\f_c}}H_\f\inv\dy \paren{\f^n}\\
\tilde\G_3&=-m\f_c^{m-1}w H_{\f_c}\inv \dy\paren{n\f_c^{n-1}v + f_n[\f_c,v]v}\\
\tilde\G_4&=  -\f_c^m H_{\f_c,a}\inv D_a \paren{f_n[\f_c,v]w}\\
\tilde\G_5&=\f_c^m H_{\f_c,a}\inv B_a\bracket{f_n[\f_c,v]w,f_m[\f_c,v] w} H_\f\inv\dy\paren{\f^n}\\
\tilde\G_6&= -\f_c^m H_{\f_c,a}\inv B_a\bracket{n\f_c^{n-1}w,m\f_c^{m-1} w}H_{\f_c}\inv\paren{H_\f-H_{\f_c}}H_\f\inv\dy\paren{\f^n}\\
\tilde\G_7&=\f_c^m{H_{\f_c,a}\inv B_a\bracket{n\f_c^{n-1}w,m\f_c^{m-1} w}H_{\f_c}\inv}\dy\paren{\f^n-\f_c^n}
\begin{split}
\end{split}
\end{align}

The difference between $A_a$ and $A_a^\infty$ may be written as
\begin{equation}
\label{eq:lineardifference}
\begin{split}
A_a - A_a^\infty &=-c m \f_c^{-1} \dy\f_c-(\f_c^m-1) H_{\f_c,a}\inv D_a\paren{n\f_c^{n-1} \cdot }\\
&\quad+c m\f_c^m H_{\f_c,a}\inv\paren{\f_c^{-1}\dy\f_c \cdot}-c n\f_c^m H_{\f_c,a}\inv D_a\bracket{(1-\f_c\inv)\cdot}+H_{1,a}\inv D_a\bracket{n(1-\f_c^{n-1})\cdot}\\
&\quad+H_{1,a}\inv \paren{\f_c^m-1}H_{\f_c,a}\inv D_a\paren{n\f_c^{n-1}\cdot}+H_{1,a}\inv D_a\paren{1-\f_c^n} D_a H_{\f_c,a}\inv D_a\paren{n\f_c^{n-1}\cdot}
\end{split}
\end{equation}
In the space weighted by $\f_c(x)^{-m}$, this difference is
\begin{equation}
\label{eq:lsdifference}
\begin{split}
\tilde{A}_a - \tilde{A}_a^\infty &= n H_{1,a}^{-1} D_a \bracket{\paren{1-\f_c^{m-1}}\cdot}+ n H_{1,a}^{-1} D_a\paren{\f_c^m-1} H_{1,a}^{-1}\paren{\f_c^{m-1}\cdot}\\
&\quad-n H_{1,a}^{-1} \bracket{D_a\paren{\f_c^m-1}} H_{1,a}^{-1}\paren{\f_c^{m-1}\cdot} -nH_{1,a}^{-1} \paren{\f_c^m-1} H_{1,a}^{-1}\paren{\f_c^m-1}H_{\f_c,a}^{-1}D_a\paren{\f_c^{m-1}\cdot}\\
&\quad +nH_{1,a}^{-1} \paren{\f_c^m-1} H_{1,a}^{-1}D_a\paren{\f_c^n-1}D_a H_{\f_c,a}^{-1}D_a\paren{\f_c^{m-1}\cdot}\\
&\quad-n H_{1,a}^{-1}D_a\paren{\f_c^n-1}D_a H_{\f_c,a}^{-1}D_a\paren{\f_c^{m-1}\cdot} +cm H_{1,a}^{-1}\paren{\f_c^{m-1}\dy\f_c \cdot}\\
&\quad+ cm H_{1,a}^{-1}\paren{\f_c^m-1}H_{\f_c,a}^{-1}\paren{\f_c^{m-1}\dy\f_c \cdot}-cm H_{1,a}^{-1}D_a\paren{\f_c^n-1}D_a H_{\f_c,a}^{-1} \paren{\f_c^{m-1}\dy\f_c \cdot}\\
&\quad+ cn H_{1,a}^{-1}D_a\bracket{\f_c^m\paren{\f_c^{-1}-1}\cdot} + cn H_{1,a}^{-1}\paren{\f_c^m-1}H_{\f_c,a}^{-1}D_a\bracket{\f_c^m\paren{\f_c^{-1}-1}\cdot}\\
&\quad-cn H_{1,a}^{-1}D_a\paren{\f_c^n-1}D_a H_{\f_c,a}^{-1}D_a\bracket{\f_c^m\paren{\f_c^{-1}-1}\cdot}
\end{split}
\end{equation}


\section{Analysis of the Characteristic Polynomial}
\label{polynomial_proof}
In this section we prove that \eqref{eq:charpoly}, $(\lambda-c\mu)(1-\mu^2)+n\mu=0$, has a unique root of minimal real part on a slit half plane
\[
 \{\lambda : \Re \lambda >-\lambda_0\} \backslash (-\lambda_0, -\tilde{\Omega}(\gamma)]
\]

There are two ways that this could be false; there could either be a multiple root or two roots with the same real part, but differing imaginary parts.  As previously noted, we will have a unique root of minimal real part for $\lambda$ in the closed right half plane, so we need only concern ourselves with $\Re \lambda <0$.

We will identify a portion of the domain $\Re \lambda <0$ for which there are neither multiple roots nor complex roots with the same real part.  Note this is a stricter condition than is needed, as the polynomial could have a double root for some $\lambda$, where the third root of $P(\mu)$ has a smaller real part than the multiple root.

Note that $P(\pm 1)$ never vanishes, hence $P(\mu)=0$ is equivalent to $R(\mu)=\lambda$, where
\begin{equation}
R(\mu)= c \mu + \frac{n \mu}{\mu^2 -1}
\end{equation}

\subsection{Roots of Order Greater than One}
We start with the possibility of a double or triple root, as this is very easy to rule out.  If $\mu$ is a multiple root, then in addition to $R(\mu) = \lambda$, we will also have
\[
\frac{d R}{d\mu} =c -n\frac{1+\mu^2}{(1-\mu^2)^2}=0
\]
which has solutions
\[
\mu=-\sqrt{\frac{2c+n\pm\sqrt{8cn +n^2}}{2c}}
\]
We have ignored the roots with a $+$ sign in front, as these will correspond to positive $\lambda$.  Note that they are all real, hence $\lambda$ will also be real.

The $\lambda$ one gets from the root 
\[
\mu_+=-\sqrt{\frac{2c+n+\sqrt{8cn +n^2}}{2c}}\leq -\frac{3\sqrt{3}}{2}n
\]
is decreasing in $c$.  The other root, 
\[
\mu_-=-\sqrt{\frac{2c+n-\sqrt{8cn +n^2}}{2c}}
\]
will map to $\lambda$ values
\begin{equation}
\lambda(\mu_-)=-\sqrt{\frac{1}{8}}\sqrt{8 c^2 +20 cn -n^2 - 8c \sqrt{n^2 + 8cn} - n \sqrt{n^2+8c n }}
\end{equation}
It can be checked that for $c>n>1$, $R\paren{\mu_+}< R\paren{\mu_-}$.
Therefore, for $\lambda > -\widetilde{\Omega}(c)$, with
\begin{equation}
\widetilde{\Omega}(c) =\sqrt{\frac{1}{8}}\sqrt{8 c^2 +20 cn -n^2 - 8c \sqrt{n^2 + 8cn} - n \sqrt{n^2+8c n }}
\label{eq:omega-constraint}
\end{equation}
$P(\mu)$ cannot have a multiple root.

\subsection{Roots of Differing Imaginary Part}
If $\mu_1 = \alpha + \imath \beta_1$ and $\mu_2= \alpha + \imath \beta_2$ are two roots of $P$, then
\[
R(\alpha+\imath\beta_1) = R(\alpha + \imath \beta_2)
\]
After matching real and imaginary parts in this expression, the three unknowns $\alpha, \beta_1, \beta_2$, must satisfy the two equations
\begin{align}
\label{eq:real-constraint}
\Re \lambda &= c \alpha + \frac{n\alpha(-1+ \alpha^2 + \beta_1^2)}{(-1+\alpha^2)^2 + 2(1+\alpha^2)\beta_1^2+ \beta_1^4}& &= c \alpha + \frac{n\alpha(-1+ \alpha^2 + \beta_2^2)}{(-1+\alpha^2)^2 + 2(1+\alpha^2)\beta_2^2+ \beta_2^4}\\ 
\label{eq:imag-constraint}
\Im \lambda &= c \beta_1 - \frac{n\beta_1(1+ \alpha^2 + \beta_1^2)}{(-1+\alpha^2)^2 + 2(1+\alpha^2)\beta_1^2+ \beta_1^4}& &=c \beta_2 - \frac{n\beta_2(1+ \alpha^2 + \beta_2^2)}{(-1+\alpha^2)^2 + 2(1+\alpha^2)\beta_2^2+ \beta_2^4}
\end{align}
Solving the \eqref{eq:real-constraint} for $\beta_2^2$ in terms of $\alpha$ and $\beta_1$, there are two families of solutions:
\begin{align}
\beta_2^2 &= \beta_1^2\\
\label{eq:beta2-beta1}
\beta_2^2 &=\frac{\paren{1-\alpha^2}\paren{3+\beta_1^2+\alpha^2}}{\alpha^2+\beta_1^2-1}
\end{align}
Without loss of generality, we assume $\beta_1 \neq 0$.  

Recall that $\lambda$ is imaginary if and only if $P(\mu)$ has a purely imaginary root, hence the condition $\Re \lambda <0$ ensures $\alpha \neq 0$.  Then, \eqref{eq:beta2-beta1} implies $0<\abs{\alpha}\leq 1$.  Furthermore, if $\abs{\alpha}=1$, then either the roots are conjugate or $\beta_2=0$.  But if $\beta_2=0$, then $\mu_2=1$, which we know is not a root of $P(\mu)$.

\subsubsection{Complex Conjugates}
When $\beta_1 = -\beta_2=\beta$, \eqref{eq:imag-constraint} implies that $\lambda$ is real and 
\begin{equation}
\label{eq:conjuate-constraint}
c - \frac{n(1+ \alpha^2 + \beta^2)}{(-1+\alpha^2)^2 + 2(1+\alpha^2)\beta^2+ \beta^4}=0
\end{equation}
which has roots $\beta^2$,
\begin{equation}
\beta^2= -1 -\alpha^2 +\frac{n}{2c} \pm \frac{\sqrt{n^2+16 c^2\alpha^2}}{2c}
\end{equation}
We may immediately rule out the negative root for $\beta^2$.  For $\beta^2>0$, $\alpha^2$ must satisfy
\begin{equation}
\label{eq:beta_constraint}
1 + \frac{n}{2c} -\frac{\sqrt{n^2+24c}}{2c} < \alpha^2<1 + \frac{n}{2c} + \frac{\sqrt{n^2+24c}}{2c}
\end{equation}

Using \eqref{eq:real-constraint}
\[
\lambda(\alpha) = \frac{n+8c\alpha^2-\sqrt{n^2+16c^2\alpha^2}}{4\alpha}
\]
Since we are only concerned with $\lambda <0$ here, $\alpha$ must, in addition to \eqref{eq:beta_constraint}, satisfy 
\begin{equation}
\label{eq:lambda_constraint}
\frac{n+8c\alpha^2-\sqrt{n^2+16c^2 \alpha^2}}{4\alpha}<0
\end{equation}

When $\alpha>0$, \eqref{eq:lambda_constraint} requires
\[
0<\alpha^2<\frac{c-n}{4 c}
\]
But $\frac{1}{4}\paren{c-n}/c<1 + n/(2c) -\sqrt{n^2+24c}/(2c)$, so complex conjugate roots with $\alpha > 0$ are not possible with $\lambda$ in the left half plane.  

When $\alpha<0$, \eqref{eq:lambda_constraint}, implies
\[
\alpha < -\frac{1}{2}\sqrt{\frac{c-n}{c}}
\]
to satisfy \eqref{eq:lambda_constraint}.  Consider $\alpha$ in the interval
\[
 \left[-\sqrt{1 + \frac{n}{2c} - \frac{\sqrt{n^2+8cn}}{2c}},-\frac{1}{2}\sqrt{\frac{c-n}{c}}\right)
\]
In this interval, there will not be complex conjugate roots, as it violates \eqref{eq:beta_constraint}.

$\lambda$, as a function of $\alpha$, is negative and increasing on this interval.  For $\lambda$ in the image of this interval, we may completely rule out complex conjugate roots.  The image of this interval is 
\[
\lambda \paren{ \left[-\sqrt{1 + \frac{n}{2c} - \frac{\sqrt{n^2+8cn}}{2c}},-\frac{1}{2}\sqrt{\frac{c-n}{c}}\right)} =\left [-\widetilde{\Omega}(c),0\right)
\]
Hence, for $0>\lambda > -\tilde{\Omega}$, one may rule out both multiple roots and complex conjugates.

\subsubsection{Non-Conjugate Complex Roots}
Consider the case of complex roots with the same real part, but imaginary parts such that $\abs{\beta_1}\neq \abs{\beta_2}$.  Squaring both sides of \eqref{eq:imag-constraint} and plugging in \eqref{eq:beta2-beta1} for $\beta_2^2$, we get a sixth order polynomial in $\beta_1^2$.  The roots, as functions of $\alpha$, are
\begin{align}
\label{eq:betaimag1}
\beta_1^2&=-1-2\alpha-\alpha^2\\
\label{eq:betaimag2}
\beta_1^2&=-1+2\alpha-\alpha^2\\
\label{eq:betarepeat1}
\beta_1^2&=1-\alpha^2-2\sqrt{1-\alpha^2}\\
\label{eq:betarepeat2}
\beta_1^2&=1-\alpha^2+2\sqrt{1-\alpha^2}\\
\label{eq:beta-root-five}
\begin{split}
\beta_1^2&=\frac{(n/c)^2+4(n/c)(1-\alpha^2)-8(1-\alpha^4)}{8(1-\alpha^2)}\\
&\quad-\frac{\sqrt{((n/c)^2-16(1-\alpha^2))((n/c)^2-8(1-\alpha^2)(2\alpha^2-(n/c)))}}{8(1-\alpha^2)}
\end{split}
\\
\label{eq:beta-root-six}
\begin{split}
\beta_1^2&=\frac{(n/c)^2+4(n/c)(1-\alpha^2)-8(1-\alpha^4)}{8(1-\alpha^2)}\\
&\quad+\frac{\sqrt{((n/c)^2-16(1-\alpha^2))((n/c)^2-8(1-\alpha^2)(2\alpha^2-(n/c)))}}{8(1-\alpha^2)}
\end{split}
\end{align}

\eqref{eq:betaimag1} and \eqref{eq:betaimag2} force $\beta_1$ to be imaginary, hence they can be ruled out.  Using \eqref{eq:beta2-beta1},  if $\beta_1$ is either \eqref{eq:betarepeat1} or \eqref{eq:betarepeat2}, then $\beta_2 = \pm \beta_1$, conjugate roots.

In the last two cases, if $\beta_1^2$ is to be real, then
\begin{equation}
\label{eq:radical-constraint}
{((n/c)^2-16(1-\alpha^2))((n/c)^2-8(1-\alpha^2)(2\alpha^2-(n/c)))}\geq0
\end{equation}
If we can find, that for $\lambda$ in the left half plane sufficiently close enough to the imaginary axis it is negative, we will be done.  \eqref{eq:radical-constraint} is negative at $\alpha =0$, so there exists a neighborhood of  the imaginary axis, such that complex non-conjugate roots may be ruled out.

The roots of the left hand side of  \eqref{eq:radical-constraint} are 
\begin{align}
\label{eq:alpha-root-one}
\alpha^2 &= 1- \frac{1}{16}\paren{\frac{n}{c} }^2= \frac{1}{16}(5-\gamma^2)(3+\gamma^2)\\
\label{eq:alpha-root-two}
\alpha^2 &= \frac{1}{2}+\frac{1}{4}\frac{n}{c} + \frac{1}{2}\sqrt{1-\frac{n}{c}}=\frac{1}{4}(3-\gamma)(1+\gamma)\\
\label{eq:alpha-root-three}
\alpha^2 &= \frac{1}{2}+\frac{1}{4}\frac{n}{c} - \frac{1}{2}\sqrt{1-\frac{n}{c}}=\frac{1}{4}(1-\gamma)(3+\gamma)
\end{align}
These are positive for $\gamma\in[0,1]$.  It may be checked that \eqref{eq:alpha-root-three} is the smallest for all $\gamma$.  Hence the root of \eqref{eq:radical-constraint} such that $\mu$ will be closest to the imaginary axis is,
\begin{equation}
\label{eq:radical-constraint-root}
\alpha=-\frac{1}{2}\sqrt{(1-\gamma)(3+\gamma)}
\end{equation}
$\alpha$ larger than \eqref{eq:radical-constraint-root} and less than zero will yield a $\lambda$ that does not have non-conjugate complex roots.

Given $\lambda$, if $\mathcal{P}(\mu;\lambda)$ is to have two roots of same real part, $\alpha$, but differing imaginary part, then, by trying any of the last four roots for $\beta_1$ (both positive and negative square roots of \eqref{eq:beta-root-five} and \eqref{eq:beta-root-six} ), in \eqref{eq:real-constraint}  the real part of $\lambda$ and $\alpha$ are related by
\begin{equation}
\label{eq:real-part-map}
\Re \lambda = c\alpha \paren{2 - \frac{(n/c)}{4(1-\alpha^2)+ (n/c)}} 
\end{equation}
Note that
\begin{equation}
\frac{d \Re \lambda}{d\alpha}=c\paren{\frac{32(1-\alpha^2)^2 + 12 (n/c) -20 \alpha^2 (n/c) + (n/c)^2}{\paren{4(1-\alpha^2) + (c/n)}^2}  } 
\label{eq:real-part-derivative}
\end{equation}
and derivative has one negative root with $\abs{\alpha}<1$ at 
\begin{equation}
\label{eq:invertibility-constraint}
\alpha=-\sqrt{1 + \frac{5}{16}\frac{n}{c}-\frac{1}{16}\sqrt{64\frac{n}{c}+17\paren{\frac{n}{c}}^2}}
\end{equation}
Comparing \eqref{eq:radical-constraint-root} with \eqref{eq:invertibility-constraint}, at $\gamma=1$
\[
  \eqref{eq:invertibility-constraint}= -1 < 0 = \eqref{eq:radical-constraint-root}
\]
and at $\gamma = 0$
\[
  \eqref{eq:invertibility-constraint}= -\sqrt{\frac{3}{4}}= \eqref{eq:radical-constraint-root}
\]
In addition, one may check that \eqref{eq:invertibility-constraint} is increasing in $\gamma \in [0,1]$ while \eqref{eq:radical-constraint-root} is decreasing on the same interval.  Therefore
\[
-\sqrt{1 + \frac{5}{16}\frac{n}{c}-\frac{1}{16}\sqrt{64\frac{n}{c}+17\paren{\frac{n}{c}}^2}}\leq-\frac{1}{2}\sqrt{(1-\gamma)(3+\gamma)}
\]
for all $\gamma \in [0,1]$.

For $\Re \mu = \alpha$ in
\begin{equation}
\paren{-\frac{1}{2}\sqrt{(1-\gamma)(3+\gamma)}, 0}
\label{eq:alpha-interval}
\end{equation}
\eqref{eq:real-part-map} will be an increasing function in $\alpha$, \eqref{eq:real-part-derivative} is positive at $\alpha=0$.  On the interval \eqref{eq:alpha-interval}, the mapping is invertible and its image is
\begin{equation}
\paren{-\frac{1}{4}c\sqrt{1-\gamma}\paren{\gamma+3}^{3/2},0}
\label{eq:lambda-interval}
\end{equation}
Therefore if $\lambda$ has real part in the interval \eqref{eq:lambda-interval}, and $\mathcal{P}(\mu;\lambda)$ is to have non-conjugate complex roots, $\alpha$ must lie in \eqref{eq:alpha-interval}.  But such an $\alpha$ violates \eqref{eq:radical-constraint}, and we may conclude that there are no such roots.  Letting $-\lambda_0$ denote the $\Re \lambda$ value at the left end point of \eqref{eq:lambda-interval},
\begin{equation}
\lambda_0 =\frac{c}{4}\sqrt{1-\gamma}\paren{\gamma+3}^{3/2} = \frac{n}{4(1-\gamma^2)}\sqrt{1-\gamma}\paren{\gamma+3}^{3/2} 
\label{eq:lambda0-constraint}
\end{equation}

Hence for $\Re \lambda > -\lambda_0$, non-conjugate complex roots are not possible for $P(\mu)$.  Together with \eqref{eq:omega-constraint}.

\section{The Zero Eigenvalue}
\label{sec:lambda-zero}
In Section \ref{sec:kernel}, $\lambda=0$ was identified as an eigenvalue of multiplicity at least two.  Using the Evans function, the order of this eigenvalue may be related to the slope with respect to $c$ of the invariant functional $\N[\f_c]$.

Using the framework from Section \ref{sec:evansapplied}, set
\begin{eqnarray}
\bold{y}^+&=&\left(\begin{array}{c} \f_c^{-m}\dx\f_c  \\
\dx\paren{\f_c^{-m}\dx\f_c} \\ 
0 \end{array}\right)
\end{eqnarray}
\begin{equation}
\bold{z}^-= \paren{c \dx\paren{\f_c^{-m}\dx\f_c},\quad-c\f_c^{-m}\dx\f_c,\quad -\int_{-\infty}^x \f_c^{-n-m}\dx\f_c}
\end{equation}
These are solutions to the dynamical systems
\begin{equation*}
\dot{\bold{y}} = B(x,\lambda=0,\gamma) \bold{y}\quad\text{and}\quad\dot{\bold{z}}=-\bold{z}B(x,\lambda=0,\gamma)
\end{equation*}
Here, $\mu_1 = -\gamma$.  Employing the notation and formulation of the Evans function of \cite{Pego92},  by Proposition 1.6, parts 2 and 3, since
\[
\bold{y}^+ = O(e^{-\gamma x}) \quad\text{as $x\to +\infty$ and}\quad\bold{z}^- = O(e^{\gamma x}) \text{as $x\to -\infty$}
\]
$\bold{y}^+$ and $\bold{z}^-$ are scalar multiples of $\zeta^+$ and $\eta^-$, respectively. $\zeta^+$ and $\eta^-$, the solutions of the dynamical systems satisfying 
\begin{equation*}
\zeta^+ e^{\gamma x } \to \bold{v}^+\quad\text{as $x \to +\infty$ and}\quad\eta^- e^{-\gamma x} \to \bold{w}^-\quad\text{as $x\to -\infty$.}
\end{equation*}
with
\begin{align*}
B^\infty \bold{v}^+ &= -\gamma \bold{v}^+& \bold{w}^-B^\infty  &= -\gamma \bold{w}^-\\
\bold{v}^+&= \threedrowvec{1}{-\gamma}{0}^T & \bold{w}^- &=\paren{2 c\gamma^2}^{-1}\paren{c\gamma^2,\quad -c\gamma,\quad -1}
\end{align*}
allowing us to define the Evans function as 
\[
D(\lambda=0) = \eta^-(x,\lambda=0)\cdot \zeta^+(x,\lambda=0)
\]

From the properties of the solitary waves, discussed in Section \ref{sec:solwave}, there exists $\beta>0$ such that 
\[
\f_c^{-m}\dx\f_c e^{\mu x} \to -\beta \text{ as $x\to +\infty$}
\]
hence 
\begin{equation*}
\zeta^+ = \frac{1}{-\beta}\bold{y}^+ \quad\text{and}\quad\eta^- = \frac{1}{-2\mu c\beta}\bold{z}^-
\end{equation*}

$D(0)=0$ by inspection.  Using $(1.22)$ from \cite{Pego92}, it is trivial to compute that $\partial_\lambda D(0)=0$.  Taking the derivative of this formula, at $\lambda=0$,
\begin{equation}
\partial_\lambda^2 D(0)=-\int_{-\infty}^\infty \eta^-_\lambda A_\lambda \zeta^+ - \int_{-\infty}^\infty \eta^- A_\lambda \zeta^+_\lambda
\end{equation}
$\zeta^+_\lambda$ and $\eta^-_\lambda$ satisfy the ODEs
\begin{equation*}
\dot{\bold{y}}_\lambda = B \bold{y}_\lambda + B_\lambda \bold{y}\quad\text{and}\quad\dot{\bold{z}}_\lambda = -\bold{z}_\lambda B -\bold{z} B_\lambda
\end{equation*}
These problems are associated with the derivatives with respect to $\lambda$ of equations \eqref{eq:eigenvalue} and \eqref{eq:adj-eigenvalue} at $\lambda=0$,
\begin{align}
\label{eq:derivative-eigenvalue}
\dx L_cY_\lambda&=\bracket{I-\dx\paren{\f_c^n\dx\paren{\f_c^{-m} \cdot}}}Y\\
\label{eq:derivative-adj-eigenvalue}
-L_c^\star \dx Z_\lambda &=\bracket{I-\f_c^{-m}\dx\paren{\f_c^n\dx \cdot}}Z
\end{align}
through the identifcations
\begin{align*}
y_\lambda^{(1)} &= \f_c^{-m} Y_\lambda & z_\lambda^{(1)} &= c \dx\paren{\f_c^n \dx Z_\lambda} \\
y_\lambda^{(2)} &= \dx\paren{\f_c^{-m} Y_\lambda}& z_\lambda^{(2)} &= -c\f_c^n \dx Z_\lambda\\
y_\lambda^{(3)} &= L_cY_\lambda +\f_c^n\dx\paren{\f_c^{-m}Y}& z_\lambda^{(3)} &= - Z_\lambda
\end{align*}

For $\lambda=0$, \eqref{eq:eigenvalue}, \eqref{eq:adj-eigenvalue}, \eqref{eq:derivative-eigenvalue} and \eqref{eq:derivative-adj-eigenvalue} are related to the generalized kernels of $A$ and $A^\star$:
\begin{align*}
Y &= \dx \f_c & Y_\lambda &= -\dc \f_c\\
Z &= \int_{-\infty}^x \frac{\dx \f_c}{\f_c^{n+m}}dx & Z_\lambda &= - \int_{-\infty}^{x}\paren{L_c^\star}^{-1}\bracket{I-\f_c^{-m}\dx\paren{\f_c^n \dx\cdot}} \int_{-\infty}^{x} \frac{\dx \f_c}{\f_c^{n+m}} dx
\end{align*}
Then,
\begin{align*}
\zeta^+_\lambda&=\frac{1}{-\beta}\threedrowvec{-\f_c^{-m}\dc\f_c}{-\dx\paren{\f_c^{-m}\dc\f_c}}{-L_c\dc\f_c +\f_c^n\dx\paren{\f_c^{-m}\dx\f_c}}^T\\
\eta^-_\lambda&= \frac{1}{2\gamma c\beta}\threedrowvec{c \dx\paren{\f_c^{n}\dx Z_\lambda}}{-c\f_c^{n}\dx Z_\lambda}{-Z_\lambda}
\end{align*}
Finally, we compute 
\begin{equation}
\partial_\lambda^2D(0)=  \frac{1}{c\gamma \beta^2}\partial_c\mathcal{N}[\f_c]
\end{equation}

\bibliography{asymptotic_stability_article}

\end{document}

%% file: grsmacros.tex
\newcommand{\abs}[1]{\lvert#1\rvert}
\newcommand{\norm}[1]{\lVert#1\rVert}
\newcommand{\paren}[1]{\left(#1\right)}
\newcommand{\bracket}[1]{\left[#1\right]}
\newcommand{\set}[1]{\left\{#1\right\}}

\newcommand{\eps}{\epsilon}
\newcommand{\vareps}{\varepsilon}

\newcommand{\dz}{\partial_{z} }
\newcommand{\dx}{\partial_{x} }

\newcommand{\even}{\mathrm{even}}

\newcommand{\comp}{\mathrm{comp.}}
\newcommand{\cut}{{\mathrm{cut}}}

\newcommand{\ess}{\mathrm{ess}}
\newcommand{\kerg}{\ker_{\mathrm{g}}}

\newcommand{\kdv}{\mathrm{KdV}}
\newcommand{\dist}{\mathrm{d}}

\newcommand{\h}[1]{{H^{#1}}}
\newcommand{\ha}{{H^1_a}}
\newcommand{\hinta}{{H^1 \cap H^1_a}}

\newcommand{\ltwo}{{L^2}}
\newcommand{\linf}{{L^\infty}}
\newcommand{\inner}[2]{\left\langle #1,#2\right\rangle}
\newcommand{\twodcolvec}[2]{\begin{pmatrix} #1\\ #2\end{pmatrix}}
\newcommand{\twobytwo}[4]{\begin{pmatrix} #1 & #2\\ #3 & #4\end{pmatrix}}
\newcommand{\threedcolvec}[3]{\begin{pmatrix} #1\\ #2\\#3\end{pmatrix}}
\newcommand{\threedrowvec}[3]{\left( #1,\quad #2,\quad#3\right)}

\newcommand{\dt}{\partial_{t}}
\newcommand{\dc}{\partial_{c}}
\newcommand{\dxi}{\partial_{\xi}}

\newcommand{\dy}{\partial_{y}}
\newcommand{\f}{\phi}
\newcommand{\g}{\gamma}

\newcommand{\thetadot}{\dot{\theta}}
\newcommand{\co}{ {c_{0}}}
\newcommand{\ct}{{c(t)}}
\newcommand{\thetat}{{\theta(t)}}
\newcommand{\dcdt}{{\dot{c}}}

\newcommand{\inv}{^{-1}}
\newcommand{\G}{\mathcal{G}}
\newcommand{\N}{\mathcal{N}}
\newcommand{\R}{\mathbb{R}}
\newcommand{\C}{\mathbb{C}}

\newcommand{\calh}{\mathcal{H}}

\newcommand{\sech}{\mathrm{sech}}
\newcommand{\sign}{\mathrm{sign}}
\newcommand{\nb}{\noindent\textbf{N.B.}\quad}
\newcommand{\qed}{\begin{flushright}{$\blacksquare$}\end{flushright}}
\newcommand{\local}{\mathrm{local}}
\newcommand{\exist}{\mathrm{exist}}
\newcommand{\proof}{\noindent \textbf{Proof}:\quad}

\newtheorem{thm}{Theorem}[section]
\newtheorem{cor}[thm]{Corollary}
\newtheorem{lem}[thm]{Lemma}
\newtheorem{prop}[thm]{Proposition}

\newtheorem{rem}[thm]{Remark}